%% file: main.tex
\documentclass[11pt]{article}
\input{preamble}
\begin{document}


\title{\textbf{\LARGE{Fighting discrimination with reputation:}\\The case of online platforms.}\thanks{ We are grateful to Charles Angelucci, Mark Armstrong, Susan Athey, Maria Elena Bontempi, Jacques Cr\'emer, Daniel Ershov, Daniel Garrett, Jonas Hjort, Alex Imas, Marc Ivaldi, Bruno Jullien, Yassine Lefouili, Andrea Mantovani, Sarit Markovich, Nicolas Pistolesi, Imke Reimers, Patrick Rey, Mathias Reynaert, Tobias Salz, Mark Schankerman, Timothy Simcoe, Steven Tadelis and Jean Tirole for their valuable comments at various stages of the paper. We also thank participants to seminar sessions at Telecom ParisTech, Columbia University, Northeastern University, Boston University, Toulouse School of Economics and the IIOC, AFSE, EARIE, Jornadas de Economia Industrial, WIPE conferences for useful comments and suggestions. }\\ \vspace{0.2cm} \normalsize }

\author{Xavier Lambin\thanks{ESSEC Business School and THEMA, xavier.lambin@essec.edu.} \and Emil Palikot\thanks{Northeastern University, e.palikot@northeastern.edu; Emil Palikot acknowledges support from the European Research Council under the Grant Agreement no. 340903.}}
\date{\today}

\maketitle
\thispagestyle{empty} 
\setcounter{page}{0}
\begin{abstract}
On a large French ridesharing platform, new minority drivers earn 11.6\% less revenue than otherwise similar nonminority drivers; the gap nearly vanishes as they accumulate reviews. Reviews drive the convergence: when a railway strike exogenously raised demand and sped up review accumulation, minority entrants gained the most. We explain the pattern with an estimated model of passenger choice and driver career concerns. Passengers hold overly pessimistic priors about minority entrants---expecting substantially lower quality before the ride than they report after it. As a result, minority drivers cut introductory prices and exert extra effort to overturn those beliefs quickly. Counterfactuals show the cost of incorrect priors is high, and the reputation system strictly benefits minority drivers.
\begin{flushright} \textbf{JEL Classification: } J15, L14, L91 \qquad \end{flushright}
\end{abstract}
\newpage{}

\section{Introduction}



On many major online marketplaces, participants choose with whom to transact, and that discretion often leads to discrimination on the basis of race, ethnicity, or gender \citep{Edelman2014, Edelman2017, ge2020racial, luca2026evolution}. Yet the same platforms are built around reputation systems that aggregate reviews and ratings into signals of individual performance \citep{tadelis2016reputation}.\footnote{A key feature underlying the success of the ``sharing economy" is the efficacy of reputation systems in building trust across social divides. See a talk by Joe Gebbia, a co-founder of Airbnb: \href{https://www.youtube.com/watch?v=16cM-RFid9U}{https://www.youtube.com/watch?v=16cM-RFid9U}, last accessed June 03, 2026. Furthermore, Frederic Mazzella, BlaBlaCar CEO, claims that the company's reputation system creates a sense of trust almost comparable to the level of trust in friends'' (\cite{blablacar2016}).} These signals can substitute individual information for group-based priors and reward high-quality types, and so attenuate disparities. This corrective power is not guaranteed; it turns on how the parties to a transaction respond to the very ratings they generate. We ask whether reputation systems in fact play this corrective role, and how buyers and sellers behave strategically where they operate.

We study this question using data from a ridesharing platform, and find that discrimination is not a fixed disadvantage but one that erodes as drivers build reputations. Ethnic minority drivers with little or no reputation earn 11.6\% less revenue than otherwise similar nonminority drivers (SE $= 2.2$\%). This gap narrows steadily as drivers accumulate reviews and becomes small and not statistically significant beyond twenty reviews. The pattern points to two mechanisms: as reviews accumulate, passengers learn about individual quality and lean less on group-level priors, and drivers invest in the reputations that make such learning possible.


We collected data from BlaBlaCar, a large French ridesharing platform whose institutional features make both discrimination and reputation unusually visible. Unlike Uber or Lyft, BlaBlaCar matches passengers to long-distance trips. Passengers choose among posted rides knowing each driver's name, photo, and full review history before they book. This combination of salient identity and consequential, hours-long contact makes ethnic discrimination feasible in a way that short, anonymous urban rides do not. The reputation system is correspondingly rich: passengers see detailed records of prior trips, and drivers shape their own standing through effort and pricing decisions. Discrimination and reputation formation are thus both first-order on the platform, which makes it a useful setting for studying how information mitigates group-based disparities.

In a market defined as a day–route combination, we observe every available driver, their characteristics, posted prices, and seats sold, along with the number of times each listing was viewed. These view counts give us a measure of how many passengers were searching for a ride on the route, which lets us model the passenger's choice problem. Outcomes differ sharply across ethnic groups, and the gap survives a rich set of driver- and listing-specific controls. The disparity operates mainly through quantities, not prices: minority drivers post slightly lower fares, yet their listings draw fewer views and sell fewer seats. The gap is concentrated early in drivers' careers: minority entrants sell nearly 10 percent fewer seats than comparable nonminority entrants, before fading to near zero as reviews accumulate.

To identify the causal effect of reputation accumulation, we exploit the 2018 French railway strike as a natural experiment. From April 3 to June 28, 2018, SNCF workers struck on a rolling schedule, two days out of every five, in protest of plans to liberalize the European railway market. Because BlaBlaCar and rail are close substitutes for intercity travel, each strike day delivered a sharp, plausibly exogenous demand shock to ridesharing: platform traffic roughly tripled and booking requests rose sixfold. Drivers who happened to be on the road during a strike filled about 50 percent more seats and, as a byproduct, accumulated reviews faster than they otherwise would have. We exploit this variation in a difference-in-differences design, estimated with the doubly-robust estimator of \cite{sant2020doubly}, comparing drivers who drove on at least one strike day to those who did not, before versus after the strike window. Treated drivers earn higher post-strike revenue, and the effect is 61\%  larger for minority drivers. Reviews, in other words, are most valuable precisely for the drivers passengers initially judge with the most skepticism and have close to no information on their individual past performance.

New drivers appear to invest in their reputations, and minority drivers appear to invest the most. We find that drivers enter at a discount and earn unusually high early ratings: discounts exceed 8\% for drivers with no reviews and fall to about 1\% by the fifteenth review, while the share of five-star ratings declines from 76\% on a driver's first review to roughly 60\% at maturity (Section~\ref{sec:strategic}). These patterns are consistent with deliberate reputation building, but the reduced-form evidence cannot establish that they are, nor can it recover the objects that matter for welfare: reviews are noisy, and the beliefs passengers hold, the incentives drivers face, and the cost of building a reputation cannot be read directly off observed prices and ratings. To recover them, we develop a structural model of career concerns—the analytical core of the paper—that generalizes \cite{Holmstrom1999} along three dimensions central to our setting: (i) passengers may hold incorrect, group-specific priors; (ii) drivers compete for passengers; and (iii) price is a strategic instrument alongside effort. A driver, characterized by an intrinsic type (initially imperfectly observed) and a marginal cost, sets prices and effort to maximize lifetime payoffs; passengers choose among available drivers to maximize expected utility, forming beliefs from group-level priors and each driver's review history; and after a ride a passenger reports overall quality, which is a function of the driver's type, her effort, and a random shock, whose components are not separately observed. Quality reports update beliefs in later periods. Estimating the model lets us separate the information in reviews from noise, recover the beliefs that drive entry behavior, and quantify both what initial prejudice costs minority drivers and what the reputation system is worth to them.


The structural model allows us to separate two forces that the outcome gap conflates: discrimination, a property of how passengers choose, and belief-based partiality, a property of the priors they hold. Holding the two apart lets us study not only the gap in outcomes but the beliefs that generate it and the role reputation plays in correcting them. The model attributes the gap between what minority drivers are expected to deliver and what they actually deliver to incorrect priors. We estimate that the market expects a minority driver with no reviews to provide quality 3.06 on a five-point scale, even though these drivers are graded 4.92 on average after the trip. The market thus underestimates minority entrants by 1.86 grades, far exceeding both the realized early rating gap and the roughly one-tenth-grade gap observed among experienced drivers. 

These incorrect priors shape incentives. Because each additional review raises posterior beliefs about minority drivers by more than it does for nonminority drivers, minority drivers gain more from selling a seat and being reviewed, and they respond by pricing low. We estimate that minority entrants set introductory prices 7.2 percent below the static profit-maximizing level, compared with 4.7 percent for nonminority entrants. Effort responds to two opposing forces. Minority entrants expect lower future profits, which weakens the incentive to exert effort, but they face greater uncertainty about their type, which strengthens it, since early reviews move beliefs by more. On net the second force dominates: minority entrants exert more effort than nonminority entrants.

We propose three counterfactuals that bracket how much of the entry gap reflects beliefs and how much reputation can undo. In the first, we give passengers correct priors. Minority entry value rises by 5.8\%. Belief correction erases roughly half of the entry gap as unwarranted pessimism; the rest is a statistical-discrimination floor, the discount a Bayesian passenger still applies given the 0.1-grade true quality difference between the groups. In the second, we make the bias permanent, so reputation never revises the group prior. Minority drivers reduce their reputation investment; they cut entry effort by 5.4\% and raise introductory prices—and their entry value falls by 6.1\%, widening the wedge in expected discounted profit from €32.7 to €49.1. The two experiments mark a ceiling and a floor: correct priors leave only the statistical-discrimination residual, persistent bias lets reputation do no work, and in the data reputation drives minority drivers from the floor toward the ceiling.

Within our observation window, BlaBlaCar itself redesigned its rating system—replacing binary up-or-down ratings with a five-star scale and later relabeling its categories—and each change shifted the distribution of grades drivers receive (Appendix~\ref{system_changes}). Our third experiment exploits this lever and asks what rating design does to discrimination. In the model, a single parameter, $h_\epsilon$, sets how much weight each review carries in updating; we estimate it at 2.74 in the current system. We re-solve the equilibrium under a sharper system ($h_\epsilon=9$ and a noisier one ($h_\epsilon=1$). Sharper ratings let the market correct its priors faster and steepen the return to investing: minority entrants raise effort by 79\%, and the wedge narrows by a fifth, to €26.0. A noisier system reverses both—effort drops by 52\% and the wedge widens to €39.1. Rating design thus shifts the discrimination wedge in the expected direction but cannot close it: even the sharper system leaves the wedge at €26.0, well above the €16.9 that correcting the prior delivers. A better rating system speeds the correction of biased beliefs; it does not substitute for getting them right from the start.

\paragraph{Relation to literature:}

Economic outcomes differ persistently across ethnic groups \citep{kuznets1955economic, alesina2016ethnic}, and discrimination imposes sizable costs \citep{banerjee2004efficiently,hjort2014ethnic}. A growing literature shows these disparities migrate onto digital platforms, where names and photos make group identity salient at the moment of transaction: in short-term rentals \citep{Edelman2014, Edelman2017, laouenan2022can, kakar2018visible}, and in ridesharing and carpooling \citep{ge2020racial, farajallah2019drives, tjaden2018ride, carol2019can,donkor2026leveling}.\footnote{Not all measured disparities on online platforms reflect discrimination; a related literature traces them to differences in preferences and choices across groups. For example, \citet{cook2021gender} find that the roughly $7\%$ gender earnings gap among {Uber} drivers is explained entirely by experience (learning-by-doing), preferences over where to drive, and driving speed, rather than by customer discrimination.} The disparities are large and context-dependent: anti-Asian discrimination spiked at the onset of COVID-19 \citep{luca2026evolution}, while platform technology that rematches riders can blunt the effect of driver prejudice \citep{cobb2024ride}. We contribute to this literature by tracing how the ethnic revenue gap evolves over a driver's career and by exploiting a natural experiment, in the form of a railway strike, to show that reputation building can mitigate racial disparities.

Whether discrimination reflects tastes \citep{becker1971economics} or beliefs \citep{Phelps1972, Arrow1973, Coate1993} matters for how it responds to information. A growing body of work stresses that the beliefs themselves may be inaccurate rather than rational \citep{bohren2025inaccurate, bohren2019dynamics}, that they can be measured and separated from tastes \citep{coffman2021role, barron2025explicit, bohren2025systemic}, and that stereotypes and confirmation bias keep them from converging even under informative feedback \citep{coffman2024stereotypes, ruzzier2023discrimination}. Our contribution to this strand is to recover the prior beliefs themselves from a reputation system and to show they are systematically too pessimistic about minority drivers: the market underestimates an entrant's quality by far more than the true difference in type across groups.

If discrimination is belief-based, information should attenuate it \citep{pallais2014inefficient, agrawal2016does, bartovs2016attention}. The closest work to ours studies this directly in the sharing economy. In field experiments on {Airbnb}, \citet{cui2019reducing} show that guests with African-American-sounding names are $19$ percentage points less likely to be accepted, but that a single positive review renders the acceptance gap statistically indistinguishable from zero; \citet{laouenan2022can} similarly find that the ethnic gap among {Airbnb} hosts is consistent with statistical discrimination and shrinks as reviews accumulate. We build on these papers but differ in three ways. First, where they establish that reviews reduce discrimination, we recover the beliefs behind it: structurally estimating the market's prior, we show it is inaccurate; specifically, we find that the prior is too pessimistic about minority drivers and, thus, we decompose how much of the disparity is unwarranted pessimism versus a statistical-discrimination floor. Second, neither paper models how the discriminated party responds; we characterize the supply-side margin, showing that minority drivers actively invest in reputation through effort and low introductory prices. Third, we move from a static information treatment to the dynamics of reputation accumulation. We identify the causal effect of reviews from a railway-strike natural experiment, and using the estimated equilibrium model to value the reputation system and to evaluate its design.

Reputation is nonetheless no free fix. On the demand side, minorities accumulate ratings
more slowly \citep{abrahao2017reputation}, and ratings inflate over time
\citep{filippas2018reputation}. It is also costly on the supply side: minority drivers
overcome their initial disadvantage only by investing in reputation, forgoing
current revenue through low introductory prices and bearing the cost of elevated effort. We estimate the welfare implications of this investment, quantifying what overcoming the
initial prejudice costs them. The design of the rating system matters as well. Coarsening a
five-star scale to a binary one eliminated a racial rating gap \citep{botelho2025scale},
quality certification disproportionately helps minority hosts \citep{alyakoob2026market},
same-race endorsements offset host bias \citep{park2023fighting}, sellers may buy reputation
to signal quality \citep{li2020buying}, and greater visibility can cut both ways
\citep{son2025does}, a margin our counterfactuals on rating-system informativeness
speak to directly.

Methodologically, we build on the theory of career concerns \citep{Holmstrom1999} and employer learning \citep{Altonji2001, chiappori1999early}, generalizing the canonical model with incorrect, group-specific beliefs, competition among drivers, and price as a strategic instrument. A recent literature studies how learning sustains or unwinds discrimination: endogenous learning can make employer biases self-perpetuating \citep{lepage2021endogenous, benson2024learning}, statistical discrimination can be a failure of social learning that temporary affirmative action corrects \citep{komiyama2026statistical}, and treating hiring as exploration improves both quality and diversity \citep{li2026hiring}. Where \citet{Coate1993} and \citet{glover2017discrimination} show discrimination can be self-fulfilling, we document the opposite margin: conditional on entry, minority drivers respond to inaccurate priors by investing; they exert effort and cut introductory prices to overturn the inaccurate beliefs.



The rest of this paper is organized as follows: Section \ref{data} describes the institutional setting, the BlaBlaCar platform, and the data collection process. Section \ref{sec:descriptive} presents reduced-form evidence on ethnic disparities in economic outcomes and documents how these disparities evolve as drivers accumulate reviews. Section \ref{train} exploits exogenous variation in review accumulation generated by a demand shock to identify the causal effect of reputation on subsequent outcomes. Section \ref{sec:strategic} analyzes drivers' strategic pricing and effort decisions. Section  \ref{career_concerns} develops a dynamic model of passenger choice, belief formation, and drivers' career concerns, and characterizes the dynamics of effort, pricing, and discrimination. Section \ref{identification} discusses identification and estimation. Section \ref{results} presents the estimation results. Section \ref{counterfactual} uses the estimated model to quantify the effects of belief-based partiality and to evaluate counterfactual changes in beliefs and rating-system informativeness. Section \ref{conclusion} concludes.

\input{data_reduced_form_sections}

\input{model}

\input{structural_section}

\section{Conclusions} \label{conclusion}

This paper contributes to a long-standing discussion about the origins and persistence of discrimination. Using data from a large ridesharing platform, we show that ethnic disparities in economic outcomes are closely linked to incomplete information. Minority drivers with five or fewer reviews earn 11.6\% less revenue than observably similar nonminority drivers, but this gap shrinks to 3.3\% among experienced drivers. This pattern suggests that information plays an important role in shaping discrimination in this market: exploiting exogenous variation in review accumulation generated by a demand shock, we find that additional reviews causally improve the subsequent performance of minority drivers.

To understand the mechanism behind these patterns, we develop and estimate a dynamic model of passenger choice and driver career concerns. Passengers form beliefs about driver quality using both group-level priors and individual reputation histories, while drivers strategically choose effort and prices because current outcomes affect future demand. The model distinguishes between discrimination, a property of passenger behavior, and belief-based partiality, a property of passengers’ prior beliefs.

The estimates indicate that passengers hold excessively pessimistic beliefs about the quality of minority entrants. While the market expects a minority entrant to provide a quality level of only 3.06, the average realized rating over the first two rides is 4.92. These beliefs create incentives to invest in reputation. Minority drivers respond by exerting more effort and offering higher price discounts than comparable nonminority entrants. As reviews accumulate, posterior beliefs converge to true quality and the influence of group-level priors disappears. 

Counterfactual experiments highlight the economic importance of this mechanism. Correcting pessimistic prior beliefs increases minority drivers’ expected lifetime value at entry by 5.8\%. The resulting discrimination wedge in expected entry values ranges from €16.9 under correct beliefs to €49.1 when biased beliefs persist. More informative rating systems further strengthen reputation-building incentives and mitigate the effects of incorrect beliefs.

The main lesson of our analysis is that discrimination on this platform is, in large part, a problem of information; and one that platform design can shape. Our analysis does not identify where these biased priors come from, whether platform design or the visibility of identity contributes to their formation is beyond what our data can settle, but it does show that the platform governs how quickly reviews correct them. The levers we are able to evaluate therefore act on the speed of that correction. Two stand out: a more informative rating system narrows the entry-stage discrimination wedge by about a fifth in our counterfactuals, and because minority entrants earn less and accumulate reviews more slowly, helping them acquire informative reviews sooner reaches the drivers who bear the largest cost. Reputation systems do not erase the cost of incorrect beliefs, but they let a market rely less on group identity and more on observed quality. How best to accelerate that learning, without unintended consequences, remains an important question for future research.

\newpage
\bibliography{biblio}
\bibliographystyle{apalike}
\newpage
\appendix
\setcounter{page}{1}
\gdef\thepage{A\arabic{page}}
\input{new_appendix}

\end{document}

%% file: preamble.tex
\usepackage[top=0.8in, bottom=0.78in, left=0.87in, right=0.87in]{geometry}
\usepackage{setspace}
\usepackage[T1]{fontenc}
\usepackage{times}
\usepackage{booktabs}
\usepackage{rotating}
\usepackage{graphicx}
\usepackage[section]{placeins} 
\usepackage[large, bf]{caption}
\usepackage{subcaption}
\newcommand{\ellt}{\ell t}
\usepackage{pdfpages}
\usepackage{palatino}
\usepackage{pdflscape}
\usepackage{textcomp}
\usepackage{longtable}
\usepackage{nicefrac}
\usepackage{adjustbox}	
\usepackage[hyphens]{url}

\usepackage{natbib}
\bibpunct{(}{)}{;}{a}{,}{,}

\usepackage{makecell}

\usepackage{amsmath, amsfonts, amssymb, amsthm}

\usepackage{mathpazo} 
\usepackage[pdftex]{hyperref}
\hypersetup{colorlinks, citecolor=blue, filecolor=blue, linkcolor=blue, urlcolor=blue}

\newtheorem{prop}{Proposition}
\newtheorem{corollary}{Corollary}

\newtheorem{definition}{Definition}
\newtheorem{ass}{Assumption}

\def\urltilda{\kern -.15em\lower .7ex\hbox{\~{}}\kern .04em}

\renewcommand{\thesubfigure}{\Alph{subfigure}}

\usepackage{titlesec}

\titlespacing*{\section}{0pt}{1.5ex plus 1ex minus .2ex}{0.8ex plus .2ex}
\titlespacing*{\subsection}{0pt}{1.2ex plus 1ex minus .2ex}{0.8ex plus .2ex}

\usepackage{dcolumn}
\newcolumntype{d}[0]{D{.}{.}{5}}
\usepackage{mathtools}
\usepackage{tikz}
\usepackage{graphicx}
\usepackage[space]{grffile}

\newtheorem*{rem*}{\protect\remarkname}
\providecommand{\remarkname}{Remark}

\usepackage[official]{eurosym}
\usetikzlibrary{arrows,positioning} 
\usetikzlibrary{decorations.pathreplacing}
\usetikzlibrary{decorations.pathmorphing}
\usetikzlibrary{intersections}
\usepackage{float}
\usepackage[american]{babel}

\usepackage{tikz}
\usetikzlibrary{patterns}
\usepackage{adjustbox}
\usepackage{hyperref}

%% file: data_reduced_form_sections.tex
\section{Empirical context and data collection}\label{data}
\begin{figure}
\centering
\caption{Price dispersion on BlaBlaCar\label{prices_trips}}
\includegraphics[scale=0.4]{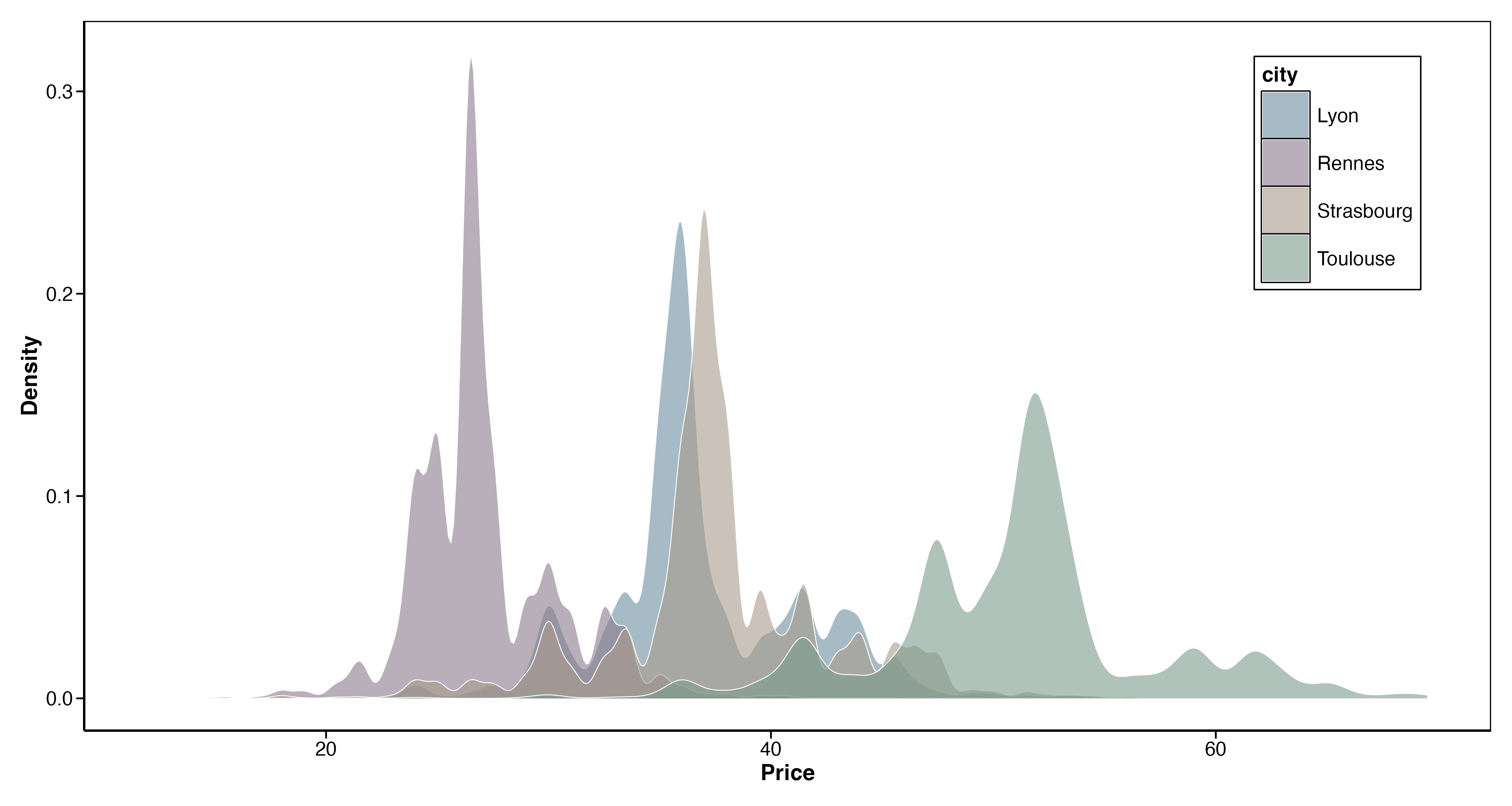}
\caption*{\textit{Notes:} Distribution of prices in euros on routes connecting Paris with Lyon, Rennes, Strasbourg, and Toulouse (bidirectional).}
\end{figure}

BlaBlaCar is an online marketplace for ridesharing established in France in 2006. The platform operates in 22 countries across Europe, as well as in Mexico, India, and Brazil, serving over 100 million active users.\footnote{\href{https://blog.blablacar.com/newsroom/news-list/blablacar-reaches-100-million-members-for-its-15th-anniversary}{https://blog.blablacar.com/newsroom/news-list/blablacar-reaches-100-million-members-for-its-15th-anniversary}} BlaBlaCar is particularly popular in France, where 1.5 million passengers use the service monthly. The platform differs from ride-hailing services such as Uber or Lyft in several important respects.
First, participation is restricted to nonprofessional drivers through limits on the number of seats and listings each driver can offer.\footnote{In 2019, after our sampling period, BlaBlaCar introduced BlaBlaBus, a separate professional bus service.} Drivers typically travel on a route for personal reasons and use the platform to defray travel costs. Second, BlaBlaCar specializes in long-distance intercity travel. In our data, the average trip spans 400 kilometers, implying several hours of interaction between drivers and passengers.
Third, drivers set their own prices. While BlaBlaCar provides a suggested price of 0.062 EUR per kilometer based solely on distance, drivers frequently deviate from this recommendation.\footnote{Prices are capped at 0.082 EUR per kilometer, though this constraint rarely binds.} Figure~\ref{prices_trips} illustrates substantial within-route price dispersion across several popular routes.

The platform's booking process provides opportunities for passengers to observe driver characteristics and reputation. When searching for a ride, passengers view a list of available drivers ranked by departure time. This initial display shows each driver's photo, name, average rating, basic ride details, and price. To access additional information, most importantly, the complete review history, passengers must click on a driver's profile.\footnote{Appendix~\ref{an_awesome_driver} provides examples of profile and listing pages.} After reviewing available options, passengers select their preferred listing and submit a booking request. Approximately half of drivers enable automatic acceptance of booking requests; the remainder manually approve or reject requests. Payment occurs upfront through BlaBlaCar's online system, with platform fees deducted from the passenger's payment.

BlaBlaCar encourages both drivers and passengers to leave reviews through multiple reminders. Each review comprises a numerical grade (1 to 5) and an optional textual comment. We collected both components for all reviews in our sample. A sentiment analysis of textual comments reveals a high correlation with numerical grades (documented in Appendix~\ref{system_changes}). Given this strong correlation, we focus our analysis on numerical grades, using the terms \emph{review}, \emph{rating}, and \emph{grade} interchangeably to refer to the 1-5 scale evaluation.\footnote{The review system features simultaneous revelation: users cannot view received reviews until they submit their own review or the two-week review window expires. Reviews become publicly visible only after both parties have submitted evaluations. BlaBlaCar has modified its reputation system several times over our sample period, affecting grading patterns. Appendix~\ref{system_changes} discusses these changes and their implications. In the analysis of passenger choice, we use all available drivers and all their grades; however, when modeling drivers' supply decisions we will focus on the subset of drivers who operated only under the latest version of the reputation system. }
A common concern with reputation systems on sharing economy platforms is grade inflation, whereby overwhelmingly positive reviews reduce system informativeness \citep{Zervas2015}. While the highest grade (5) is modal on BlaBlaCar, sufficient variation exists in lower grades to preserve system informativeness. The mean grade in our sample is 4.6.

\paragraph{Data collection:}
We collected our dataset using a web crawler on \href{https://www.blablacar.fr}{www.blablacar.fr} between July 1, 2017 and March 18, 2019. The program randomly selected pairs of cities from a predefined list of the 111 largest cities in France. All trips either originated from or terminated in Paris or its vicinity, with the other endpoint in one of the remaining 110 largest French cities.

The program collected all information accessible to prospective passengers. For each driver available on a given route, we accessed their profile and extracted all displayed characteristics, including name, age, photo, biography, and number of Facebook friends. We also extracted the complete history of ratings and textual comments received by each driver. We also observed the number of clicks and sold seats for each listing at the time of data collection. Clicking on a listing is necessary to book a trip and reveals a detailed description of the ride, though passengers can decline to book at no cost. We calculate revenue per listing as the product of sold seats and price.

Our scraping methodology captures listings at various stages of their lifecycle. Since drivers post listings at different times and our crawler visits the platform independently of posting timing, observed listings vary in their time since posting and time remaining until departure. We control for both dimensions—listing age and hours until departure—in our empirical specifications.\footnote{This sampling approach explains why many observations have zero sold seats and zero revenue. To verify this does not bias our results, we used the BlaBlaCar API to collect final outcomes for a subset of listings. Results using this validation sample are similar to our main findings.}

We identify driver gender and ethnicity using two complementary methods. A growing body of empirical work has established the practice of using names and photographs to infer demographic characteristics in online marketplaces \citep{bertrand2004emily, fryer2004causes, pope2011s, Edelman2014, Edelman2017, gaddis2017black, athey2022smiles}. Following this literature, we employ both name-based and image-based classification. First, we match driver names against a database of ethnic name origins published by the French government, supplemented with other publicly available sources.\footnote{Names of foreign origin translated into French exhibit considerable spelling variation. We phonetically encode name lists and allow for minor spelling variations to improve classification accuracy.}  Second, we employ facial recognition software to refine our classifications.\footnote{\href{www.kairos.com}{www.kairos.com}} Facial recognition has proven effective in identifying demographic characteristics in contexts where visual information is available to market participants \citep{jaeger2020automated, zhang2022reducing, athey2022smiles}. Appendix~\ref{classification} provides detailed documentation of our identification procedure and demonstrates how name-based and facial recognition methods complement each other. We define minority drivers as those with names of Arabic or African origin or those identified as such through facial recognition. Our combined approach improves upon studies that rely solely on name-based classification by incorporating visual information that passengers observe when making booking decisions.

We augment our dataset with several additional sources. We proxy vehicle quality using average prices for the same vehicle model from eBay Germany, a prominent online marketplace with publicly accessible pricing data. We calculate fuel efficiency by matching vehicle models to a dataset of long-distance fuel consumption. We also collect daily city-level fuel prices and highway tolls to construct instrumental variables for ride prices. We calculate distances and expected travel times by car and public transportation for each route at the scheduled departure time using the Google Maps API. Finally, we incorporate city-specific characteristics for each origin and destination, including population, median income, crime index, and share of foreign-born residents. Our data also include information on railway strikes that occurred in spring 2018, which we exploit as a natural experiment in our identification strategy. Appendix~\ref{glossary} provides detailed variable definitions and supplementary data sources.

Our dataset contains 224,749 unique drivers, of whom 112,851 appear at least twice, enabling panel analysis. Among drivers observed more than once, the median driver appears 3 times (mean 4.2). We employ three measures of driver outcomes. First, the number of clicks proxies for listing popularity. On average, passengers can choose from 30 available drivers for a given route, and the typical listing receives 20 clicks. The number of clicks also captures the intensity of passenger search activity in each market. Second, we observe the number of seats sold at the time of data collection. On average, drivers had sold 0.24 seats when we observed their listing. Drivers may adjust prices before the first booking, but once one seat sells, the price is fixed for all subsequent passengers. Third, we measure revenue as the product of price and sold seats. In our structural model, we recover marginal costs, enabling us to calculate economic profits.\footnote{Our sampling approach may undersample highly successful listings that sold out quickly and were removed from display. This would bias our estimates only if listing fill rates differ systematically between minority and nonminority drivers. Appendix~\ref{scraping_bias} explores this potential bias and finds it is likely negligible. Nevertheless, if nonminority drivers' listings fill faster, our output gap estimates should be interpreted as lower bounds.}
Table~\ref{descriptive_stats} presents summary statistics for key variables used in our analysis.
\begin{table}
\centering
\caption{Summary Statistics}
\label{descriptive_stats}
\resizebox{0.7\textwidth}{!}{
\begin{tabular}{@{}lrrrrr@{}}
\toprule
\toprule
Variable & N & Mean & Std. Dev. & Min & Max \\
\midrule
Ride price (EUR) & 557,031 & 31.77 & 16.97 & 2.00 & 148.50 \\
Number of reviews & 557,031 & 34.80 & 57.59 & 0 & 416 \\
Taken seats & 557,031 & 0.24 & 0.54 & 0 & 4 \\
Revenue (EUR) & 557,031 & 6.41 & 15.21 & 0.00 & 82.50 \\
Minority driver & 557,031 & 0.14 & 0.35 & 0 & 1 \\
Male & 557,031 & 0.73 & 0.45 & 0 & 1 \\
Driver age & 556,938 & 38.03 & 13.43 & 18 & 103 \\
Published ads & 557,031 & 49.52 & 81.84 & 0 & 649 \\
Reputation & 488,317 & 0.92 & 0.06 & 0.20 & 1.00 \\
Seniority (months) & 551,414 & 42.56 & 28.70 & 1 & 118 \\
Posts per month & 557,031 & 1.49 & 2.48 & 0.00 & 29.86 \\
Has picture & 557,031 & 0.97 & 0.18 & 0 & 1 \\
Bio length (words) & 556,432 & 9.84 & 15.19 & 0 & 93 \\
Car price (thousands EUR) & 458,412 & 6.07 & 5.02 & 0.60 & 24.40 \\
Fuel consumption (L/100km) & 484,188 & 4.99 & 0.81 & 3.57 & 9.11 \\
Automatic acceptance & 557,031 & 0.42 & 0.49 & 0 & 1 \\
Hours until ride & 500,416 & 96.26 & 108.01 & 0.00 & 502.57 \\
Posted since (days) & 556,488 & 6.30 & 9.64 & 0.00 & 133.15 \\
Public transport duration (hours) & 536,524 & 3.97 & 2.41 & 0.14 & 15.24 \\
Distance (km) & 553,767 & 398.92 & 200.03 & 8.30 & 944.67 \\
Travel cost (EUR) & 449,864 & 57.96 & 30.08 & 0.00 & 174.97 \\
SNCF strike & 557,031 & 0.04 & 0.19 & 0 & 1 \\
Ride description length (words) & 551,889 & 23.01 & 35.36 & 0 & 178 \\
Median income (thousands EUR) & 524,032 & 18.98 & 2.13 & 13.06 & 30.90 \\
Weekday & 557,031 & 0.67 & 0.47 & 0 & 1 \\
Luggage size & 116,696 & 1.94 & 0.39 & 1 & 3 \\
Detour & 116,162 & 1.76 & 0.45 & 1 & 3 \\
Allows pets & 220,127 & 0.22 & 0.41 & 0 & 1 \\
\bottomrule
\bottomrule
\end{tabular}
}
\begin{minipage}{\textwidth}
\vspace{0.5em}
\footnotesize
\textit{Notes:} This table presents summary statistics for the main variables used in the analysis. See Appendix~\ref{glossary} for variable definitions and sources of supplementary data.
\end{minipage}
\end{table}

\section{Outcome disparities and the role of reputation}\label{sec:descriptive}

\begin{figure}[t]
\centering
\caption{Outcome gap narrows with reputation accumulation}\label{decile_graphs}
\begin{subfigure}[b]{0.49\linewidth}
\centering
\includegraphics[width=\linewidth]{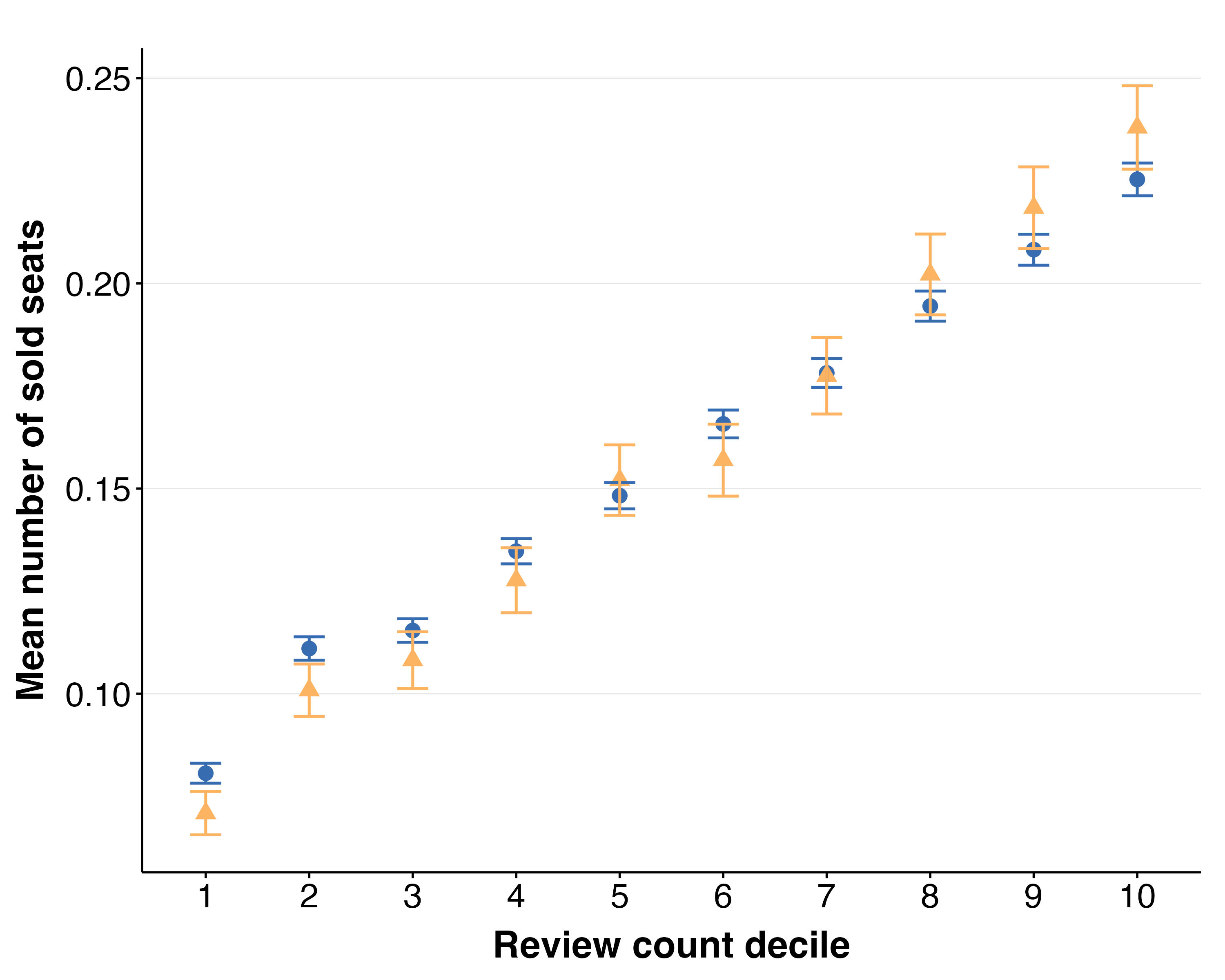}
\caption{Sold seats}
\end{subfigure}
\begin{subfigure}[b]{0.49\linewidth}
\centering
\includegraphics[width=\linewidth]{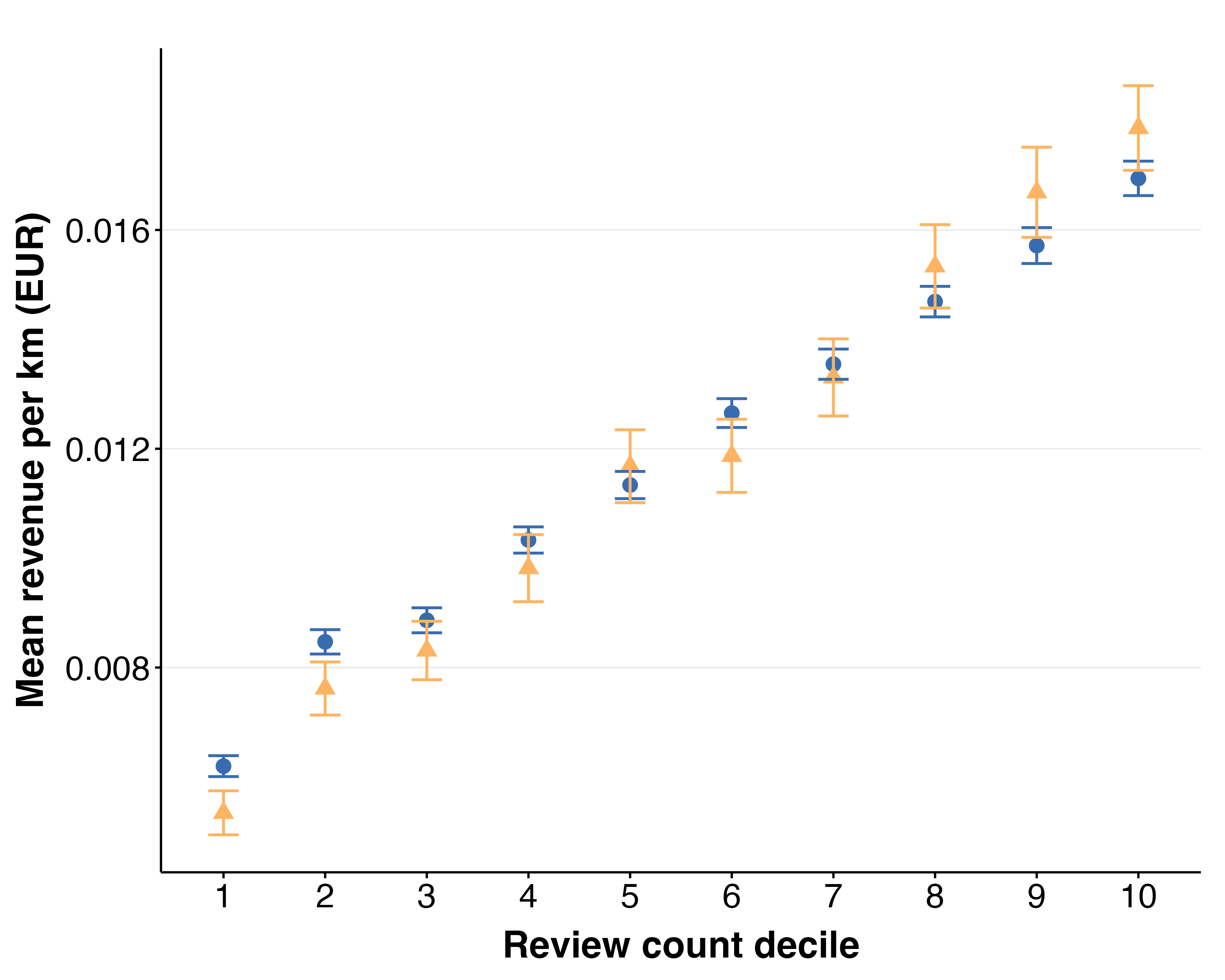}
\caption{Revenue per kilometer}
\end{subfigure}
\vspace{0.3em}
\caption*{\footnotesize\textit{Notes:} Mean outcomes by driver ethnicity and review count decile. Blue circles indicate nonminority drivers; yellow triangles indicate minority drivers. Error bars represent 95\% confidence intervals.}
\end{figure}

There are substantial outcome disparities between minority and nonminority drivers. Although minority drivers post about 3.5\% lower prices (30.9 EUR versus 32.0 EUR), their listings receive roughly 10\% fewer clicks (18.3 versus 20.3) and generate 11\% lower revenue (5.79 EUR versus 6.48 EUR).\footnote{These unconditional differences may reflect variation in route selection, trip timing, or vehicle characteristics. Appendix~\ref{ols} presents estimates of the minority outcome gap controlling for observable characteristics using ordinary least squares, and Table~\ref{tab:aipw} presents doubly-robust estimates using the augmented inverse probability weighting (AIPW) estimator.} However, this outcome gap differs substantially across drivers with different levels of experience. To illustrate, we partition drivers by ethnicity and review count deciles, then compute mean sold seats and revenue per kilometer within each decile. Figure~\ref{decile_graphs} presents the results.

Two patterns emerge. First, drivers with more reviews achieve substantially higher outcomes. The economic magnitude is large: drivers in the tenth decile earn more than 2.5 times the revenue of first-decile drivers. Second, outcome disparities between minority and nonminority drivers vary systematically with reputation. Among drivers with few reviews, minority drivers significantly underperform their nonminority counterparts; differences are statistically significant in the first two deciles. Among drivers with more reviews, ethnic differences become statistically insignificant. Minority drivers with extensive review histories even exhibit slightly higher outcomes than comparable nonminority drivers, though these differences lack statistical significance.

\subsection{Doubly-robust estimates of the minority gap}

Figure~\ref{decile_graphs} documents a robust correlation between minority status and economic outcomes that diminishes with reputation. However, minority and nonminority drivers may differ along observable dimensions that correlate with both the number of reviews and outcomes. To address potential selection on observables, we estimate the effect of minority status using augmented inverse probability weighting (AIPW), a doubly-robust estimator that combines propensity score weighting with outcome regression \citep{robins1994estimation}.\footnote{The AIPW estimator achieves double robustness: it yields consistent estimates if either the propensity score model or the outcome model is correctly specified, though not necessarily both. Nevertheless, the causal interpretation rests on the assumption of unconfoundedness, which we cannot verify. Section~\ref{train} exploits a natural experiment to document the causal impact of reviews on outcomes. We implement AIPW using the causal forest algorithm \citep{wager2018estimation, athey2019estimating} as implemented in the \texttt{grf} package \citep{tibshirani2025grf}. Additionally, Appendix~\ref{reputation_effect} presents conditional gap estimates controlling for observable driver and trip characteristics across reputation levels using OLS, and Appendix~\ref{panel} presents estimates using panel-data methods.} We report two estimands: the overlap-weighted average treatment effect (ATE), which reweights observations by $\hat{e}(X)(1-\hat{e}(X))$ to focus on regions of good covariate balance \citep{crump2009dealing}, and the average treatment effect on the treated (ATT), which estimates the effect for minority drivers specifically. Appendix~\ref{app:propensity} displays the propensity score distributions and confirms substantial common support.

Table~\ref{tab:aipw} presents the results. We estimate effects separately for drivers at different experience levels to examine how the minority outcome gap evolves with reputation accumulation. Panel~A reports estimates for sold seats, and Panel~B reports estimate for revenue, along with the effect expressed as a percentage of the majority baseline (the mean outcome for nonminority drivers in each experience group).

\begin{table}[t]
\centering
\caption{Effect of Minority Status on Driver Outcomes: AIPW Estimates}
\label{tab:aipw}
\resizebox{0.75\textwidth}{!}{%
\begin{tabular}{l@{\hskip 1.5em}c@{\hskip 1.5em}c@{\hskip 1.5em}c@{\hskip 1.5em}c}
\hline\hline
 & \multicolumn{1}{c}{(1)} & \multicolumn{1}{c}{(2)} & \multicolumn{1}{c}{(3)} & \multicolumn{1}{c}{(4)} \\
 & All & Entrants & Intermediate & Experienced \\
 & Drivers & (0--5 reviews) & (6--15 reviews) & (15+ reviews) \\
\hline
\\[-1.8ex]
\multicolumn{5}{l}{\textit{Panel A: Sold Seats}} \\[0.8ex]
ATE (Overlap) & $-0.015^{***}$ & $-0.018^{***}$ & $-0.018^{***}$ & $-0.001$ \\
 & (0.003) & (0.004) & (0.006) & (0.007) \\[0.3em]
ATT & $-0.015^{***}$ & $-0.017^{***}$ & $-0.020^{***}$ & $-0.006$ \\
 & (0.003) & (0.004) & (0.006) & (0.006) \\[0.3em]
\quad Effect (\% of baseline) & $-6.0\%$ & $-9.6\%$ & $-8.3\%$ & $-1.7\%$ \\
\quad Majority baseline & 0.255 & 0.181 & 0.235 & 0.336 \\[1em]
\midrule
\multicolumn{5}{l}{\textit{Panel B: Revenue (EUR)}} \\[0.8ex]
ATE (Overlap) & $-0.476^{***}$ & $-0.612^{***}$ & $-0.503^{***}$ & $-0.217$ \\
 & (0.075) & (0.124) & (0.162) & (0.160) \\[0.3em]
ATT & $-0.488^{***}$ & $-0.620^{***}$ & $-0.544^{***}$ & $-0.271^{**}$ \\
 & (0.070) & (0.118) & (0.153) & (0.134) \\[0.3em]
\quad Effect (\% of baseline) & $-7.1\%$ & $-11.6\%$ & $-8.2\%$ & $-3.3\%$ \\
\quad Majority baseline & 6.86 & 5.35 & 6.62 & 8.26 \\[0.5em]
\hline
Observations & 317,328 & 98,316 & 67,961 & 81,240 \\
\hline\hline
\end{tabular}
}
\vspace{0.8em}
\begin{minipage}{0.95\textwidth}
\footnotesize
\textit{Notes:} AIPW estimates of the effect of minority driver status using causal forests. The sample is trimmed to observations with estimated propensity scores in $[0.05, 0.95]$. ATE (Overlap) reweights by $\hat{e}(X)(1-\hat{e}(X))$; ATT is the average treatment effect for treated minority drivers. Effect (\% of baseline) reports the ATT divided by the majority baseline. Covariates include driver age, gender, platform seniority, posting frequency, biography length, car value, automatic acceptance status, time until departure, time since posting, competition, distance, strike indicator, ride duration, and price. Standard errors in parentheses. $^{*}p<0.10$, $^{**}p<0.05$, $^{***}p<0.01$.
\end{minipage}
\end{table}

The estimates are consistent with the reduced-form findings. For entrant drivers with zero to five reviews, minority status leads to a substantial reduction in economic outcomes. The ATT indicates that minority entrants sell 9.6\% fewer seats and earn 11.6\% less revenue compared to observably similar nonminority entrants. These effects are precisely estimated and statistically significant at the 1\% level. The minority outcome gap diminishes monotonically with experience. For intermediate drivers (6--15 reviews), the revenue gap narrows to 8.2\%, and for experienced drivers (15+ reviews), it falls to 3.3\%. While the experienced-driver effect remains marginally significant for revenue, it is statistically indistinguishable from zero for sold seats. These large, significant effects for entrants that attenuate toward zero for experienced drivers, suggests that reputation building enables minority drivers to overcome initial disparities. The overlap-weighted ATE estimates, which emphasize observations with propensity scores near 0.5 where both groups are well represented, yield similar conclusions.
 
\paragraph{Selection and survivorship.} Differential attrition by ethnicity could mechanically generate the convergence pattern in Figure~\ref{decile_graphs} if low-performing minority drivers disproportionately exit the platform. The data do not support this concern. The minority share varies modestly across reputation deciles, from 16.4\% in the first decile to 13.0\% at the minimum (sixth decile) to 14.9\% in the tenth decile, and exhibits a U-shaped rather than monotonically declining pattern. To further investigate potential selection effects, we tracked driver activity over time by revisiting driver profiles months after initial data collection. Appendix~\ref{appendix_exit} presents logit estimates showing that minority entrants do not exit the platform at higher rates than nonminority entrants. Together, these findings indicate that selection cannot explain the reputation effect.

\section{Railway strike as a natural experiment}\label{train}
The evidence presented in Figure~\ref{decile_graphs} shows that minority drivers who are new to the platform experience larger increases in sales from each review than majority drivers, allowing them to gradually narrow the outcome gap. In this section, we exploit a natural experiment to establish a causal relationship between reviews and outcomes, and to demonstrate that this effect is particularly pronounced for minority entrants.

\subsection{The 2018 French railway strike}
French railway workers conducted a national strike during our sample period in opposition to plans to liberalize the European railway market.\footnote{The strike protested proposed reforms that would open the French railway market to competition and restructure the state-owned railway company SNCF.} The strike followed a pattern of two consecutive days of disruptions every five days over three months (April--June 2018). Since BlaBlaCar and railways are direct substitutes for intercity travel, the negative railway supply shock generated a positive demand shock for ridesharing. Platform usage surged dramatically: in April 2018, 5 million passengers traveled on BlaBlaCar, more than three times the typical monthly volume of 1.5 million, and booking requests increased sixfold.\footnote{Source: \href{https://www.lemonde.fr/economie/article/2018/04/03/les-transports-alternatifs-grands-gagnants-de-la-greve-a-la-sncf_5279932_3234.html}{Le Monde, April 3, 2018}.}

\begin{figure}[!htbp]
\centering
\caption{Railway strike as a demand shock\label{fig7}}
\begin{subfigure}[b]{0.49\linewidth}
\centering
\includegraphics[width=\linewidth]{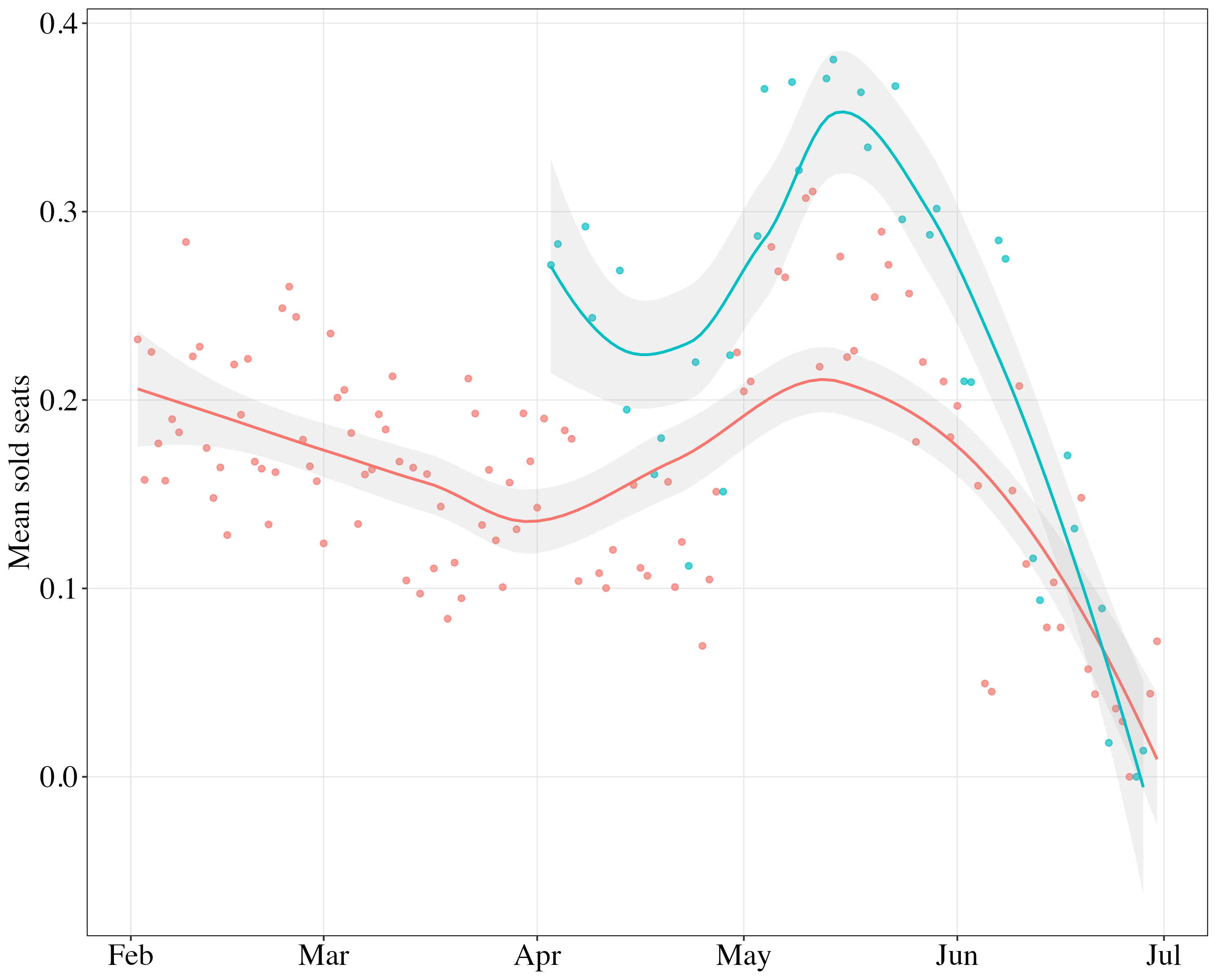}
\caption{Sold seats}
\label{fig7:a}
\end{subfigure}
\begin{subfigure}[b]{0.49\linewidth}
\centering
\includegraphics[width=\linewidth]{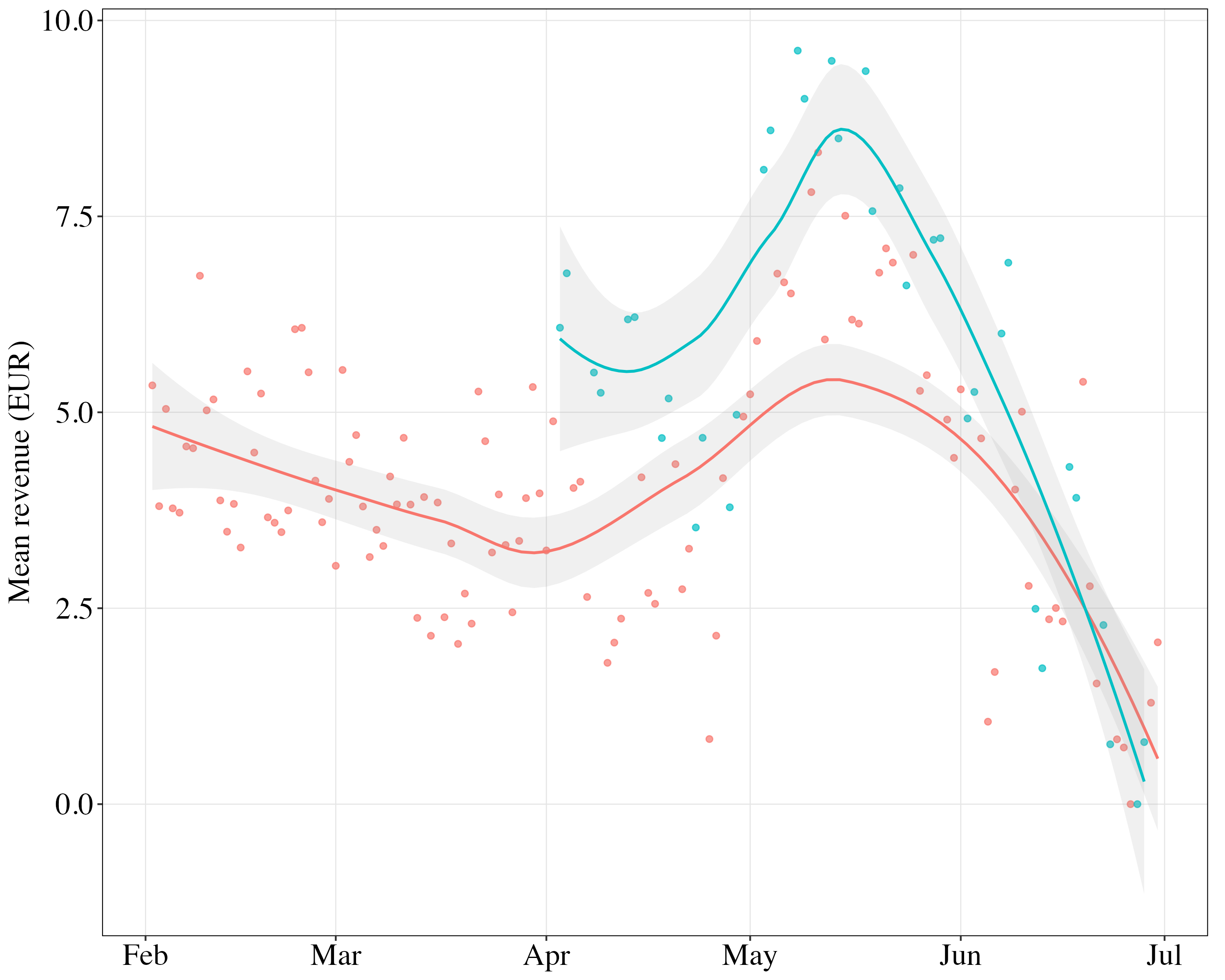}
\caption{Revenue}
\label{fig7:b}
\end{subfigure}
\vspace{0.3em}
\caption*{\footnotesize\textit{Notes:} Time series of driver outcomes during the strike period. Red circles indicate non-strike days; blue circles indicate strike days. The vertical axis measures mean outcomes; the horizontal axis shows dates from February through June 2018.}
\end{figure}

Figure~\ref{fig7} visualizes the impact of the strike on driver outcomes. The figure plots daily mean sold seats (Panel A) and revenue (Panel B) from February through June 2018, with strike days marked in blue and non-strike days in red. The data reveal a clear discontinuous increase in both outcomes on strike days. During strike periods, drivers sold approximately 50\% more seats and earned correspondingly higher revenue compared to adjacent non-strike days. The effect is most pronounced in April and May when strike intensity peaked. The two outcome measures move together, as expected if the demand surge fed directly into sales. The smooth curves represent locally weighted regression fits, showing that outcomes are systematically higher on strike days while accounting for underlying seasonal trends.

\subsection{Identification strategy}
\begin{figure}[!htbp]
\centering
\caption{No evidence of selection into treatment\label{fig8}}
\begin{subfigure}[b]{0.49\linewidth}
\centering
\includegraphics[width=\linewidth]{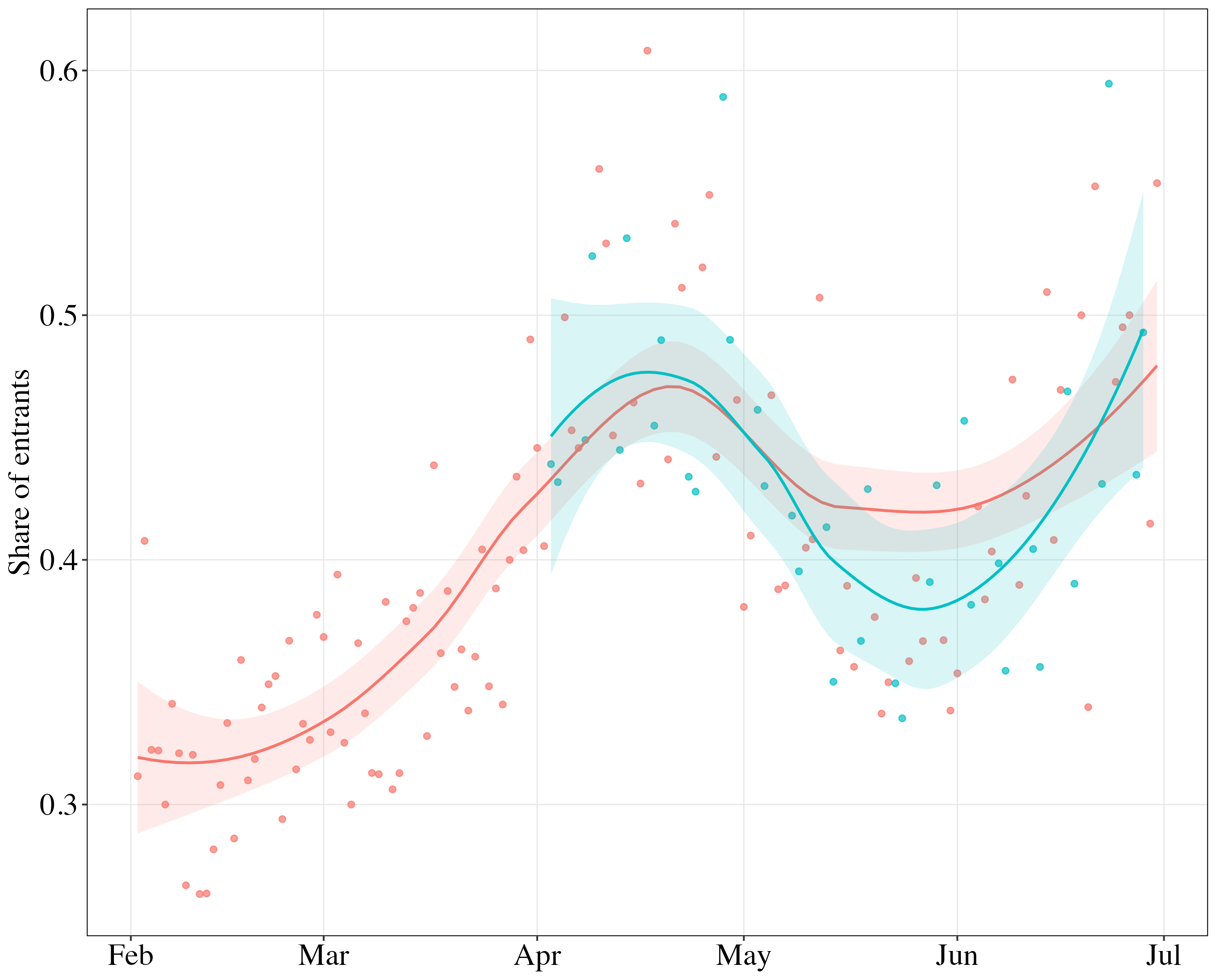}
\caption{Share of entrants}
\label{fig8:a}
\end{subfigure}
\begin{subfigure}[b]{0.49\linewidth}
\centering
\includegraphics[width=\linewidth]{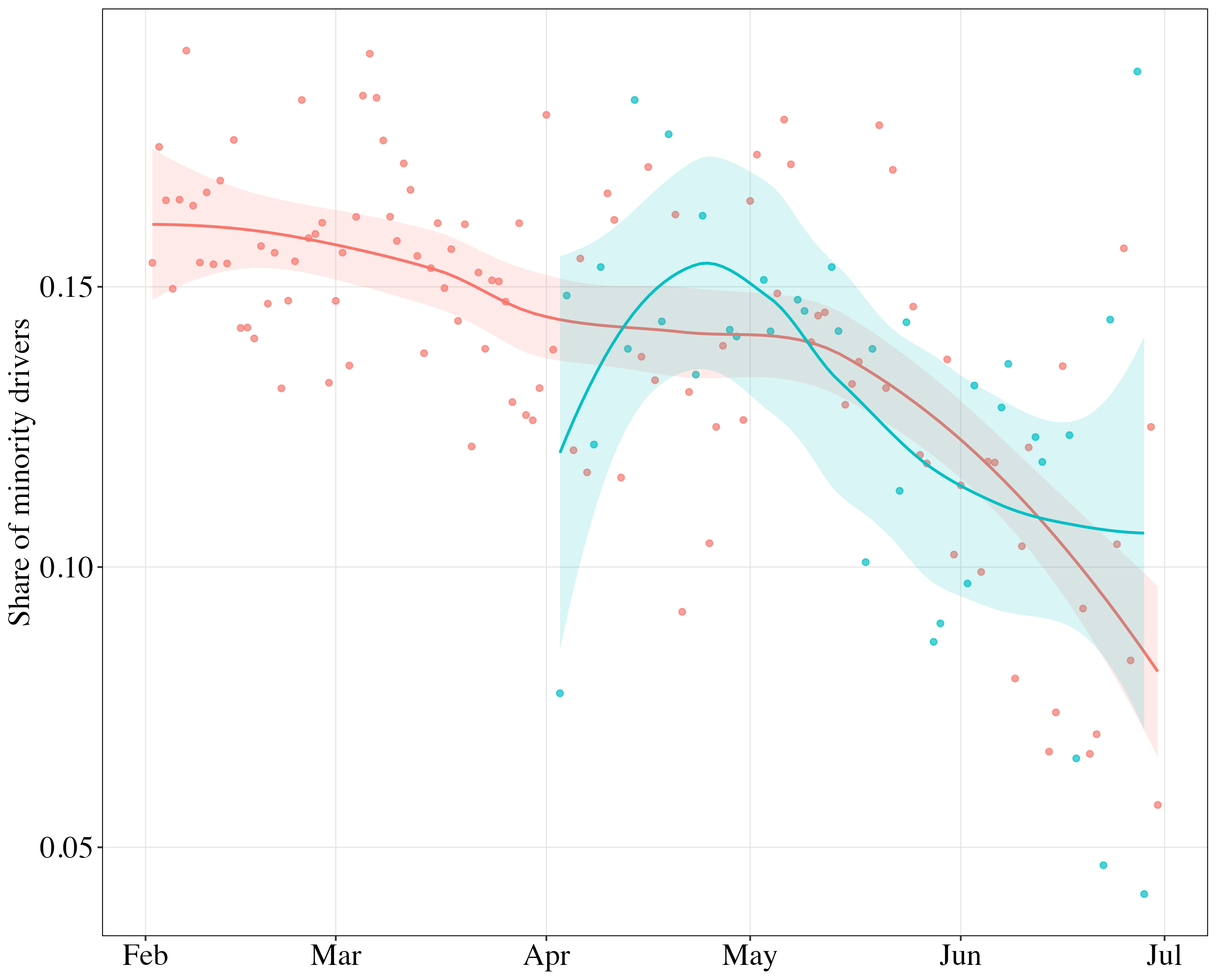}
\caption{Share of minority drivers}
\label{fig8:b}
\end{subfigure}
\vspace{0.3em}
\caption*{\footnotesize\textit{Notes:} Composition of drivers over time. Red circles indicate non-strike days; blue circles indicate strike days. No discontinuous changes in driver composition are evident on strike days.}
\end{figure}

We interpret the strike as a natural experiment in which the treatment—driving on a strike day—generates exogenous variation in review accumulation. Drivers who happened to be traveling on strike days sold more seats and consequently received more reviews; in contrast, drivers who drove on non-strike days during this period faced a typical demand level. The key identifying assumption is that drivers did not select into treatment based on the strike schedule. This assumption is plausible because BlaBlaCar drivers are nonprofessionals who travel for personal reasons and typically plan trips in advance, making it unlikely that they would alter travel plans in response to railway disruptions.
We provide two pieces of evidence supporting the exogeneity assumption. First, if drivers selected into strike days opportunistically, we would observe an influx of new drivers on those days. Figure~\ref{fig8:a} shows no significant difference in the share of entrants between strike and non-strike days. Second, if minority drivers were more responsive to the demand surge, their representation would increase on strike days. Figure~\ref{fig8:b} demonstrates that minority drivers comprised 14.7\% of active drivers on strike days and 14.8\% on non-strike days during this period—a statistically insignificant difference. Appendix~\ref{sec:app_strikes} (Table~\ref{tab:strike_balance}) compares additional driver and trip characteristics across strike and non-strike days, finding no systematic differences.
\subsection{Estimation and results}
We estimate the causal effect of review accumulation using a difference-in-differences design with the doubly robust estimator developed by \cite{sant2020doubly}. This estimator is consistent if either the propensity score model or the outcome regression model is correctly specified. The treated group comprises drivers who traveled on at least one strike day and the control group comprises drivers who drove at least once in this period, but not on a strike day. \footnote{Because the treatment is binary exposure to driving on at least one strike day, the estimand is the reduced-form (intent-to-treat) effect of strike-day exposure on post-strike outcomes, not the effect of an additional review as such. We read it as operating through review accumulation for three reasons. First, the demand shock is transitory: it ends with the strike on June~28, 2018, after which treated and control drivers face a common demand environment, so a persistent post-strike gap cannot reflect the shock itself. Second, the two groups are balanced on pre-strike characteristics, including the number of prior reviews and reputation (Table~\ref{tab:strike_balance}), so the gap does not reflect pre-existing differences. Third, the only durable driver-level state that strike-day exposure changes is the stock of reviews accumulated while driving during the window: treated drivers sold roughly $50\%$ more seats on strike days (Figure~\ref{fig7}) and were reviewed accordingly. We therefore interpret the estimates as the effect of strike-induced reputation building; because we do not instrument review count directly, they should not be read as the effect of a single review.}

Our parameter of interest is the effect of reviews accumulated during strike days on post-strike outcomes. Based on the heterogeneous reputation effects documented in Figure~\ref{decile_graphs}, we estimate treatment effects separately for four groups: minority entrants ($\leq$15 reviews), nonminority entrants (
$\leq$15 reviews), minority experienced drivers (>15 reviews), and nonminority experienced drivers (>15 reviews). Table~\ref{tab:att_revenue_seats} presents the results.

\begin{table}[!htbp]
\centering
\caption{Average treatment effects on revenue and seats sold\label{tab:att_revenue_seats}}
\resizebox{0.6\textwidth}{!}{%
    \begin{tabular}{lccccc}
      \toprule      \toprule
      & \makecell{All \\ drivers} & \makecell{Non-minority \\ experienced} & \makecell{Minority \\ experienced} & \makecell{Non-minority \\ entrant} & \makecell{Minority \\ entrant} \\
      \midrule
      \multicolumn{6}{l}{\textit{Panel A: Revenue}} \\
      ATT & 1.110*** & $-$0.394 & $-$0.148 & 1.775*** & 2.855*** \\
          & (0.197) & (0.343) & (0.676) & (0.278) & (0.579) \\[0.5em]
      \multicolumn{6}{l}{\textit{Panel B: Seats sold}} \\
      ATT & 0.054*** & $-$0.004 & $-$0.007 & 0.080*** & 0.119*** \\
          & (0.008) & (0.014) & (0.030) & (0.010) & (0.022) \\
      \bottomrule      \bottomrule
    \end{tabular}%
}
\vspace{0.3em}
\caption*{\footnotesize\textit{Notes:} Doubly robust difference-in-differences estimates of average treatment effects on the treated (ATT). Treatment is defined as driving on at least one strike day. Experienced drivers have more than 15 reviews; entrants have 15 or fewer reviews. Standard errors in parentheses. $^{*}p<0.10$, $^{**}p<0.05$, $^{***}p<0.01$.}
\end{table}

The results reveal substantial heterogeneity in treatment effects. Minority entrants who drove during the strike earned 2.86 EUR more per trip in the post-strike period compared to minority entrants who did not drive during the strike (an effect significant at the 1\% level). Nonminority entrants experienced a smaller but still significant effect of 1.78 EUR. The difference between these estimates (1.08 EUR, or 61\% larger for minorities) is economically meaningful and statistically significant at the 10\% level. Panel B shows corresponding effects on seats sold: minority entrants sold 0.119 additional seats per listing (a 49\% increase relative to the baseline mean of 0.24), compared to 0.080 for nonminority entrants.
In contrast, treatment effects for experienced drivers are small and statistically indistinguishable from zero for both ethnic groups. This pattern aligns with the diminishing marginal returns to reputation documented in Figure~\ref{decile_graphs}: additional reviews provide minimal information about experienced drivers whose quality is already well-established.

The strike-induced demand surge allowed drivers to fill seats and accumulate reviews at an accelerated pace. In subsequent periods, passengers responded to these additional reviews by increasing their propensity to book with treated drivers. The exogenous variation in review accumulation provides causal evidence that reviews improve outcomes, particularly for minority drivers with limited reputational capital. The 61\% larger treatment effect for minority entrants relative to nonminority entrants indicates that reviews are especially valuable for drivers whom passengers initially evaluate with greater skepticism. These findings support the interpretation that reputation systems enable minority drivers to overcome initial prejudice by credibly signaling quality through accumulated reviews.

\section{Strategic behavior of drivers}\label{sec:strategic}
The reputation effects documented in the previous section create incentives for drivers to invest in reputation building. Drivers possess two strategic instruments to accelerate reputation accumulation: they can offer discounted prices to increase the probability of selling seats (and thus receiving reviews), and they can exert effort to secure higher ratings. In this section, we provide descriptive evidence that drivers employ both strategies, with the investment concentrated in early career stages when marginal returns to reputation are highest.

\subsection{Sample construction and measurement}
To study strategic pricing and effort provision over driver careers, we construct a panel of drivers observed from platform entry through maturity. We retain drivers with at least 50 reviews, ensuring we observe complete career trajectories including the reputation-building phase.\footnote{Because this sample conditions on drivers who ultimately reach $50$ reviews, the within-career patterns could also reflect survivorship, lower-rated drivers exiting before maturity, or changes in route and trip mix over the career, rather than early-career discounting and effort alone. Reassuringly, Appendix~\ref{appendix_exit} provides evidence consistent with no disparate survivorship across ethnic groups: minority and nonminority entrants exit the platform at statistically indistinguishable rates (Table~\ref{exits}). More fundamentally, our structural estimates do not rely on these descriptive profiles being causal: the equilibrium effort and introductory-discount magnitudes reported in Section~\ref{results} are recovered from the estimated demand and supply primitives and the solved equilibrium, not from the within-career patterns plotted here, which serve only to motivate the model. The figures inform a single modeling choice, the ``burnout" cutoff $t^{*}=20$, whose value Appendix~\ref{appendix_sensitivity} shows is not pivotal.}  For each driver, we extract their full review sequence and prices for the trips that appeared in our dataset.

We define drivers as mature after accumulating 20 reviews, the threshold beyond which Figure~\ref{decile_graphs} shows the minority outcome gap becomes statistically insignificant and marginal returns to reputation diminish. We assume that prices and ratings after this threshold reflect drivers' intrinsic characteristics absent strategic reputation-building incentives. Mature prices thus represent profit-maximizing prices given established reputation; mature ratings reflect baseline service quality without additional effort.

We measure strategic pricing through discounts: the percentage deviation of early-career prices from each driver's mature price (mean price per kilometer with 20+ reviews). Positive discounts indicate low pricing consistent with investing in accelerated review accumulation. We measure strategic effort through the share of 5-star ratings (the maximum) received at each career stage. While drivers cannot directly control ratings, they influence them through service quality—punctuality, courtesy, accommodation, and journey comfort. We refer to such strategic quality enhancement as ``effort.” Comparing early-career to mature 5-star shares therefore reveals how much effort drivers exert during reputation building.

\subsection{Strategic behavior: pricing and effort}

\begin{figure}[t]
\centering
\caption{Strategic behavior declines with reputation accumulation}
\label{fig:strategic_behavior}
\begin{subfigure}[b]{0.48\textwidth}
\centering
\includegraphics[width=\textwidth]{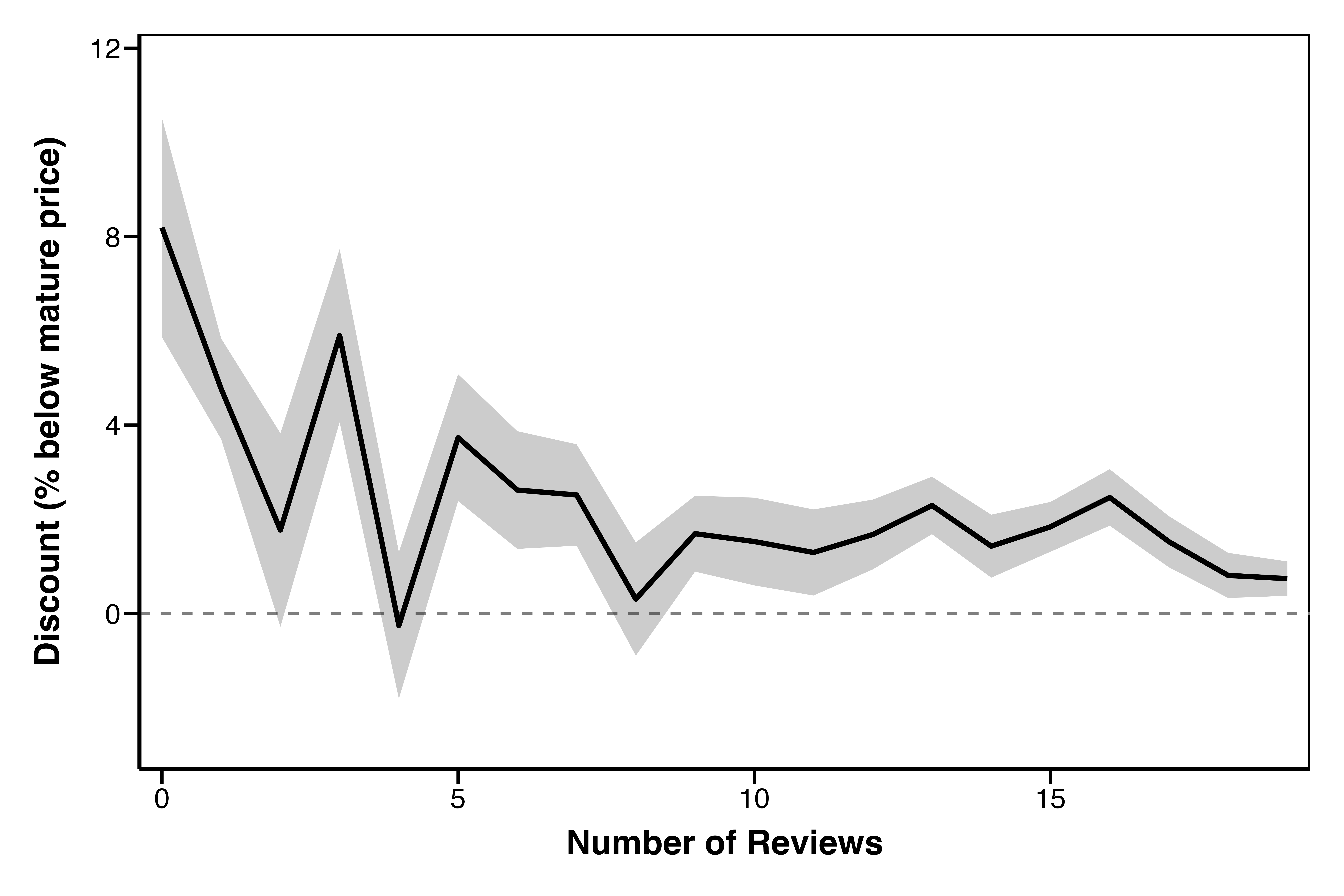}
\caption{Pricing discounts}
\label{fig:pricing_discount}
\end{subfigure}
\hfill
\begin{subfigure}[b]{0.48\textwidth}
\centering
\includegraphics[width=\textwidth]{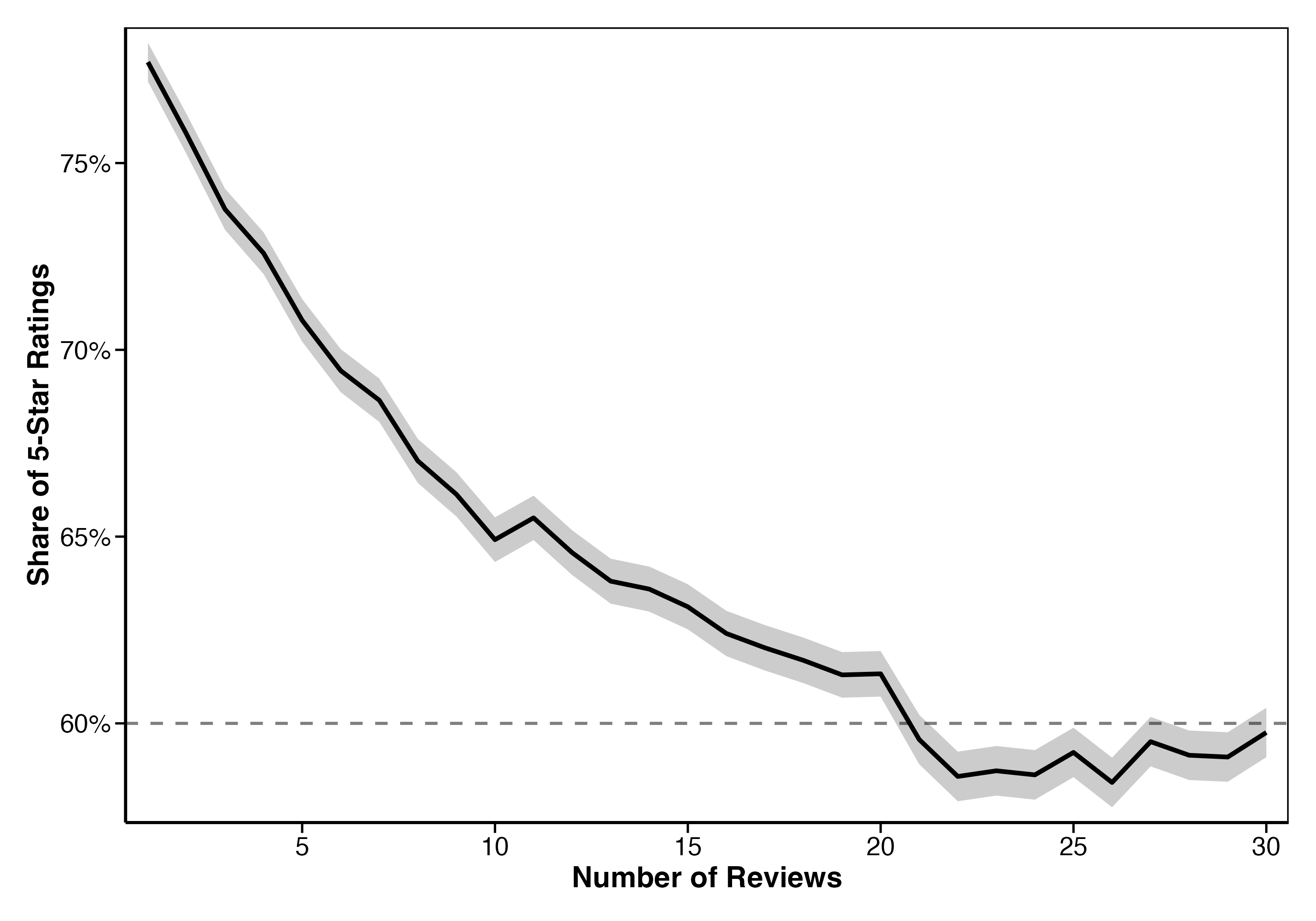}
\caption{Share of 5-star ratings}
\label{fig:five_star_share}
\end{subfigure}
\vspace{0.5em}
\caption*{\footnotesize\textit{Notes:} Panel~(a) displays mean percentage discount relative to mature price (average price per kilometer for drivers with 20+ reviews). Discounts are bounded at $\pm$50\% to limit the influence of outliers. Panel~(b) displays the share of ratings equal to 5 (maximum). Both panels restrict the sample to drivers with at least 50 total reviews to ensure complete early-career observation. Shaded bands represent 95\% confidence intervals.}
\end{figure}

Drivers on BlaBlaCar have two strategic instruments for building reputation: pricing and effort. Lower prices increase the probability of selling seats and receiving reviews, while higher effort improves the ratings received. Both instruments are costly, so rational drivers should deploy them intensively early in their careers---when the marginal value of reputation is highest---and phase them out as returns diminish. Figure~\ref{fig:strategic_behavior} tests these predictions.

Panel~(a) plots pricing discounts over the first 19 reviews. The vertical axis measures the percentage discount relative to each driver's mature price; the horizontal axis tracks review accumulation. Drivers offer substantial discounts early in their careers, with mean discounts exceeding 8\% for drivers with zero reviews. Discounts decline rapidly as drivers accumulate reviews: by the tenth review, they approach 2\%. After 15 reviews, discounts stabilize near 1\%, effectively converging to mature pricing.

Panel~(b) presents complementary evidence on effort provision. The figure plots the share of ratings that are 5 stars (the maximum) as a function of review number. If drivers exert higher effort early in their careers to secure positive reviews, we should observe elevated 5-star shares initially, with convergence to each driver's intrinsic quality level as careers progress. The data strongly support this prediction. Among drivers receiving their first review, 76\% receive 5 stars. This share declines monotonically, falling to approximately 72\% by the fifth review and stabilizing near 60\% after 20 reviews. The 16-percentage point decline from first to mature reviews suggests substantial early-career effort provision.

The two panels are nearly symmetric. Both strategic instruments are deployed intensively during the first 10--15 reviews and phased out thereafter. The convergence of pricing to mature levels and ratings to stable shares after 15--20 reviews suggests that drivers perceive diminishing marginal returns to reputation investment beyond this threshold. The plateau around 60\% five-star ratings plausibly reflects drivers' intrinsic service quality absent strategic effort provision.

These patterns are consistent with drivers responding rationally to reputation-building incentives. However, reduced-form relationships cannot identify the precise magnitudes of strategic responses or quantify how they vary across driver types. Optimal pricing and effort strategies depend on drivers' unobserved marginal costs and baseline service quality, and both the incentives and these unobservable characteristics may differ across ethnic groups. The magnitude of optimal discounts depends on demand elasticities, the informational content of reviews, and marginal costs---primitives that require additional structure to recover. Similarly, quantifying effort provision requires separating driver heterogeneity in intrinsic quality from strategic quality enhancement. Section~\ref{career_concerns} develops a structural model that addresses these identification challenges, enabling us to recover the primitives governing strategic behavior and evaluate counterfactual policies.

%% file: model.tex
\section{Dynamic model of discrimination - demand and supply}\label{career_concerns}
We develop a model of discrimination in which sellers (drivers) compete in a market. Buyers (passengers) value both price and quality and learn about drivers' quality from group identity and past performance. Drivers belong to either a minority ($m$) or a majority ($n$).

\subsection{Set-up of the model}

\paragraph{Drivers.} - Consider a driver who has observable group identity $g \in \left\{m, n\right\}$ and unobservable ability $\eta \sim N(\mu_{g},1/\tau_{g})$, with mean $\mu_{g} \in \mathcal{R}$ and precision $\tau_{g} >0$. Drivers have a cost of offering the service to a passenger equal to $c$. Drivers participate in the market for periods $t = 1, ..., T $, where $T$ may be finite or infinite. In every period, drivers set prices $p \in \mathcal{R}$ for their service and exert effort $a \in \mathcal{R}_{+}$; providing effort costs the driver $f(a)$, where $f(a)$ is increasing, convex, and the derivative is strictly increasing. If a driver serves passengers in period $t$, service  quality is $q_{t} = \eta + a_{t} + \epsilon_{t}$, where $\epsilon_{t} \sim N(0, 1/\tau_{\epsilon})$ is an independent random shock with precision $\tau_{\epsilon}$. At time $t$ driver's problem writes

\begin{align}
\max_{p_{t},a_{t}}\left\{\sum_{s=t}^{T} \delta^{s-t}(p_{s}-c)\mathbf{E}_t\left[\mathcal{S}_{s}(p_{s},\mathbf{q}_{s})\right] - f(a_{t}) \right\}, \label{driver_max}
\end{align}
where $\mathcal{S}_{s}(p_{s},\mathbf{q}_{s})$ is the number of sold seats in period $s$, which depends on the history of quality reports $\mathbf{q}_{s}=(q_{1},....,q_{s-1})$, and $\delta \in (0,1) $ is a discount factor. We refer to $\pi_{t} \equiv (p_t-c)\mathbf{E}_t\left[\mathcal{S}(p_t,\mathbf{q}_{t})\right]$ as the expected driver's profit for a given price $p_{t}$ and history of reviews $\mathbf{q}_{t}$.

\paragraph{Passengers.} - Passengers are active in one period; they observe available drivers, either pick one of them or decide not to trade. If they trade, they report quality $q_{t}$ afterward.  Before choosing the driver, passengers observe drivers' groups $g$, histories of quality reports and prices $p_{t}$. Passengers hold a prior belief that $\eta \sim N(\hat{\mu}_{g},1/\tau_{g})$, and $\hat{\mu}_{g}$ need not coincide with $\mu_{g}$.

\paragraph{Market.} - In period $t$, there are $M_{t}$ passengers, where $M_{t}$ is a random variable such that $\mathbf{E}\left[M_{t}\right] = \mathcal{M}_{t}$, and $N_{t}$ drivers. The market structure $\Omega_{t}$ summarizes the number of drivers and their characteristics (costs, histories of quality reports, and quality types). The market entry process is assumed to be exogenous and results in $\mathbf{E}\left[\Omega_{t}\right] = \mathbf{\Omega}$. For analytical tractability, we assume the following:

\begin{ass}
We assume that the set of potential drivers that can enter the market in a given period is large enough so drivers do not expect to compete against each other in subsequent periods. Thus, when setting $p_{t}$ and $a_{t}$, drivers do not consider their impact on quality reports of other drivers.
\end{ass}

\paragraph{Timing.} - The timing of the game is as follows: \textit{i)} Drivers set prices that maximize their discounted sums of utility subject to histories of quality reports, costs, and expected market structure. \textit{ii)} Passengers arrive to the market and observe available drivers. Each passenger either chooses the driver that maximizes her utility or decides not to trade, which gives a payoff of zero.  \textit{iii)} Drivers exert effort to maximize their discounted sums of utility. \textit{iv)} Passengers observe quality and report it.


\subsection{Passengers' belief formation and choice problem}

\paragraph{Belief formation and updating.} - A passenger has a belief-based partiality toward nonminority drivers if she believes that the average ability of nonminority drivers is higher than the average ability of minority drivers. Belief-based partiality can be biased or unbiased, depending on whether it coincides with the true population average for each group.

\begin{definition}
A passenger has a belief-based partiality toward nonminority if $\hat{\mu}_{n}>\hat{\mu}_{m}$. This partiality is unbiased if $\hat{\mu}_{n} =\mu_{n}$ and  $\hat{\mu}_{m} =\mu_{m}$, and otherwise is biased.
\end{definition}

Passengers learn about drivers' ability from the evaluation history. Their posterior belief is derived using Bayes' rule, given the prior belief on the average ability in group $g$. Passengers observe the quality $q_{t}$ from the evaluation history. However, they do not distinguish the individual components (ability, effort, noise). To interpret the evaluation history passengers need to form accurate belief about the level of effort exerted by the driver in the past. Let $a_{t}^{*}$ be an equilibrium level of effort in period $t$. Thus, from the quality report $q_{t}$ passengers learn $z_{t} \equiv q_{t} -  a_{t}^{*} = \eta + \epsilon_{t}$ and the posterior belief about the driver's ability writes

\begin{align}
\mathbf{E}_{t+1}\left[\eta|\mathbf{q}_{t}\right] = \frac{\tau_{g}\hat{\mu}_{g}}{\tau_{g}+t\tau_{\epsilon}}+\frac{\tau_{\epsilon}}{\tau_{g}+t\tau_{\epsilon}}\sum_{s=1}^{t}z_{s}. \label{posterior}
\end{align}

Increasing the precision of the quality reports (or decreasing the randomness of the outcome of the task) enlarges the weight assigned to the past performance, while increasing the precision of the distribution of ability boosts the importance of the prior belief.

\paragraph{Passengers' choices.} - A passenger $j$ chooses between drivers and the outside option of not trading to maximize her utility $u_{ijt}$ ; the utility depends on quality, price, and passenger-driver- specific shock $\varepsilon_{ijt}$,  
\begin{align}
u_{ijt} = \alpha \mathbf{E}_t\left[q_{it}|\mathbf{q}_{i,t-1}\right]+ \gamma p_{it} + \varepsilon_{ijt} \label{utility}
\end{align}

, where $\alpha$, and $\gamma$ are the marginal values of expected quality and income respectively. Thus, passenger $j$ chooses driver $i$, when 

\begin{align}
\mathbf{E}_t\left[u_{ijt}|\mathbf{q}_{i,t-1}\right] = \max \left \{\mathbf{E}\left[u_{kjt}|\mathbf{q}_{k,t-1}\right] , 0\right\} \forall k \in N_{t}. \label{choicepass}
\end{align}
\paragraph{Discrimination.} - Discrimination is the disparate treatment of drivers based on the group to which the driver belongs, rather than individual attributes. In our framework, a passenger discriminates against minority drivers, when a minority driver is chosen with a lower probability than a nonminority driver with the same price and history of quality reports. Let

\begin{align}
D(p,\mathbf{q}) \equiv \mathbf{E}\left[\mathcal{S}(p,\mathbf{q})|n\right] -\mathbf{E}\left[\mathcal{S}(p,\mathbf{q})|m\right]
\end{align}
denote the difference between the expected number of sold seats of a nonminority and minority driver conditional on the same histories of quality reports and prices. Note that $D(p,\mathbf{q})$ does depend on the number of reviews $t$ through the length of history $\mathbf{q}$.

\begin{definition} 
A minority (nonminority) driver faces discrimination if $D(p,\mathbf{q}) > 0$ ($D(p,\mathbf{q}) < 0).$
\end{definition}
In our framework, discrimination is a property of passengers' behavior, whereas biased beliefs are a property of the primitives of the model.
\subsection{Dynamics of effort}
Drivers' effort is noncontractible and it is exerted after passengers choose drivers; the incentive to exert a nonzero level of it is driven by the impact of quality evaluation on future profits. From equation (\ref{driver_max}) The first order condition of drivers' maximization problem writes

\begin{align}
\sum_{s=t}^{\infty}\delta^{s-t}\mathbf{E}\left[\frac{\partial \pi_{s}}{\partial a_{t}} \right] - f'(a_{t}) =0.
\end{align}

To obtain the utility maximizing level of effort a driver equates the marginal benefit, which is the increase in future profits, with marginal cost, the derivative of the cost of effort function.

\begin{prop}\label{opt_effort}
Suppose that, along the equilibrium path, the marginal effect of passenger utility on current profits is positive and bounded, and that this marginal profit effect does not increase with the number of reviews. The equilibrium sequence of effort $\left\{a^{*}\right\}_{t}$ decreases as $t$ increases, so $a^{*}_{t} < a^{*}_{t'}$ for $t> t'$.
\end{prop}
The proof of Proposition \ref{opt_effort}  is provided in Appendix \ref{proof1}; it follows from two assumptions made earlier: the positive impact of past effort on profits, and the functional form assumption on $f()$.

Drivers exert effort to increase future profits. Since initial reviews have a substantial impact on posterior beliefs, the level of effort is high when the number of reviews is low. As more reviews become available, the residual uncertainty about the driver's type tends to zero; thus the incentive to exert effort decreases too.

\subsection{Dynamics of pricing}
While exerting effort increases grades, changing the price affects reputation building through two channels. First, it changes current profits by affecting the probability of selling. Second, by changing the probability of selling, it also affects the probability of receiving a grade and hence the speed at which buyers learn about the driver. As drivers accumulate experience, however, the informational value of an additional grade vanishes. In the limit, only the standard pricing motive remains.
\begin{prop}\label{prop_prices}
The equilibrium sequence of prices $\left\{p^{*}\right\}_{t}$ tends towards $\tilde{p}$, where 
\begin{align}\label{objective_in_limit}
\tilde{p} \equiv \arg\max \left\{(p-c)\mathbf{E}\left[\mathcal{S}(p,\eta)\right] \right\}
\end{align}
is the profit maximizing price under complete information, where $\mathcal{S}(p,\eta)$ is the market share of a driver who has an infinite history of quality reports $q_t = \eta+a^*_t$ for all $t$.
\end{prop}

Proof of Proposition \ref{prop_prices} is in Appendix \ref{proof_prices}. In the proof, we show that the limit of expected quality, given the equilibrium level of efforts, is the driver's true type $\eta$ and that the conditional variance shrinks to the variance of the quality reports.

\begin{corollary}\label{cor_prices}
For an entrant whose expected realized quality exceeds the market’s prior belief, a larger pessimistic belief gap $ \mu_{g} - \hat{\mu}_{g}$ increases the informational value of a match and lowers the optimal introductory price.
\end{corollary}

 Proof of Corollary \ref{cor_prices} is in Appendix \ref{cor_beliefshift}. The corollary indicates that the minority drivers have an incentive to \emph{invest} in reputation, by offering low introductory prices. When the driver's expectation of the grade is higher than that of the market, the driver has the incentive to reduce the price in order to increase the probability of selling and benefiting from having the market beliefs revised upwards in the next period. The larger the biased belief-based partiality the higher the incentive to reduce the introductory price.

\subsection{Dynamics of discrimination}

In our framework, discrimination arises due to a combination of incomplete information and the belief that mean ability differs across groups ($\hat{\mu}_{m}\neq \hat{\mu}_{n}$). The impact of the beliefs about the group mean ability is gradually losing importance as drivers gain reviews. Proposition \ref{disc_dyn} formalizes it.
\begin{prop}\label{disc_dyn}
As drivers gain quality reports discrimination tends to zero
\begin{align}
\lim_{t \rightarrow \infty}D(p, \mathbf{q}) = 0.
\end{align}
\end{prop}
Proof of Proposition \ref{disc_dyn} is in Appendix \ref{disc_dyn_proof}; it is a direct consequence of the beliefs updating via Bayes Rule and the efforts following equilibrium sequence $\left\{a^{*}\right\}_{t}$. Note that the outcomes of individual drivers might diverge as drivers receive reviews; however, conditioned on these reviews the outcomes converge.

\begin{corollary}\label{rev_beliefs}
Let $\tilde{\Delta}$ denote the complete-information gap in average seats sold between 
nonminority and minority drivers:
$\tilde{\Delta} \equiv \mathbf{E}_{n}\!\left[\mathcal S(p,\eta)\right] -\mathbf{E}_{m}\!\left[\mathcal S(p,\eta)\right],$ where the expectations are taken over the true group-specific type distributions.:
\begin{itemize}
\item \emph{Convergence.} $\lim_{t\to\infty}\mathbf{E}\!\left[\Delta(\mathcal{A}_{t})\right]=\tilde{\Delta}$,
whether or not priors are biased: as reviews accumulate, the influence of group priors vanishes and the
gap is governed by true types alone.
\item \emph{Direction of bias.} Let $\mathbf{E}\!\left[\Delta(\mathcal{A}_{t})\mid\hat\mu_g=\mu_g\right]$ be
the expected gap that would arise under unbiased priors at the same experience level. Pessimistic beliefs
about minority drivers raise the expected gap above this benchmark at every finite $t$, i.e.
$\mathbf{E}\!\left[\Delta(\mathcal{A}_{t})\right]>\mathbf{E}\!\left[\Delta(\mathcal{A}_{t})\mid\hat\mu_g=\mu_g\right]$,
with the wedge vanishing as $t\to\infty$.
\end{itemize}
\end{corollary}

Corollary \ref{rev_beliefs} follows directly from Bayesian learning. If beliefs are unbiased, average demand differences across groups reflect only differences in true ability distributions. If passengers underestimate the quality of minority drivers, minority drivers initially receive less demand than warranted by their true quality. As reviews accumulate, the influence of these incorrect priors vanishes and beliefs converge to actual quality. Therefore, the excess demand gap generated by incorrect beliefs disappears over time, leaving only the gap implied by differences in true ability distributions.

The direction of adjustment depends on the sign of the bias: when the prior is overly pessimistic, the outcome gap is initially inflated and declines toward the difference justified by the underlying ability distributions, whereas when the prior is overly optimistic the gap is initially understated and rises toward the same limit.

%% file: structural_section.tex
\section{Identification and estimation}\label{identification}

We estimate and solve the model in four steps, each pinning down a distinct block of primitives or equilibrium objects. First, demand is estimated by conditional logit with a control function for price. We include the bin-specific minority interactions to estimate the belief gaps about the quality of minority and nonminority drivers across different stages of their careers. Second, driver types, the review-noise precision, and effort are recovered from late-career grades, after the phase of strategic reputation building. Third, we recover marginal costs by inverting the static pricing first-order condition on the experienced drivers subsample; Finally, the dynamic stage is solved as an oblivious equilibrium (OE). The order is forced by the model: the conversion of utility coefficients into grade-scale beliefs runs through the quality coefficient, effort is the within-cell residual of grades net of type, and the OE inherits all primitives from the static steps.

\subsection{Demand}\label{demand_side}

Passenger $j$ choosing driver $i$ in market $m$ at time $t$ has utility
\begin{equation}\label{eq:utility}
u_{ijt\ell}=\alpha \,\mathbb{E}\!\left[q_{it}\mid \mathbf{q}^{it}, g_{i}\right]+\gamma p_{it}+X_{it}'\theta+\varepsilon_{ijt},
\end{equation}
where $\mathbb{E}[q_{itm}\mid \mathbf{q}^{it}, g_{i}]$ is the expected quality of driver $i$ given her reputation history $\mathbf{q}^{it}$ and group identity $g_{i}$, $p_{it}$ is the posted price, $X_{it}$ is a vector of driver and trip characteristics, and $\varepsilon_{ijt}$ is Type~I extreme value. The dependence on $g_{i}$ enters through the group-specific prior $\hat{\mu}_g$ in the belief-updating rule. The outside option has utility zero \citep{mcfadden1974frontiers}. Choice probabilities are multinomial logit.

Equation~\eqref{eq:utility_minority} is the empirical counterpart of Equation \eqref{eq:utility}. We operationalize $\mathbb{E}[q_{itm}\mid \mathbf{q}^{it}, g_{i}]$ in two pieces: the displayed average rating, $\text{reputation}_{it}$, captures the posterior mean conditional on the review history, and $\ln(1+t)$ captures the demand response to review count beyond the rating itself. The reduced-form evidence in Section~\ref{sec:descriptive} shows that the minority penalty is concentrated in the first reviews and dissipates with experience, so we let it vary across three bins: $0$--$5$ reviews, $6$--$20$, and $21+$, with $21+$ as the omitted reference. The estimating equation is

\begin{align}\label{eq:utility_minority}
u_{ijt\ell}=&\alpha \cdot \text{reputation}_{it}+\gamma p_{it}+\beta_{\text{resid}}\,\widehat{\nu}_{it}+\psi \ln(1+t) \notag
\\ & +\kappa_{1}\left(minority_{i}\times \mathbf{1}\{t\le 5\}\right)+\kappa_{2}\left(minority_{i}\times \mathbf{1}\{6\le t\le 20\}\right)+X_{it}'\theta+\varepsilon_{ijt},
\end{align}

where $\text{reputation}_{it} \in [0,1]$ is the normalized average rating, $t$ the number of reviews, $\widehat{\nu}_{it}$ the first-stage price residual that enters as the control function (see the ``Price'' paragraph below), and $minority_{i}$ a minority indicator taking value 1 if the individual is a minority. Identification of $\kappa_{1}$ and $\kappa_{2}$ uses variation in minority status across drivers within an experience bin, conditional on $X_{it}$, the rating, and the price residual defined below. Any persistent gap among $21+$ drivers is absorbed into the reference category. We assume that the shock $\varepsilon_{ijt}$ is independent across choice situations conditional on observed characteristics and the instruments. The cost shifters affect choice only through price. Conditional on the rating, the number of reviews, and $X_{it}$, no driver-specific unobservable enters utility.

\paragraph{Price.} Posted prices reflect driver-specific unobservables, vehicle quality, route-specific experience, time flexibility, that also enter passenger utility. We instrument with two cost shifters: \emph{smartstop}, an indicator for the BlaBlaCar SmartStop feature, and origin and destination fuel prices on the day of the ride. A SmartStop is an algorithmically generated offer for a sub-segment of a driver's declared route: the platform creates it without the driver's explicit input and inflates its price above the driver's usual per-kilometer rate to raise the chance the driver accepts the resulting booking request. Because SmartStops are displayed to passengers identically to regular rides, they generate price variation that is plausibly orthogonal to demand yet 
unobservable to passengers at booking (see Appendix~\ref{app:iv} for details).

The first stage regresses $p_{it}$ on the instruments and observables. Following \citet{petrin2010control}, the first-stage residual enters \eqref{eq:utility_minority} alongside price as a control function: $\gamma$ is then identified from variation in price orthogonal to the unobservable summarized by the residual. Appendix~\ref{app:iv} reports the first stages and alternative specifications.

\paragraph{Market size.} A market is a route-day. We define market size as the largest click count received by any listing on that route-day; this is our proxy for the number of potential passengers that searched for drivers on that route. The difference between market size and seats sold is the outside-option mass.

\paragraph{From coefficients to beliefs.}\label{beliefs_identification} The bin-specific interactions in Equation \eqref{eq:utility_minority} identify the residual minority discount in expected quality within each experience bin. We impose two normalizations. First, the omitted-reference choice for the $21+$ bin sets the residual minority coefficient there to zero, so the market's expected quality for an experienced minority driver equals her displayed grade. This is consistent with the model: given the noise precision recovered in Section~\ref{identification_supply}, by $t=21$ the posterior on driver type concentrates on the displayed grade, and the residual difference due to the population prior is minor. Second, we assume the market is unbiased about nonminority drivers at every experience bin, so the omission of nonminority experience-bin fixed effects sets their expected quality equal to the displayed grade throughout the career.

On their own, the two normalizations and the demand coefficients identify a relative object: the market's belief about minority drivers measured against the nonminority and experienced-driver benchmarks, expressed on the grade scale. With $\alpha_{\text{grade}}=\alpha/4$, the ratio $\widehat{\kappa}_{1}/\alpha_{\text{grade}}$ is the grade-scale belief gap between a minority entrant and a nonminority entrant at the same displayed grade, and the same map applied to $\widehat{\kappa}_{2}$ delivers the gap for emerging drivers. Because the market is normalized to be unbiased about nonminority drivers, the nonminority entrant belief is anchored at the average displayed grade among nonminority drivers with $1$--$5$ reviews, and the minority entrant belief is that anchor minus $\widehat{\kappa}_{1}/\alpha_{\text{grade}}$. Recovering the bias itself, $\hat\mu_g-\mu_g$, defined as the gap between this belief and the true population mean, requires one further input beyond the demand normalizations: the true type means $\bar\mu_g$, estimated separately from post-burnout grades in Section~\ref{identification_supply}. We obtain $\hat\mu_g-\mu_g$ by combining the demand-side belief in levels with that supply-side type mean.

\subsection{Supply-side primitives: types, noise, effort}\label{identification_supply}

Driver type is the post-burnout grade mean. The model predicts (Section~\ref{sec:strategic}) that grades rise with effort early in the career and then flatten at the driver's type. We set the burnout cutoff at $t^{*}=20$ reviews, the point at which the within-driver paths of grades and prices have both flattened in the data (Figure~\ref{fig:strategic_behavior}).\footnote{To mitigate survivorship bias, the post-$t^*$ sample is restricted to drivers observed long enough past $t^*$ to estimate a stable mean; Section~\ref{sec:descriptive} shows that low grades raise exit, so restricting to stayers removes the induced correlation between observed grades and survival.} For driver $i$,
\begin{align}
\eta_{i}=\frac{1}{T-t^{*}}\sum_{t=t^{*}+1}^{T}q_{it},
\end{align}
where $T$ is the last period $i$ is observed.

The review-noise precision $\tau_{\epsilon}$ is the inverse of the mean within-driver variance of $q_{it}-\eta_{i}$ for $t>t^{*}$. Population moments follow: $\hat{\mu}_{g}$ is the sample mean of $\eta_{i}$ in group $g$, and $\tau_{g}$ is the inverse sample variance. Together with $\tau_{\epsilon}$ these moments parametrize the Bayesian updating rule of Section~\ref{career_concerns}.

Effort is identified from the mean gap between grades and estimated types within state cells. 
Let $\mathcal C_{g,t,x}$ denote the set of observations for drivers in group $g$, at review count $t$, 
and with discretized observable characteristics $x$, and let $N_{g,t,x}=Card(\mathcal C_{g,t,x})$. We estimate
\begin{align}
a^{*}_{g,t,x}
=
\frac{1}{N_{g,t,x}}
\sum_{(i,s)\in \mathcal C_{g,t,x}}
\left(q_{is}-\widehat\eta_i\right).
\end{align}
Under the model, grades equal type plus effort plus a mean-zero review shock. Subtracting 
$\widehat\eta_i$ removes persistent type differences, and averaging within the cell removes the 
mean-zero shock. The within-cell mean residual therefore identifies the equilibrium effort level 
for state $(g,t,x)$.

\paragraph{Identifying assumptions.} Grades depend on type, effort, and a mean-zero review shock; prices do not enter.\footnote{In Appendix~\ref{gradesprices} we provide evidence suggestive that within-driver variation in grades does not depend on prices. However, due to a small sample of drivers whom we can match to multiple observations of listings with prices and subsequent grades we also cannot rule out moderate dependency.} The shock is independent across drivers and across time. Post-$t^{*}$ effort is negligible, so within-driver grade variation after the cutoff is pure noise. This last restriction is the substantive one. Appendix~\ref{appendix_sensitivity} varies $t^{*}$ from $10$ to $30$; $h_{\epsilon}$ and the population moments move within a narrow band. We test the zero-effort restriction directly by regressing post-$t^{*}$ grade residuals on contemporaneous log competitor count, within-route--day relative log price, log hours-to-departure, and a strike indicator with driver fixed effects (Appendix~\ref{appendix_sensitivity}); the joint $F$-statistic is $0.37$ ($p=0.83$), consistent with the zero-effort restriction.

\paragraph{Cell sizes.} Identifying effort additionally requires that the cells used in the effort residual above (population, reputation history, and observables) are large enough to average out the shock.
Discretizing the state by group, exact review count $t \in \{0,\ldots,30\}$, and a tercile of driver age, the median cell contains $1{,}183$ listings and the 10th percentile $162$; no cell falls below $10$ observations, so we do not collapse the cells further.

The marginal cost of effort $f'(\cdot)$ is not used to recover any of the structural objects below: the identified effort sequence is sufficient to compute the return to effort, the implied reputation-building price discount, and the comparative statics across groups. Appendix~\ref{costeffort} reports a nonparametric approximation of $f(\cdot)$, a functional-form horse race, and out-of-sample checks.

\subsection{Marginal costs}\label{identification_pricing}

Marginal cost is identified from prices through the static pricing first-order condition, applied on the experienced subsample where the continuation value of an additional review is assumed to be zero. Two reduced-form facts support this: the return to reviews has flattened by $t=21$ (Section~\ref{sec:descriptive}), and the Bayesian update on type is essentially degenerate at the displayed grade. On this subsample observed prices solve a static Bertrand-Nash problem,
\begin{align}
p_{i_{\ell t}}^{*} \;=\; c_{i} + \frac{1}{|\gamma|\,(1-s_{i_{\ell t}})},\label{static}
\end{align}
and inverting the first-order condition gives
\begin{align}
\widehat{c}_{i} \;=\; p_{i\ellt} - \frac{1}{|\widehat{\gamma}|\,(1-\widehat{s}_{i\ellt})},\label{mc_inversion}
\end{align}
where $\widehat{\gamma}$ and $\widehat{s}_{imt}$ come from the demand estimates on the experienced-only sample. For $t \ge 21$ the share $\widehat{s}_{i\ellt}$ is read directly from the realized choice menu, so $\widehat{c}_{i}$ does not depend on the equilibrium concept used at the dynamic stage.

\subsection{Dynamic equilibrium}\label{identification_dynamic}

For $t \le t^{*}=20$ observed prices include a shading term. Drivers cut prices to raise the probability of selling a seat and accumulating a review, whose informational value is highest at the start of the career and falls as the posterior tightens. We model dynamic prices and effort as the policy of an oblivious equilibrium following \citet{WeintraubBenkardVanRoy2008,WeintraubBenkardVanRoy2010}, with two adaptations: entry to a route is Poisson at group-specific rate $\lambda_g$ per route--period and exit is a constant hazard $\zeta$; no free-entry condition is imposed.

The OE consistency requirement applies to drivers' conjectures about the competitor distribution, not to passengers' beliefs about types. Passengers update their beliefs about an individual driver's type via the Bayesian rule of Section~\ref{career_concerns}; the market prior $\hat\mu_g$ is held at its estimated value in the baseline equilibrium and perturbed only in the counterfactuals of Section~\ref{counterfactual}.

The driver state is $x = (g, t, \eta, \tilde\mu)$, where $g$ is the group, $t$ the review count, $\eta$ the driver's type, and $\tilde\mu$ the market's posterior mean on the driver's type given her review history. The market state is the marginal distribution $\bar{s}$ of $(g, t, \tilde\mu)$ over active drivers per route--period. Demand depends on $\bar{s}$ only through the inclusive value of competitors,
\begin{equation}\label{eq:Dbar}
\bar{D}(\bar{s}) \;=\; \bar{N}_{\text{route}} \cdot \mathbb{E}_{x' \sim \bar{s}}\!\left[ \exp(V_{\text{nonprice}}(x') + \gamma\, p^{*}(x')) \right] + \exp(V_0),
\end{equation}
where $\bar{N}_{\text{route}}$ is the expected number of competitors per route--period and $V_{\text{nonprice}}$ collects the non-price utility components. An OE is a pair $(\sigma^*, \bar{s}^*)$ with $\sigma^* = (p^*, a^*)$ such that drivers' policies are best responses to $\bar{D}(\bar{s}^*)$ and $\bar{s}^*$ is the long-run distribution induced by $\sigma^*$, $\lambda_g$, and $\zeta$. In words: each driver plays a best response to a fixed long-run distribution of competitors, and that distribution is the one her strategy generates. Appendix~\ref{app:oe} states the equilibrium definition formally and the algorithm used to compute it. The terminal Bertrand stage at $t = t^{*}$ is recomputed against $\bar{D}(\bar{s}^*)$ inside the fixed-point loop, not held at the data inversion. The OE adds no parameters beyond those identified above.

The model carries two distinct time units. The market period is a route--day: it sets the passenger mass that governs realized seat sales and the expected stock of competing listings $\bar N_{\text{route}}$, which enters demand through the competitors' inclusive value $\bar D(\bar s)$ in Equation~\eqref{eq:Dbar}. The driver's decision clock, by contrast, runs in listing-periods: one period elapses each time a driver posts a listing, so a driver with $n$ reviews occupies the same Bellman state regardless of calendar time since entry. Within a listing-period a driver advances from $t$ to $t+1$ only if she sells a seat and is reviewed, and she exits with per-listing hazard $\zeta$. The implicit assumption is that drivers discount and exit per listing rather than per calendar unit, and that posting frequency does not respond to the equilibrium objects.

\paragraph{Calibration of $\lambda_g$ and $\zeta$.} The two clocks meet in the steady-state competitor distribution $\bar s$. Each group enters at rate $\lambda_g$, the average daily count of $t=0$ listings posted by drivers in group $g$ on the route, computed from the panel of listings with the censoring correction in Appendix~\ref{app:oe:entryexit}; the listing-period transitions (advance-on-sale and exit at hazard $\zeta$) then map this inflow into the occupancy of each $(t,\tilde\mu)$ state. Because $\bar s$ is normalized to a probability distribution over competitor states, only the \emph{ratio} $\lambda_{\text{min}}:\lambda_{\text{non}}$ enters (it fixes the group composition and the induced spread over reviews and posteriors) while the absolute daily scale of $\lambda_g$ cancels in the normalization. The \emph{level} of competition is set not by $\lambda_g$ but by the separately calibrated stock $\bar N_{\text{route}}$, the expected number of competing listings per route--day, which multiplies $\bar s$ in $\bar D$. The route--day units of $\lambda_g$ therefore never have to be reconciled with the listing-period clock: they enter only as relative weights. The exit hazard $\zeta$ is platform-wide. We measure it from the data in calendar time and then convert it to the listing clock.

\paragraph{Identifying assumptions.} On the $t \ge 21$ subsample, observed prices are static-FOC best responses, with no remaining option value of reviews. The OE fixed point exists and is stable: the iteration converges within the tolerances $\text{TV}(\bar s)<10^{-3}$, $\max|\Delta p|<5\times10^{-2}$, and $\max|\Delta a|<5\times10^{-3}$ in at most 30 outer iterations from a marginal-cost initialization. Each driver acts on a fixed long-run distribution of competitors $\bar s^*$, and the entry and exit processes are independent of the realization of any individual driver's type or reputation.

\subsection{Estimation algorithm and inference}\label{estimation_algorithm}

Estimation proceeds in four steps. (i)~We estimate the demand coefficients $(\widehat\gamma,\widehat\alpha,\widehat\psi,\widehat\kappa_1,\widehat\kappa_2,\widehat\theta)$ by conditional logit with the price control function. (ii)~We construct the supply-side primitives $(\widehat\eta_i,\widehat\tau_\epsilon,\widehat{\mu}_m,\widehat\tau_m,\widehat a^*_{imt})$ from post-burnout grades. (iii)~We invert marginal costs $\widehat c_i$ from the static first-order condition on the $n \ge 21$ subsample. (iv)~We solve the oblivious equilibrium by iterating between drivers' best responses and the long-run competitor distribution $\bar s^*$ until both stabilize; the algorithm and convergence diagnostics are detailed in Appendix~\ref{app:oe}.

\section{Results}\label{results}

The market underestimates the quality of minority entrants by $1.86$ grades on a $1$--$5$ scale, against a $0.108$-grade gap in true types. In the resulting oblivious equilibrium, minority entrants exert $22\%$ more effort, post a $7.2\%$ introductory price discount against $4.7\%$ for nonminority entrants, and earn $11.6\%$ less expected discounted profit at entry. The remainder of this section reports the demand estimates, the implied entry beliefs, the supply-side primitives, and the equilibrium paths of effort, prices, and value.

\subsection{Demand estimates}

Table~\ref{tab:demand_experience} reports the demand estimates. The minority utility penalty is $\widehat{\kappa}_{1}=-0.145$ (s.e.\ $0.019$) at $0$--$5$ reviews, $\widehat{\kappa}_{2}=-0.088$ (s.e.\ $0.015$) at $6$--$20$, and zero by construction at $21$ or more. The emerging-bin penalty is $61\%$ of the entry-stage penalty, and both are more than four standard errors away from zero. Adding diesel prices to the instrument set leaves the coefficients essentially unchanged, at $-0.140$ and $-0.084$.

The price coefficient is $\widehat{\gamma}=-0.106$ in the smartstop column and $-0.094$ with fuel added. Reputation enters with $\widehat{\alpha}=0.352$ (respectively $0.433$), and $\ln(1+t)$ with $\widehat{\psi}=0.108$. The object that maps the minority utility penalties into belief gaps below is the per-grade utility weight $\alpha_{\text{grade}}=\widehat{\alpha}/4=0.088$ (respectively $0.108$), which rescales the coefficient on the $[0,1]$ rating to the $1$--$5$ grade scale; Section~\ref{results_beliefs} divides each penalty by it, $\widehat{\kappa}/\alpha_{\text{grade}}$, to express the penalty in grades. First stages are strong ($F=16{,}796$ and $F=6{,}245$).

\begin{table}[!htbp] \centering
\caption{Demand estimates: minority penalty by driver experience\label{tab:demand_experience}}
\resizebox{0.6\textwidth}{!}{%
\begin{tabular}{lcc}
\toprule\toprule
 & IV: Smartstop & IV: Smartstop + Fuel \\
 & (1) & (2) \\
\midrule
Price (EUR)                         & $-0.106^{***}$ & $-0.094^{***}$ \\
                                    & (0.001) & (0.001) \\[0.3em]
Price residual (control fn.)        & $0.106^{***}$ & $0.094^{***}$ \\
                                    & (0.001) & (0.001) \\[0.3em]
Reputation (0--1)                   & $0.352^{***}$ & $0.433^{***}$ \\
                                    & (0.051) & (0.051) \\[0.3em]
$\ln(1+t)$             & $0.108^{***}$ & $0.110^{***}$ \\
                                    & (0.002) & (0.002) \\[0.3em]
Minority $\times$ $\mathbf{1}\{t\le 5\}$        & $-0.145^{***}$ & $-0.140^{***}$ \\
                                    & (0.019) & (0.019) \\[0.3em]
Minority $\times$ $\mathbf{1}\{6\le t\le 20\}$  & $-0.088^{***}$ & $-0.084^{***}$ \\
                                    & (0.015) & (0.015) \\
\midrule
Controls                            & Yes & Yes \\
First-stage $F$                     & 16{,}796 & 6{,}245 \\
N (choice situations)               & 1{,}949{,}074 & 1{,}949{,}074 \\
\bottomrule\bottomrule
\end{tabular}
}
\vspace{0.5em}
\caption*{\footnotesize\textit{Notes: Conditional logit with a control-function correction for price endogeneity. The reference category for the minority interaction is drivers with $21+$ accumulated reviews. Column~(1) instruments price with an indicator for SmartStop; Column~(2) adds origin and destination diesel prices on the day of the ride. Neither $\text{reputation}_{it}$ nor $\ln(1+t)$ enters the first stage. Controls include driver age, advance notice, time since posting, automatic acceptance, profile picture, and an outside-option indicator. Standard errors in parentheses. $^{***}$p$<$0.01, $^{**}$p$<$0.05, $^{*}$p$<$0.1. Appendix~\ref{app:iv} reports the 2-piece variant (entry vs.\ 6+) as a robustness check and the full first-stage regressions.}}
\end{table}

\subsection{Market prior beliefs about entrants}\label{results_beliefs}

\begin{figure}[!htbp]
\centering
\caption{Market prior beliefs and realized entrant grades}\label{fig:beliefs_entrant}
\includegraphics[scale=0.8]{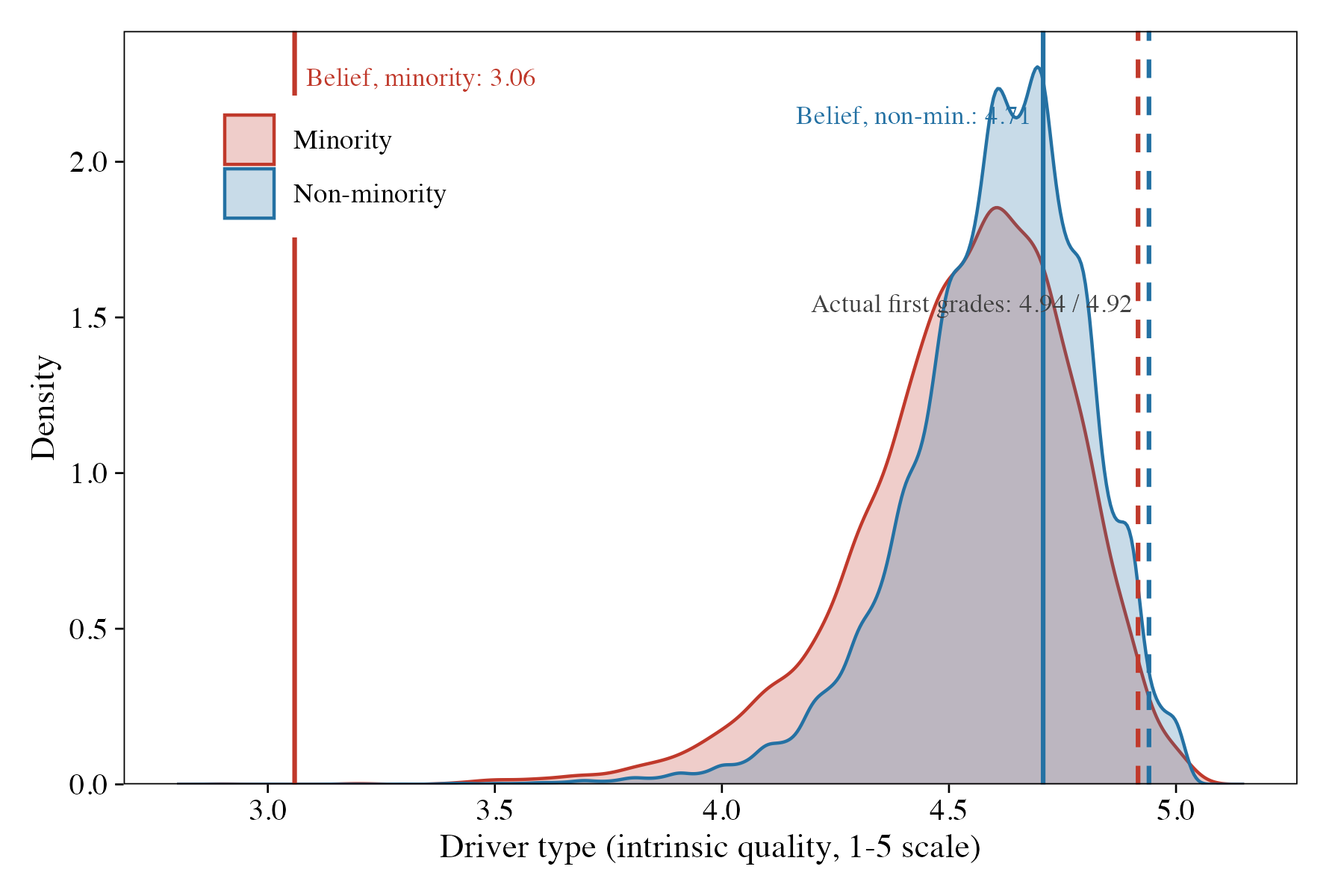}
\caption*{\footnotesize\textit{Note: Shaded densities -- distribution of driver intrinsic types (post-burnout mean grade, $t>t^*$) for minority (red) and nonminority (blue) drivers. Solid vertical lines -- market prior belief about entrant quality for each group, computed from the final demand specification. Dashed vertical lines -- mean of the first two grades received by entrants of each group. The gap between solid and dashed lines is the market's pessimism relative to realized early quality.}}
\end{figure}

The market underestimates minority entrant quality by $1.86$ grades and nonminority entrant quality by $0.23$ grades. Realized first-grade means are $4.94$ for nonminority and $4.92$ for minority entrants, a gap of $0.02$ grades in realized early ratings---distinct from the intrinsic-type gap of $0.108$ grades (Section~\ref{results_supply}), since early grades embed effort and noise as well as type. Expected quality at entry is $4.71$ for nonminority drivers, anchored at the average displayed grade among nonminority drivers with $1$--$5$ reviews,  and $4.71-1.65=3.06$ for minority drivers, where $\widehat{\kappa}_{1}/\alpha_{\text{grade}}=-1.65$ converts the entrant utility penalty to the grade scale. The residual penalty at the $6$--$20$ bin is $-1.00$ grade, measured against the same $21+$ reference. The grade-scale penalty thus declines from $1.65$ at entry to $1.00$ at $6$--$20$ and, by the normalization, to zero at $21+$. Figure~\ref{fig:beliefs_entrant} plots the two priors and the two realized first-grade means against the type density.

\subsection{Supply-side estimates}\label{results_supply}

\paragraph{Type distributions.} The post-burnout type means are $\bar{\mu}_{\text{non}}=4.21$ for nonminority drivers and $\bar{\mu}_{\text{min}}=4.10$ for minority drivers, a $0.108$-grade gap that is $17\times$ smaller than the $1.86$-grade entry-belief gap. The corresponding type precisions are $\tau_{\text{non}}=27.32$ and $\tau_{\text{min}}=18.86$. Throughout this section, $\bar{\mu}_g$ denotes the estimate of the true population mean, distinct from the market belief $\hat{\mu}_g$ defined in Section~\ref{career_concerns}; the same convention is used in Appendix~\ref{appendix_sensitivity}.

\paragraph{Noise precision.} The within-driver variance of post-burnout grades gives $\widehat{\tau}_{\epsilon}=2.74$. The first review shifts the posterior mean by $\tau_\epsilon/(\tau_g+\tau_\epsilon)$, equal to $9.1\%$ for nonminority and $12.7\%$ for minority drivers; the weight on each subsequent review falls as the posterior tightens. Over $t^{*}\in[10,30]$, $\widehat{\tau}_{\epsilon}$ ranges between $2.80$ and $2.92$ on the restricted estimation sample used for that exercise and the type gap between $0.103$ and $0.120$ (Appendix~\ref{appendix_sensitivity}).

\begin{figure}[!htbp]
\caption{Marginal costs}\label{mc}
\centering
\includegraphics[scale=0.6]{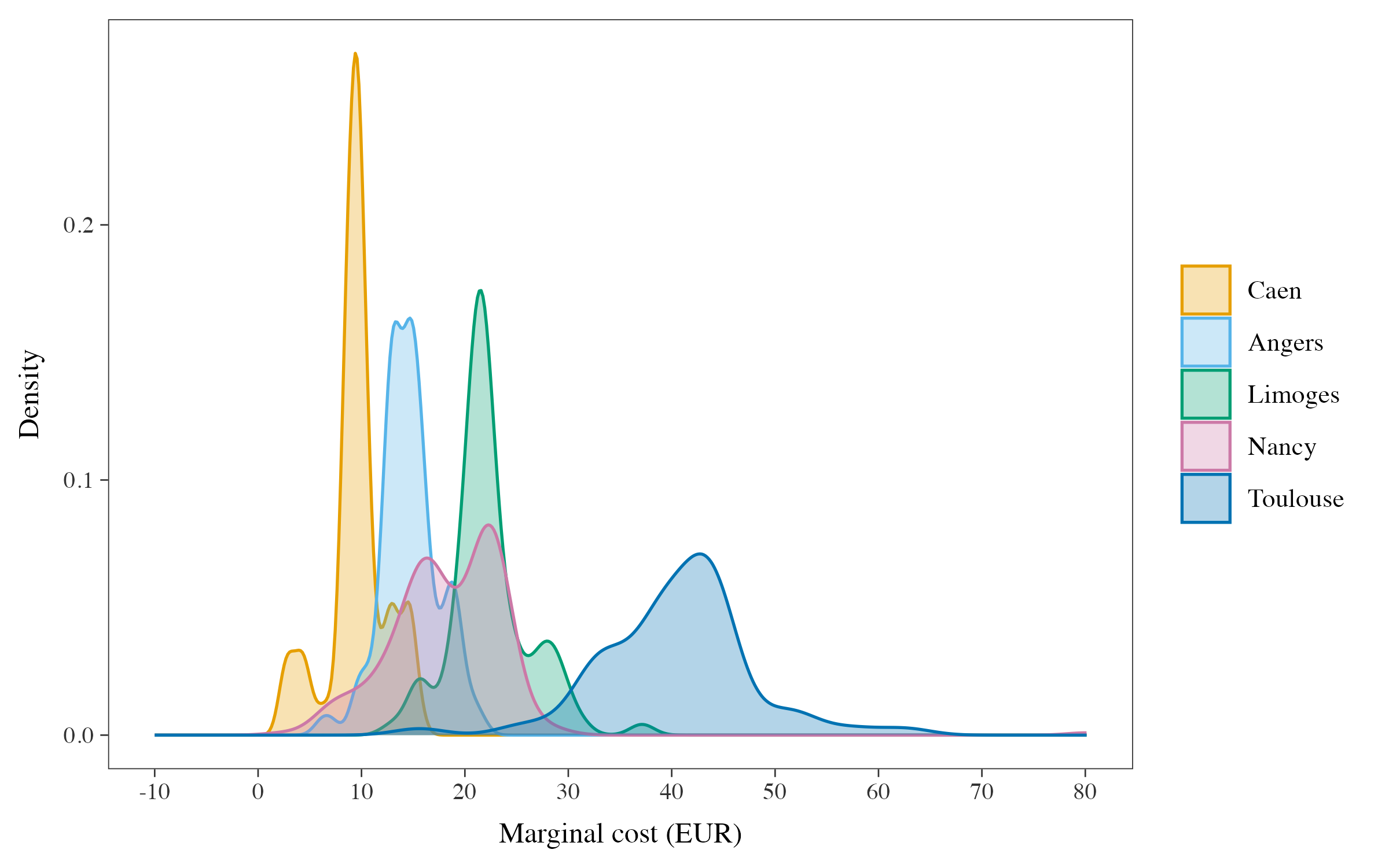}
\caption*{\footnotesize\textit{Note: Marginal costs recovered from equation~\eqref{mc_inversion} on markets where every listed driver has at least $21$ reviews. Selected routes from Paris.}}
\end{figure}

\paragraph{Marginal costs.} Average marginal cost from the Bertrand inversion is $23.3$~EUR for nonminority drivers and $22.5$~EUR for minority drivers, a $3.6\%$ gap. Costs scale with trip length. Figure~\ref{mc} plots the distribution.

\subsection{Equilibrium prices, effort, and value}\label{results_incentives}

\begin{figure}[!htbp]
\centering
\caption{Equilibrium effort and prices over a driver's career}\label{fig:oe_paths}
\begin{subfigure}[b]{0.49\linewidth}
\centering
\includegraphics[width=\linewidth]{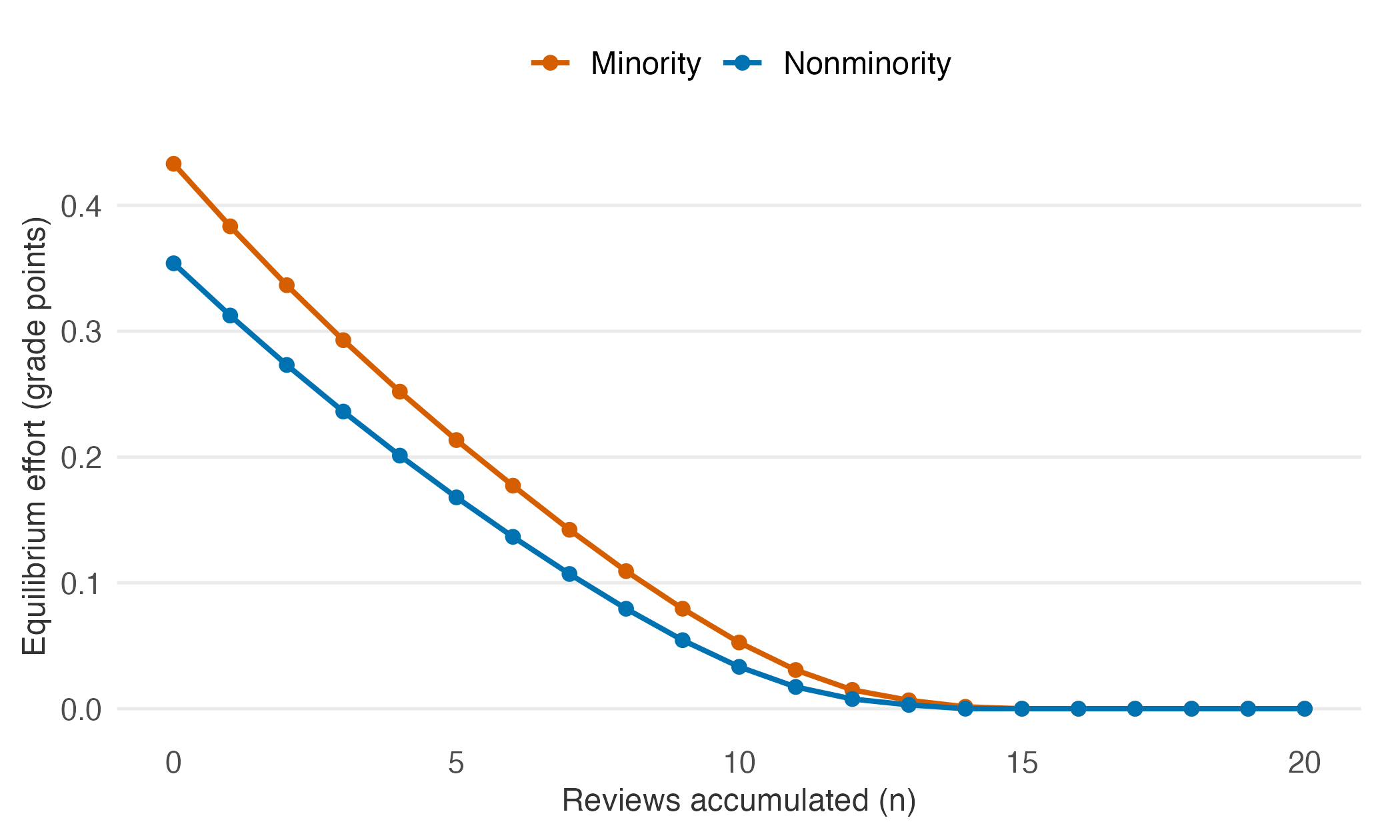}
\caption{Effort}\label{impact}
\end{subfigure}
\begin{subfigure}[b]{0.49\linewidth}
\centering
\includegraphics[width=\linewidth]{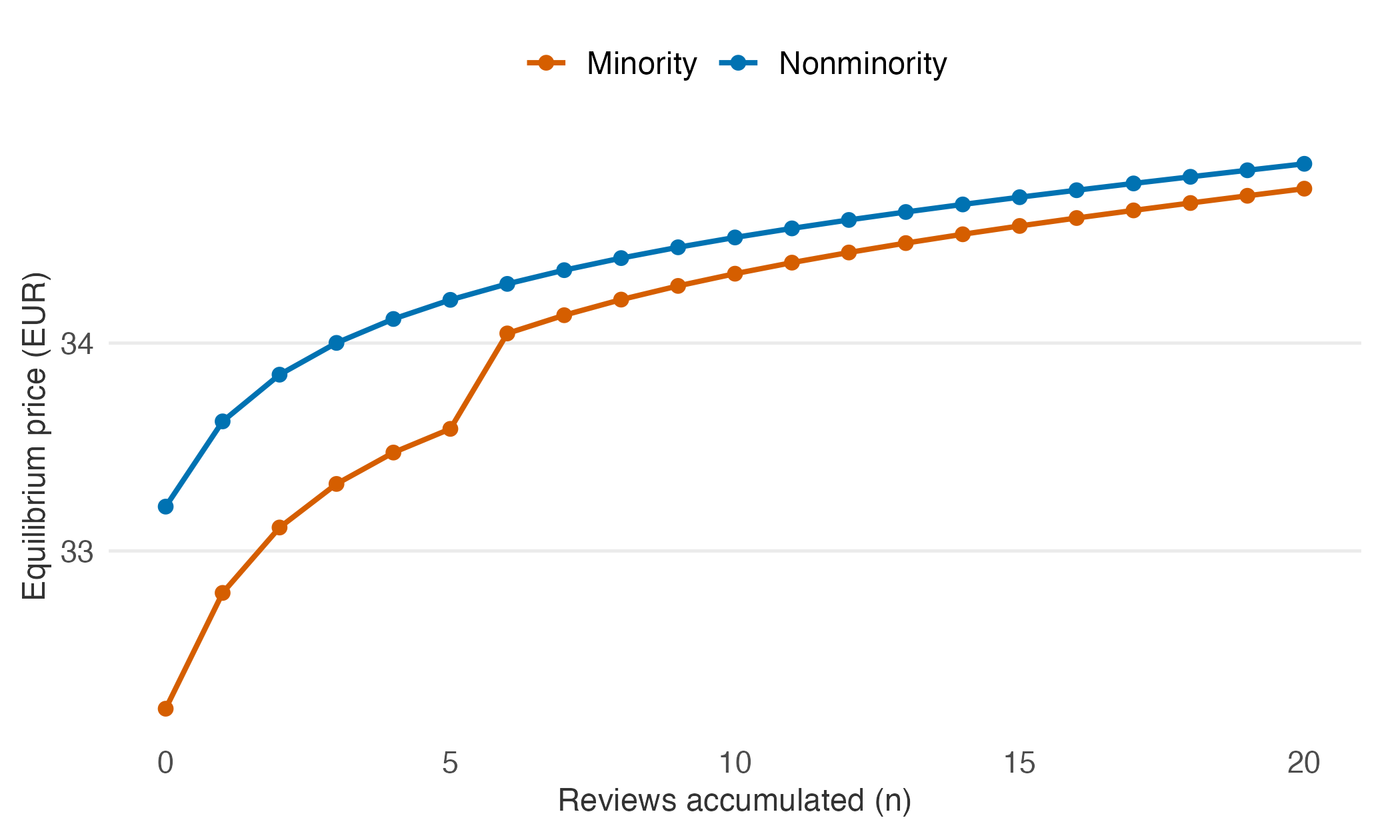}
\caption{Price}\label{avg_prices_by_review}
\end{subfigure}
\vspace{0.3em}
\caption*{\footnotesize\textit{Notes:} Group-flow-weighted equilibrium effort $a^{*}_{g,t}$ (left panel) and prices $p^{*}_{g,t}$ (right panel) across 128 representative markets at the estimated OE, $t=0,\ldots,20$. Red --- minority; blue --- nonminority.}
\end{figure}

\paragraph{Effort over the career.} Equilibrium effort at entry is $a^{*}_{\text{min},0}=0.43$ against $a^{*}_{\text{non},0}=0.35$ grade points, a $22\%$ minority gap. Averaged over $t\in[0,5]$, minority effort is $0.32$ against $0.26$ ($+24\%$); it is largely exhausted by the emerging bin, averaging $0.04$ against $0.03$ over $t\in[6,19]$. The left panel of Figure~\ref{fig:oe_paths} reports the full path.

\paragraph{The introductory price discount.} Both groups shade their entry price below the terminal Bertrand price; the discount is $7.2\%$ for minority drivers (\texteuro $32.24$ versus \texteuro $34.74$) and $4.7\%$ for nonminority drivers (\texteuro $33.21$ versus \texteuro $34.86$). The price gap narrows from \texteuro $0.97$ at $t=0$ to \texteuro $0.12$ at $t=20$. The right panel reports the path. The estimated equilibrium also reproduces the within-driver pattern of prices in the data: sorted into quintiles and deciles, mean observed and model-predicted within-driver price deviations rise together (Appendix~\ref{app:oe:fit}).

\paragraph{The discrimination wedge in entry value.} Expected discounted profit at entry is $V_{\text{min},0}=248.15$ for minority entrants against $V_{\text{non},0}=280.83$ for nonminority entrants, a gap of \euro$32.67$, or $11.6\%$ of the nonminority entry value. Section~\ref{counterfactual} decomposes this gap into a portion correctable through belief revision and a residual statistical-discrimination floor consistent with the $0.108$-grade gap in true type means.

\section{Counterfactual experiments} \label{counterfactual}

We run three counterfactuals on the OE estimated in Section~\ref{results}. Each perturbs a single primitive --- the market prior $\hat\mu_g$, its persistence, or the rating-system precision $\tau_\epsilon$ --- and re-solves the equilibrium. Demand and cost primitives are held fixed, so the comparison isolates the role of the perturbed primitive in the dynamic stage. Because each perturbation shifts the steady-state distribution of competitors, the OE machinery of Section~\ref{identification_pricing} is required: holding the empirical $\Omega_t$ fixed at its baseline value would misstate prices and welfare.

Table~\ref{cfs} summarizes the three experiments. The minority entry value is the most responsive outcome: it gains $5.8\%$ when the prior is corrected and falls $6.1\%$ under persistent bias, while the nonminority entry value moves by less than half a percent in every counterfactual. In euros, the discrimination wedge $V^*_{\text{non}, 0} - V^*_{\text{min}, 0}$ ranges from \euro$16.88$ when the prior is correct to \euro$49.05$ under persistent bias, a factor of three around a baseline of \euro$32.67$. Rating-system precision moves the wedge in the expected direction without closing it.

\begin{table}[!htbp]\centering
\caption{Summary of counterfactuals\label{cfs}}
\begin{tabular}{lcccc}
\toprule\toprule
 & Effort & Intro price & \multicolumn{2}{c}{Entry value} \\
\cmidrule(lr){4-5}
 & $\Delta a^{*}_{\text{min},0}$ & $\Delta p^{*}_{\text{min},0}$ & $\Delta V_{\text{min},0}$ & $\Delta V_{\text{non},0}$ \\
\midrule
Correct prior                               & $+4.61\%$  & $+1.74\%$ & $+5.84\%$ & $-0.47\%$ \\
Persistent bias                             & $-5.40\%$  & $+1.09\%$ & $-6.14\%$ & $+0.41\%$ \\
\addlinespace
High-precision rating ($\tau_\epsilon = 9$) & $+79.11\%$ & $-0.43\%$ & $+2.47\%$ & $-0.19\%$ \\
Low-precision rating ($\tau_\epsilon = 1$)  & $-52.01\%$ & $+0.42\%$ & $-2.39\%$ & $+0.17\%$ \\
\bottomrule\bottomrule
\end{tabular}
\vspace{0.5em}
\caption*{\footnotesize\textit{Notes: Percentage changes against the baseline OE estimated in Section~\ref{results}. Column~1: equilibrium effort at $t=0$ for minority drivers. Column~2: equilibrium intro price for minority drivers. Columns~3 and 4: expected discounted profit at entry for minority and nonminority drivers respectively. Each row solves a new oblivious equilibrium under the perturbed primitive; prices, effort, and the long-run distribution $\bar{s}^*$ all adjust.}}
\end{table}

\subsection{Cost of the incorrect prior}\label{cf:correct_prior}

The first counterfactual sets $\hat\mu_g = \mu_g$ for both groups. The minority entrant belief moves from $3.06$ to $4.53$ on the displayed-grade scale, a $1.47$-grade upward shift. Minority entry value rises by $5.8\%$, and the wedge halves from \euro$32.67$ ($11.6\%$ of nonminority entry value) to \euro$16.88$ ($6.0\%$). Roughly half of the lifetime entry-profit gap between minority and nonminority drivers is therefore attributable to the component of the prior gap that is not warranted by true type means; the residual is a statistical-discrimination floor consistent with the $0.108$-grade gap in true type means, against which a Bayesian passenger optimally still discounts a minority entrant.

Three forces combine to produce this number. Posteriors shift upward at every review count for minority entrants, so the demand they face at any given price rises. Equilibrium intro prices for minority drivers rise by $1.7\%$, and the introductory discount falls from $7.2\%$ to $5.7\%$: with less reputation deficit to overcome, the early-career penalty thins. Demand expansion outweighs the dampened incentive to chase reviews, so equilibrium effort at entry rises by $4.6\%$; the effort first-order condition scales with expected seats per period, which shifts up at every reputation level. Nonminority drivers re-optimize against the new minority pricing and the new long-run competitor distribution, and their entry-state value falls by $0.5\%$.

Because the OE re-solves jointly for policies and the steady-state competitor distribution $\bar{s}^*$, the $5.8\%$ headline incorporates a general-equilibrium adjustment that a single-agent recursion would miss.

\subsection{Persistent bias}\label{cf:persistent_bias}

The second counterfactual makes the bias durable. Under the baseline Bayesian rule of Section~\ref{career_concerns}, a biased group prior eventually washes out: with informative signals and finite prior precision, each driver's posterior converges to her true type $\eta$ as reviews accumulate, so the group-level discount unwinds in the long run. To keep it from unwinding, we re-center the signal the market reads. Passengers apply the same Bayesian recursion, but update each minority driver's posterior toward $\eta - (\mu_{\text{min}} - \hat\mu_{\text{min}})$ rather than toward $\eta$, interpreting observed quality net of effort as if her type were drawn from a distribution centered $\mu_{\text{min}} - \hat\mu_{\text{min}} = 1.47$ grades below the truth. Learning still proceeds and each driver still accumulates an individual reputation, but the posterior now converges to a biased asymptote $1.47$ grades low, so the group-level discount does not unwind.

Minority entry value falls by $6.1\%$, and the wedge widens to \euro$49.05$. The mechanism mirrors Section~\ref{cf:correct_prior} with the sign reversed: the static disadvantage at entry is no longer correctable through review accumulation. The mechanical weight a review carries in updating, $\tau_\epsilon/(\tau_{\text{min}} + t\,\tau_\epsilon)$, is unchanged---a review moves the individual posterior exactly as in the baseline. What falls is the review's economic value: because the posterior now converges $1.47$ grades below the truth rather than to it, accumulating reviews no longer delivers the upward belief revision that drove the return to reputation building. The weaker payoff blunts investment: equilibrium effort at entry falls by $5.4\%$, intro prices rise by $1.1\%$, and nonminority entry value rises by $0.4\%$.

Read against Section~\ref{cf:correct_prior}, the experiment bounds the share of the discrimination wedge that is recoverable through learning rather than fixed by preferences. The $5.8\%$ welfare gain available under correct priors disappears when the bias is structural; the $6.1\%$ loss under persistent bias does not. Reputation building is the active margin distinguishing the two regimes.

\subsection{Rating-system informativeness}\label{cf:rating_system}

The third counterfactual perturbs the signal precision $\tau_\epsilon$. The baseline value $\tau_\epsilon = 2.74$ is a primitive of the platform's rating technology; higher $\tau_\epsilon$ raises the per-review weight in posterior updating and accelerates convergence to the truth. We re-solve the OE at two alternative values: a high-precision regime $\tau_\epsilon = 9$, motivated by the older BlaBlaCar rating system documented in Appendix~\ref{system_changes}, and a low-precision regime $\tau_\epsilon = 1$.

The discrimination wedge shrinks to \euro$26.01$ ($-20.4\%$) under $\tau_\epsilon = 9$ and widens to \euro$39.08$ ($+19.6\%$) under $\tau_\epsilon = 1$. Across the three regimes, the wedge response in welfare units is approximately linear in $\log \tau_\epsilon$ --- a visual reading across three points, but consistent with the direction the model predicts. Sharper ratings shift incentives toward more rapid posterior correction: under $\tau_\epsilon = 9$, minority equilibrium effort at entry rises by $79\%$, intro prices fall by $0.4\%$, and entry value rises by $2.5\%$; nonminority entry value falls by $0.2\%$. The low-precision regime delivers the opposite sign in the four columns: effort falls by $52\%$, intro prices rise by $0.4\%$, and minority entry value falls by $2.4\%$.

The effort response is large and asymmetric --- a $79\%$ jump when the signal sharpens against a halving when it dulls --- because the effort first-order condition scales with $\tau_\epsilon / (\tau_g + t \tau_\epsilon)$, which is steeply nonlinear in $\tau_\epsilon$ at low $t$.

\paragraph{Composition effects.} The model takes entry as exogenous, so we do not solve for the platform-level driver mix. The changes in entry-state value still indicate the direction of any composition response. Minority drivers face a stronger entry incentive when the prior is correct or when ratings are more informative, and a weaker one under persistent bias and a noisier rating system; nonminority incentives move in the opposite direction. We leave the equilibrium implications of endogenous entry to future work.

%% file: new_appendix.tex
\section{Navigation on Blablacar.fr} \label{an_awesome_driver}
\begin{figure}[htbp]
\centering
\caption{Navigating BlaBlaCar: search, listing, and driver profile}\label{fig:navigation}
\begin{subfigure}[t]{0.48\linewidth}
\centering
\includegraphics[width=\linewidth]{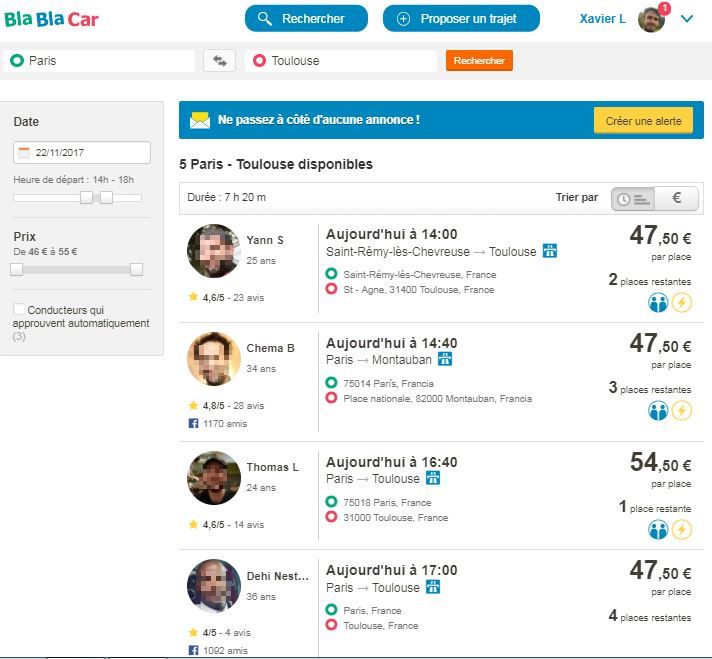}
\caption{Listing offered on a given route}\label{trip_sample}
\end{subfigure}
\hfill
\begin{subfigure}[t]{0.48\linewidth}
\centering
\includegraphics[width=\linewidth]{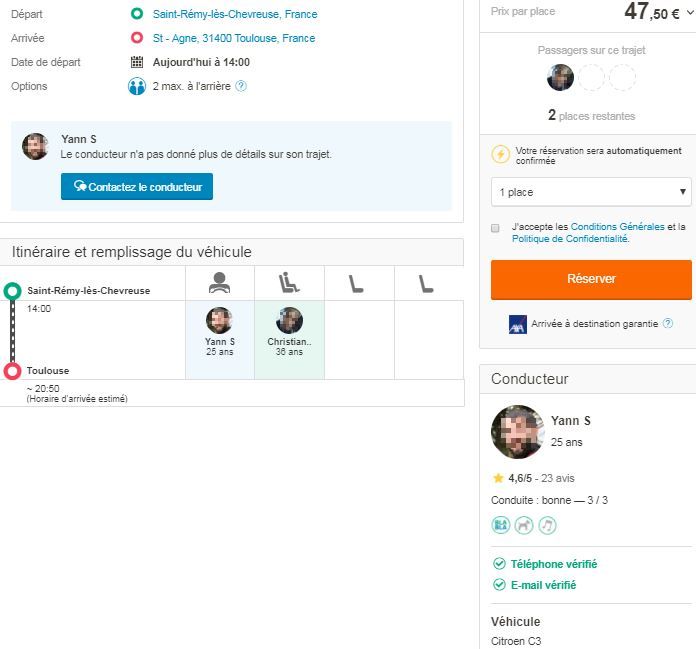}
\caption{Details of a posting}\label{listing_sample}
\end{subfigure}

\vspace{0.6em}

\begin{subfigure}[t]{0.48\linewidth}
\centering
\includegraphics[width=\linewidth]{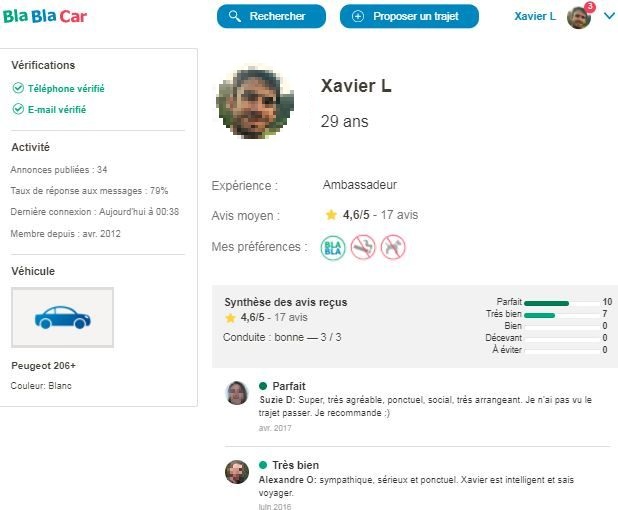}
\caption{A driver's profile}\label{profile_sample}
\end{subfigure}
\end{figure}

A passenger searching for a ride first enters an origin, destination, and date, and is shown a ranked list of matching listings (Figure~\ref{trip_sample}). Clicking a listing opens a page with full trip details (Figure~\ref{listing_sample}). From there, the passenger can either open the driver's profile, which displays the full review history, photo, and biography (Figure~\ref{profile_sample}), or proceed directly to payment. BlaBlaCar's service fee is a function of the posted price.

\section{Changes in the BlaBlaCar reputation system} \label{system_changes}
\begin{figure}[!htbp]
\centering
\caption{Mean latent score and average actual rating by month}\label{informativeness}
\includegraphics[width=0.8\linewidth]{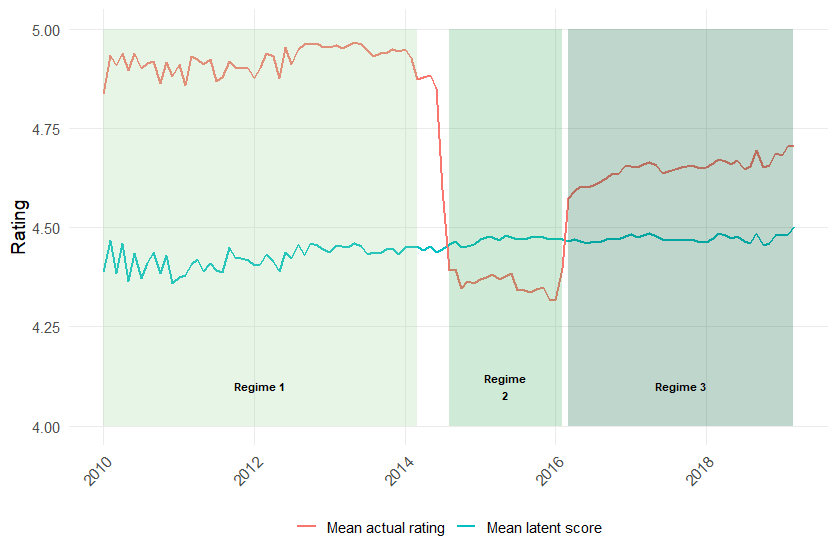}
\vspace{0.3em}
\caption*{\footnotesize\textit{Notes:} Monthly means of the displayed star rating (\emph{mean actual rating}) and of the review-text sentiment score (\emph{mean latent score}), 2010--2019. Shaded bands mark the three reputation-system regimes. Displayed ratings shift sharply at each regime change while latent sentiment stays flat.}
\end{figure}

Our study of demand focuses on the period from July 2017 through March 2019, during which the reputation system remained stable. However, the platform underwent several important changes beforehand. Until February 2014, BlaBlaCar used a binary rating system in which users were simply asked whether they would travel again with the person. The platform subsequently moved into a five-star system. The wording of the rating categories was revised in 2016: \textit{Extraordinaire} (``Extraordinary'') became \textit{Parfait} (``Perfect''), while \textit{Excellent} became \textit{Très bien} (``Very good''). These seemingly minor changes had a substantial impact on the distribution of ratings and even on average ratings, as illustrated in Figure~\ref{informativeness}. Intuitively, users may be more willing to describe a ride as \textit{perfect} than as \textit{extraordinary}, shifting ratings upward even in the absence of any change in underlying ride quality.

In Figure~\ref{informativeness}, we denote by ``Regime 1'' the binary system that prevailed until February 2014, ``Regime 2'' the system in place from August 2014 to January 2016, and ``Regime 3'' the system used from March 2016 onward. For comparability, and consistent with the information displayed on user profiles, reviews from the binary system are mapped into either 1 (negative) or 5 (positive). Between these regimes, transitional periods occurred during which the new rating systems were progressively rolled out.

Average ratings remain relatively stable within each regime but vary substantially across regimes. To assess whether these differences reflect changes in the rating system itself rather than changes in underlying driver quality, we construct a sentiment measure based on the textual content of reviews.

We measure textual sentiment using the Hugging Face Transformers framework (PyTorch backend) and a pretrained French-language sentiment model based on CamemBERT. Each review is tokenized and passed through the model to obtain a probability distribution over five sentiment classes corresponding to one through five stars. The model is a distilled CamemBERT architecture fine-tuned by its authors for supervised sentiment classification; in our implementation, it is used exclusively in inference mode, without any additional fine-tuning. For each review, we construct a latent quality measure equal to the expected-star score,

\[
E[\text{stars}] = \sum_{k=1}^{5} k\,P(k).
\]

\begin{figure}
\centering
\caption{Distribution of model-implied sentiment (expected stars) by review grade.}\label{fig:sentiment_vs_rating}
\includegraphics[width=0.6\linewidth]{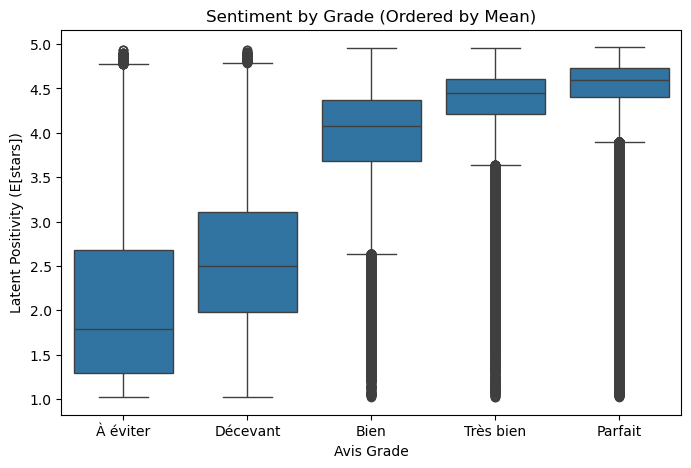}
\vspace{0.3em}
\caption*{\footnotesize\textit{Notes:} Boxes span the interquartile range with the median line; whiskers extend to $1.5\times$ the interquartile range, and points beyond are outliers. Review-grade categories run from \textit{À éviter} (worst) to \textit{Parfait} (best).}
\end{figure}

Figure~\ref{fig:sentiment_vs_rating} illustrates the relationship between NLP-implied sentiment and observed ratings using boxplots by rating category. Sentiment is strongly increasing in the reported rating, indicating that the textual measure is both monotonic and highly discriminative across rating levels.

The green line in Figure~\ref{informativeness} reports the average sentiment extracted from review text. In contrast to observed ratings, sentiment exhibits relatively little variation across regimes, aside from a small but smooth upward trend over time. This pattern suggests that the large changes in observed ratings were primarily driven by changes in the rating system itself rather than by substantial changes in the underlying quality of rides.

To avoid confounding the effects of review accumulation with potential differences in the interpretation of ratings across regimes, all reduced-form and structural analyses in the remainder of the paper, including the estimation of driver types and effort responses, are conducted using data generated under Regime~3 only.

These historical changes also provide an opportunity to quantify the informational content of alternative reputation systems, which motivates the counterfactual analysis in Section \ref{cf:rating_system}. Ideally, one would like to observe how the same ride would have been rated under different reputation systems. Since such counterfactual ratings are unavailable, the precision of a rating system cannot be measured directly. Instead, we exploit textual reviews to construct a proxy for underlying ride quality.

Our identifying assumption is that review sentiment contains information about the quality of the ride that is only imperfectly captured by the reported star rating. Figure \ref{fig:sentiment_vs_rating} provides support for this assumption: model-implied sentiment is strongly increasing in the reported rating category. We therefore interpret the NLP-based sentiment score as a noisy proxy for the driver's latent quality. Because the measure is noisy, the resulting estimates should be viewed as conservative measures of informational content.

To assess the informativeness of each rating regime, we examine the extent to which a single rating predicts future perceived quality. Specifically, for each review received after a driver's twentieth review, we compute the average sentiment of the next ten reviews received by that driver. We then regress this measure of future latent quality on the current rating. The resulting $R^2$ has a natural interpretation: it measures the fraction of variation in future perceived quality that can be explained by a single rating. A more informative rating system should therefore generate a larger $R^2$, since ratings convey more information about persistent underlying quality.

The results are reported in Table \ref{regime_changes}. The omitted category is a rating of 1, so the constant corresponds to the expected future latent quality following the lowest possible rating. Across all regimes, higher ratings predict higher future sentiment, indicating that ratings contain meaningful information about underlying quality. The primary object of interest, however, is the explanatory power of the rating system as a whole. Under the binary system (Regime 1), ratings explain only 0.5\% of the variation in future latent quality. The explanatory power increases more than fourfold under Regime 2 ($R^2=0.021$) and reaches 3.7\% under Regime 3. Relative to Regime 2, the final regime increases predictive power by approximately 75\%. A likely explanation is that the upper end of the Regime 2 scale provided limited discrimination. In particular, the two highest categories, \textit{Extraordinaire} and  \textit{Parfait}, are semantically very similar, yet together account for approximately 93\% of all ratings. As a result, users may have found it difficult to consistently distinguish between these categories. By contrast, the wording adopted in Regime 3 appears to generate more informative distinctions among highly rated rides.

These differences are economically meaningful despite the modest absolute magnitude of the $R^2$ values. Star-rating systems are inherently coarse and discrete measures of quality, while the outcome variable is itself a noisy proxy for latent quality. Consequently, even a highly informative reputation system cannot be expected to explain a large share of the variation in future outcomes. The relevant comparison is therefore relative rather than absolute. From this perspective, the evidence suggests that the later reputation systems, and particularly Regime 3, allow users to communicate substantially more information about ride quality than the original binary system.

\begin{table}[!htbp]\centering
\caption{Informativeness of the rating system, by regime\label{regime_changes}}
\begin{tabular}{lccc}
\toprule\toprule
 & \multicolumn{3}{c}{\textit{Dependent variable:} future latent quality} \\
\cmidrule(lr){2-4}
 & Regime 1 & Regime 2 & Regime 3 \\
 & (1) & (2) & (3) \\
\midrule
Star rating $=2$ &               & $0.070^{***}$ & $0.034^{***}$ \\
                 &               & (0.015)       & (0.006)       \\[0.3em]
Star rating $=3$ &               & $0.148^{***}$ & $0.142^{***}$ \\
                 &               & (0.014)       & (0.006)       \\[0.3em]
Star rating $=4$ &               & $0.199^{***}$ & $0.220^{***}$ \\
                 &               & (0.014)       & (0.007)       \\[0.3em]
Star rating $=5$ & $0.093^{***}$ & $0.215^{***}$ & $0.255^{***}$ \\
                 & (0.026)       & (0.014)       & (0.007)       \\[0.3em]
Constant         & $4.390^{***}$ & $4.273^{***}$ & $4.231^{***}$ \\
                 & (0.027)       & (0.014)       & (0.007)       \\
\midrule
Observations     & 7{,}603       & 221{,}599     & 1{,}960{,}368 \\
$R^{2}$          & 0.005         & 0.021         & 0.037         \\
\bottomrule\bottomrule
\end{tabular}
\vspace{0.5em}
\caption*{\footnotesize\textit{Notes: OLS regressions of future latent quality---the mean review-text sentiment of a driver's next ten reviews---on the current star rating, estimated separately for each reputation regime. The omitted category is a rating of $1$; Regime~1 is the binary system, so only the $5$-star coefficient is identified. Standard errors clustered at the driver level in parentheses. $^{*}$p$<$0.1; $^{**}$p$<$0.05; $^{***}$p$<$0.01.}}
\end{table}

\section{Classification method for gender and ethnicity} \label{classification}
Driver-specific characteristics are key determinants in our model. Hence, the drivers' type must be identified as accurately as possible. Specifically, gender and ethnicity are critical to our analysis. To identify these characteristics, both prospective riders and the econometrician consider two relevant sources of information: the first name and the profile picture. We use both sources to infer gender and ethnicity.

\subsection{Classification of gender}
As a first source of information, we use the name of the driver. We match our dataset of driver names with those of various sources relating first names with ethnicity. The French Government repository of names (\href{https://www.data.gouv.fr/fr/datasets/liste-de-prenoms/}{www.data.gouv.fr/fr/datasets/liste-de-prenoms}) constitutes our main source of information. We complement it with data from other sources.\footnote{\href{http://www.signification-prenom.net/}{www.signification-prenom.net}, \href{http://madame.lefigaro.fr/prenoms/origine/}{www.madame.lefigaro.fr/prenoms/origine}} This data enables us to identify the gender of almost 80\% of drivers, along with 3\% unisex names.

We then use facial recognition to identify gender whenever a picture is available. This process also enables us to identify 80 \% of the dataset. By combining these two processes, we can directly identify gender for 95\% of the dataset.

Further, we use facial recognition to enrich and correct our name database.
Rare or misspelled names (either because the driver registered under a nickname or because of translation variations if the name is not originally French) can be re-classified. This process can identify the gender of some drivers whose names are not listed in our inventories and who do not have a picture (or for pictures where gender is not easily identified) because other drivers with the same name may have posted identifiable pictures. This method brings the precision of our gender identification as high as 99\%.
Panel~A of Figure~\ref{fig:classification} summarizes our identification process.

\subsection{Classification of ethnicity}
Our methodology for the identification of ethnicity follows the same steps and uses the same sources as those for gender classification. First, we collect the origins of names from the data sources mentioned above. This provides the ethnicity of approximately 81\% of our sample. However, names might not be a perfect indicator of ethnicity. Indeed, many visible minorities have a French name for various historical reasons or because they have foreign origins but were born in France. In that  case, a simple name analysis would classify them as non-minorities while they might belong to a minority on the basis of their skin color.

Hence, we  use facial recognition to identify ethnicity whenever a picture is available. The algorithm proposes an ethnicity for 80 \% of the dataset. However, only ``white'', ``black'', ``Asian'' , and ``Latino'' ethnicities are proposed. People of Arabic origin are classified as ``white''. Hence, facial recognition is useful only to classify drivers more accurately between African origin, and majority or Arabic origin.

\begin{figure}[!htbp]
\centering
\renewcommand{\thesubfigure}{\Alph{subfigure}}
\caption{Classification process for gender and ethnicity}\label{fig:classification}
\begin{subfigure}[t]{0.48\linewidth}
\centering
\includegraphics[width=\linewidth]{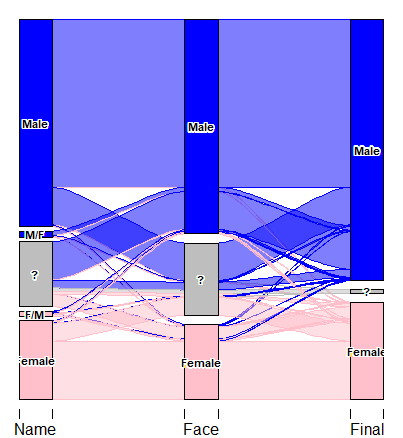}
\caption{Gender}\label{gender_class}
\end{subfigure}
\hfill
\begin{subfigure}[t]{0.48\linewidth}
\centering
\includegraphics[width=\linewidth]{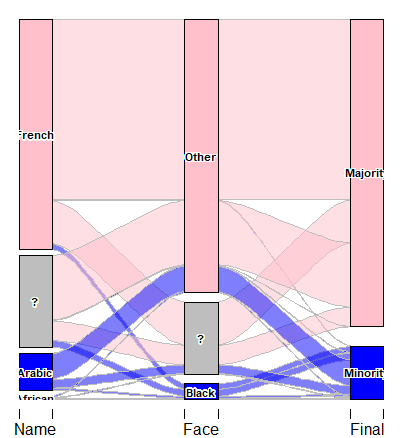}
\caption{Ethnicity}\label{ethnicity_class}
\end{subfigure}
\vspace{0.3em}
\caption*{\footnotesize\textit{Notes:} Alluvial diagrams of the classification flow across three stages---by name (\emph{Name}), by facial recognition (\emph{Face}), and the combined \emph{Final} label---for gender (Panel~A) and ethnicity (Panel~B). Band widths are proportional to the number of drivers; the \emph{?} category collects unclassified or ambiguous cases.}
\end{figure}

We also  use facial recognition to enrich and correct our name repository and to better identify ethnicity.
Overall, facial recognition reclassifies 2.5\% of drivers with a French name and 5\% of drivers with Arabic names (predominantly Muslim names) into Sub-Saharan ethnicity. Including facial recognition increases the sample size for minorities from 11\% to 14\% of our sample. Panel~B of Figure~\ref{fig:classification} summarizes our identification process.

\section{Oversampling of minorities for short-notice rides}\label{scraping_bias}

Our scraper takes snapshots of the listings displayed on the website at a given point in time. Because BlaBlaCar removes a listing once its seats are filled, the scraper observes only listings that have not yet sold out. Even though we conduct multiple scripts each day, very attractive listings, and listings posted into high-demand markets, might be missed. This sampling rule biases the minority gap toward zero: if minority listings are on average less attractive, they remain visible longer and enter the sample more often, while the most attractive nonminority listings drop out before the scraper sees them. The minority gaps we report should thus be read as lower bounds on the true gaps.

\begin{figure}[!htbp]
\centering
\caption{Minority share of visible listings by days until departure}\label{minorities_notice}
\includegraphics[scale=0.6]{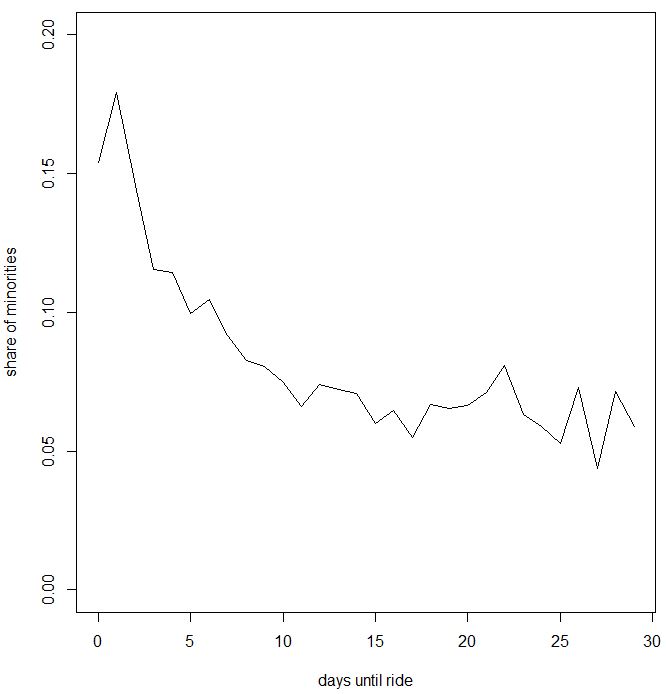}
\caption*{\footnotesize\textit{Note: Share of minority drivers among listings still visible to the scraper, plotted against the number of days until departure (the ride date) at the time each listing is observed. The share peaks close to departure --- the high-demand window in which attractive listings are most likely to have already sold out and left the sample --- and falls to a flat, markedly lower level as the horizon lengthens and sell-out selection weakens.}}
\end{figure}

Figure~\ref{minorities_notice} is consistent with this mechanism. It plots the minority share of visible listings against the number of days until departure at the moment each listing is observed. The share is highest close to departure --- approaching 18\% a day or two before the ride --- precisely the high-demand window in which attractive, disproportionately nonminority listings are most likely to have already sold out and left the sample. 

Two further observations indicate that this selection does not drive our results. First, BlaBlaCar tells drivers that most passengers book only a few days before departure, so most listings still have empty seats when the scraper observes them. Second, the selection works against finding a gap: minority drivers use automatic booking confirmation more often than nonminority drivers (18\% of drivers with automatic confirmation are minorities, versus 12\% with manual confirmation), which speeds up their sell-out rate and pushes in the opposite direction.

\section{Notation glossary and definition of variables}\label{glossary}

\subsection{Notation Glossary}

\begin{table}[htbp]
\centering
\caption{Notation Glossary}
\label{tab:notation_glossary}
\resizebox{\textwidth}{!}{
\begin{tabular}{lll}
\toprule
\textbf{Symbol} & \textbf{Meaning} & \textbf{Comments} \\
\midrule
\multicolumn{3}{l}{\textit{Indices and sets}} \\
$i,j,k$ & Driver, passenger, alternative indices & \\
$t,s$ & Review-count / dynamic-state indices & $t$ used as review count \\
$\ell$ & Market (route-day) index & \\
$g\in\{m,n\}$ & Group label (minority/nonminority) & Structural group identifier \\
$d_{im}$ & Minority indicator in demand equation & Equals 1 for minority driver \\
$N_t$ & Number of active drivers at $t$ & \\
$M_t$ & Number of passengers at $t$ & \\
\midrule
\multicolumn{3}{l}{\textit{Primitives: types, costs, shocks}} \\
$\eta_i$ & Driver $i$ intrinsic type (quality) & Latent, time-invariant \\
$\bar{\mu}_g$ & True mean type in group $g$ & Estimated from data \\
$\hat{\mu}_g$ & Market prior mean belief for group $g$ & May be biased \\
$\tau_g$ & Precision of group-$g$ type distribution & Inverse variance \\
$\tau_\epsilon$ & Precision of review noise & Inverse variance of $\epsilon$ \\
$\epsilon_t$ & Idiosyncratic quality shock & Mean zero \\
$c_i$ & Marginal service cost of driver $i$ & Recovered from pricing FOC \\
$f(a)$ & Effort cost function & Increasing, convex \\
\midrule
\multicolumn{3}{l}{\textit{Quality, history, effort}} \\
$q_{it}$ & One-shot realized quality/review outcome & Quality realization \\
$\mathbf{q}^{it}$ & Review history of driver $i$ up to $t$ & Bold $\mathbf{q}$ denotes history \\
$a_{it}$ & Effort of driver $i$ at $t$ & \\
$a_{it}^*$ & Equilibrium effort & Policy function \\
$\tilde{\mu}$ & Posterior mean belief about driver type & OE state component \\
$t^*$ & Burnout cutoff for strategic behavior & Baseline $t^*=20$ \\
\midrule
\multicolumn{3}{l}{\textit{Demand and utility}} \\
$u_{ij}$ & Utility of passenger $j$ from driver $i$ &  \\
$\varepsilon_{ijt}$ & Logit utility shock & Type-I extreme value \\
$p_{it}$ & Posted price & Driver decision \\
$\alpha$ & Utility loading on expected quality & \\
$\gamma$ & Utility loading on price & Typically negative \\
$\psi$ & Coefficient on $\ln(1+t)$ & Review-count effect \\
$\kappa_1,\kappa_2$ & Minority utility penalties by experience bin & For $t\le5$ and $6\le t\le20$ \\
$X_{it}$ & Observed listing/driver controls & \\
$s_{imt}$ & Choice probability / market share & Logit share \\
\midrule
\multicolumn{3}{l}{\textit{Seats, profits, discrimination}} \\
$\mathcal{S}(p,\mathbf{q})$ & Expected seats sold at $(p,\mathbf{q})$ & Demand outcome \\
$\pi_t$ & Per-period expected profit & $(p_t-c)\,\mathbb{E}[\mathcal{S}(p_t,\mathbf{q}_t)]$ \\
$\delta$ & Discount factor & Intertemporal discounting \\
$D(p, \mathbf{q})$ & Discrimination gap in expected seats sold & Majority minus minority at same $(\mathbf{q},p)$ \\
$\mathcal{A}_{g,t}$ & Mean seats sold in group $g$ at $t$ & \\
$\Delta(\mathcal{A}_t)$ & Gap $\mathcal{A}_{n,t}-\mathcal{A}_{m,t}$ & \\
\midrule
\multicolumn{3}{l}{\textit{Dynamic equilibrium (OE) objects}} \\
$x=(g,t,\eta,c,\tilde{\mu})$ & Driver state vector in OE & \\$\bar{s}$ & Long-run distribution of competitor states & OE population state \\
$\sigma^*=(p^*,a^*)$ & Equilibrium policy functions & Price and effort \\
$\bar{D}(\bar{s})$ & Demand aggregator given $\bar{s}$ & Inclusive-value object \\
$\lambda_g$ & Entry rate of group $g$ drivers & Route-period level \\
$\zeta$ & Exit hazard per listing-period & \\
$V_{g,0}$ & Entry value for group $g$ & Expected discounted profit \\
\bottomrule
\end{tabular}
}
\end{table}

\FloatBarrier

\subsection{Definition of variables}\label{varsdef}
\begin{table}\centering
\caption{Definition of main variables}
\resizebox{\textwidth}{!} {
\begin{tabular}{@{\extracolsep{5pt}}ll}
\\[-1.8ex]\hline
\hline\\[-1.8ex]
\textbf{name of a variable} & \textbf{description}\\
\hline
price & price set by the driver in EUR; has to be lower than maximum price: 0.082 per km\\
age & age of the driver in years\\
reviews & number of reviews received by the driver\\
male & gender defined based on photo recognition and name\\
minority & takes the value of one when the driver is of Arabic or African origin, and zero otherwise;\\
	& defined based on photo recognition and name (see Appendix~\ref{classification} for details)\\
picture & takes the value of one when driver added a picture, and zero otherwise\\
talkative & categorical variable (bla, blabla, blablabla) indicating how talkative the driver is\\
bio & number of words in driver's description\\
ride description & number of words in ride's description\\
reputation & mean of grades received by the driver\\
published rides & number of rides ever published by the driver\\
number of clicks & number of clicks a given listing has received; clicking is necessary for booking a ride\\
& but not sufficient; measured at the moment of data collection\\
sold seats & number of seats already sold; measured at the moment of data collection\\
revenue & sold seats multiplied by price\\
posts per month & mean number of listings posted by the driver since she joined the platform\\
seniority & number of months since the driver joined the platform\\
competition& number of listings available on the same day on the same route\\
median revenue& mean of median revenues in cities of departure and arrival; source: INSEE\\
public transport & travelling time by public transport on the route at listings' departure time; source: Google API\\
train strike & SNCF official strike implicating a given route\\
value of car & price of a comparable car model in thousands of EUR; when a model of a car is not available \\
& mean price of a brand; source: eBay Germany (Kaggle dataset)\\
fuel consumption & mean fuel consumption of a model of a car; when model of a car is not available\\
& mean consumption of a brand; source: ADEME\\
length (km) & distance in km between cities of departure and arrival; source: Google API\\
length (hours) & estimated driving time by a car on a given route and time; source: Google API\\
hours until departure & number of hours between data collection and a ride departure\\
posted since & number of hours between the posting of the listing and data collection\\
automatic acceptance & takes the value of one if booking requests are automatically accepted and zero if the driver chose to \\
&accept/reject requests manually\\
to fuel price & average price of a litre of diesel in a city of arrival in cents\\
from fuel price & average price of a litre of diesel in a city of departure in cents\\
toll viamich & total toll costs on a given route in EUR; source: https://www.viamichelin.com/\\
travel costs & mean of fuel costs multiplied by fuel consumption plus toll fees\\
weekday& takes a value of 1 on weekdays and zero on weekends\\
pets & takes a value of 1 if the driver accepts pets and zero otherwise\\
music & takes a value of 1 if the driver listens to music in the car and zero otherwise\\
smoke & takes a value of 1 if the driver accepts smoking in the car and zero otherwise\\
detour & categorical variable: 1 if no detour, 2 if some detour (up to 15 min), and 3 if more than 15 minutes detour\\
luggage & categorical variable: 1 if no luggage, 2 if small bags, 3 if big bags are allowed\\

\hline\\[-1.8ex]\hline\\[-1.8ex]

\end{tabular}
}

\end{table}

\paragraph{Sources of supplementary data}
\begin{itemize}
\item Database of names constructed from French government statistics (\href{https://www.data.gouv.fr/fr/datasets/liste-de-prenoms/}{data.gouv.fr/datasets/liste-de-prenoms}) and supplementary public sources.
\item Used-car prices from a public Kaggle dataset of eBay Germany listings: \href{https://www.kaggle.com/orgesleka/used-cars-database}{kaggle.com/orgesleka/used-cars-database}.
\item Fuel consumption of cars: French environment and energy management agency (ADEME).
\item City-specific population, median income, crime index, and share of foreign-born residents: French national statistics office (INSEE).
\end{itemize}

\section{Outcomes Gap Between Minority and Majority Drivers}\label{ols}

We estimate the association between minority status and driver outcomes using the following specification:
\begin{equation}
y_{itr} = \alpha + X_{it}'\beta + Z_{i}'\gamma + \tau_{t} + \xi_{r} + \varepsilon_{itr},
\end{equation}
where $i$ indexes drivers, $t$ indexes time, and $r$ indexes routes. The outcome $y_{itr}$ is one of three measures: listing clicks (a proxy for passenger interest), seats sold, or revenue (EUR). The vector $X_{it}$ contains time-varying listing characteristics (auto-acceptance, posting timing, etc.), $Z_{i}$ contains time-invariant driver-level attributes (minority status, gender, age, seniority, reputation), $\tau_{t}$ denotes time fixed effects, and $\xi_{r}$ denotes route fixed effects. We report standard errors that are robust to heteroskedasticity.

Table~\ref{tab:ols_outcomes} presents the results. Minority status is associated with significantly worse outcomes across all three measures. Minority drivers receive 0.6 fewer clicks per listing, sell 0.02 fewer seats, and earn EUR~0.63 less in revenue, all else equal. These differences are economically meaningful: evaluated at the sample mean, minority drivers earn approximately 10 percent less revenue than observationally equivalent majority drivers.

\begin{table}[!htbp]
\centering
\caption{Effect of minority status on driver outcomes}\label{tab:ols_outcomes}
\begin{tabular}{lccc}
\toprule\toprule
 & \multicolumn{3}{c}{\textit{Dependent variable}} \\
\cmidrule(lr){2-4}
 & Clicks & Seats sold & Revenue \\
 & (1) & (2) & (3) \\
\midrule
Minority      & $-0.602^{***}$ & $-0.019^{***}$ & $-0.629^{***}$ \\
              & (0.099)        & (0.003)        & (0.070)        \\[0.3em]
Reviews       & $0.035^{***}$  & $0.002^{***}$  & $0.038^{***}$  \\
              & (0.002)        & (0.000)        & (0.001)        \\[0.3em]
Reviews$^{2}$ & $-0.000^{***}$ & $-0.000^{***}$ & $-0.000^{***}$ \\
              & (0.000)        & (0.000)        & (0.000)        \\
\midrule
Driver and listing characteristics & Yes & Yes & Yes \\
Route FE      & Yes       & Yes       & Yes       \\
Time FE       & Yes       & Yes       & Yes       \\
Observations  & 389{,}211 & 392{,}965 & 388{,}644 \\
$R^{2}$       & 0.250     & 0.074     & 0.076     \\
\bottomrule\bottomrule
\end{tabular}
\vspace{0.5em}
\caption*{\footnotesize\textit{Notes:} OLS estimates. Dependent variables are listing clicks (1), seats sold (2), and revenue in EUR (3). ``Driver and listing characteristics'' comprise gender, driver age, posts per month, bio length, car value, platform seniority, profile photo, automatic acceptance, hours until departure, days since posted, an SNCF strike indicator, and ride-description length. All specifications include route and time fixed effects. Heteroskedasticity-robust standard errors in parentheses. $^{*}$p$<$0.1; $^{**}$p$<$0.05; $^{***}$p$<$0.01.}
\end{table}

The number of reviews is positively associated with all outcomes, with diminishing returns as indicated by the negative coefficient on the quadratic term. This pattern is consistent with passengers valuing reputation and with the informativeness of additional reviews declining as profiles accumulate more feedback. Conditional on the number of reviews, platform seniority is negatively associated with outcomes, suggesting that active reputation-building, rather than mere tenure, drives performance improvements.

Several listing characteristics predict outcomes in intuitive directions. Listings with automatic acceptance generate substantially higher revenue, reflecting passengers' preference for booking certainty. Listings posted further in advance of departure receive more clicks but convert at lower rates. The SNCF railway strike period is associated with large increases in all outcomes, consistent with a positive demand shock to the ridesharing platform during disruptions to rail service.

\section{Minority Output Gap Across Reputation Levels}\label{reputation_effect}

When a driver has no reviews, passengers must rely on observable characteristics, including name and photograph, which reveal ethnicity, to form expectations about service quality. As drivers accumulate reviews, this individual-specific information increasingly shapes passenger beliefs, attenuating the role of group-level priors.

Under statistical discrimination with biased priors, the minority-majority gap should narrow as reviews reveal that minority drivers provide higher quality than initially expected. Under taste-based discrimination, by contrast, the gap should persist regardless of reputation, since passengers would continue to avoid minority drivers even after observing their performance.

To examine how the minority penalty evolves with reputation, we estimate our baseline specification separately for three experience groups: entrants (0--5 reviews), intermediate (6--15 reviews), and experienced drivers (40+ reviews). Table ~\ref{tab:ols_experience} reports results.

\begin{table}[htbp]
\centering
\caption{Effect of Minority Status on Driver Outcomes by Experience Level}
\label{tab:ols_experience}
\bigskip
\resizebox{1\textwidth}{!}{
\begin{tabular}{lccccccccc}
\toprule\toprule
                        & \multicolumn{3}{c}{Clicks} & \multicolumn{3}{c}{Seats Sold} & \multicolumn{3}{c}{Revenue (EUR)} \\
                        \cmidrule(lr){2-4} \cmidrule(lr){5-7} \cmidrule(lr){8-10}
                        & 0--5 & 6--15 & 40+ & 0--5 & 6--15 & 40+ & 0--5 & 6--15 & 40+ \\
\midrule
Minority                & $-0.702^{***}$ & $-0.138$ & $0.001$ & $-0.023^{***}$ & $-0.016^{***}$ & $-0.004$ & $-0.598^{***}$ & $-0.543^{***}$ & $-0.366^{***}$ \\
                        & (0.184) & (0.225) & (0.182) & (0.004) & (0.006) & (0.006) & (0.119) & (0.158) & (0.136) \\[0.6em]
Reviews                 & $-0.182^{***}$ & $0.124^{***}$ & $0.016^{***}$ & $0.008^{***}$ & $0.004^{***}$ & $0.001^{***}$ & $0.220^{***}$ & $0.083^{***}$ & $0.012^{***}$ \\
                        & (0.039) & (0.026) & (0.001) & (0.001) & (0.001) & (0.000) & (0.025) & (0.019) & (0.001) \\[0.6em]
Male                    & $-2.550^{***}$ & $-2.310^{***}$ & $-1.350^{***}$ & $-0.008^{***}$ & $-0.006$ & $0.005$ & $-0.384^{***}$ & $-0.281^{**}$ & $-0.015$ \\
                        & (0.151) & (0.168) & (0.166) & (0.003) & (0.004) & (0.005) & (0.093) & (0.115) & (0.122) \\[0.6em]
Driver age              & $-0.106^{***}$ & $-0.092^{***}$ & $-0.029^{***}$ & $-0.001^{***}$ & $-0.001^{***}$ & $-0.001^{***}$ & $-0.030^{***}$ & $-0.025^{***}$ & $-0.006$ \\
                        & (0.005) & (0.005) & (0.005) & (0.000) & (0.000) & (0.000) & (0.003) & (0.004) & (0.004) \\[0.6em]
Posts per month         & $-0.274^{***}$ & $-0.551^{***}$ & $-0.784^{***}$ & $0.002$ & $-0.006^{***}$ & $-0.016^{***}$ & $0.046^{*}$ & $-0.144^{***}$ & $-0.284^{***}$ \\
                        & (0.039) & (0.049) & (0.025) & (0.001) & (0.001) & (0.001) & (0.024) & (0.031) & (0.015) \\[0.6em]
Seniority (months)      & $-0.022^{***}$ & $-0.019^{***}$ & $-0.045^{***}$ & $-0.000^{***}$ & $-0.001^{***}$ & $-0.001^{***}$ & $-0.008^{***}$ & $-0.015^{***}$ & $-0.024^{***}$ \\
                        & (0.003) & (0.003) & (0.003) & (0.000) & (0.000) & (0.000) & (0.002) & (0.002) & (0.002) \\[0.6em]
Photo                   & $1.810^{***}$ & $1.060^{**}$ & $-0.614$ & $0.001$ & $0.003$ & $-0.023^{*}$ & $0.093$ & $0.162$ & $-0.641^{*}$ \\
                        & (0.333) & (0.493) & (0.445) & (0.007) & (0.012) & (0.014) & (0.226) & (0.359) & (0.346) \\[0.6em]
Auto-accept             & $-0.278^{**}$ & $-1.260^{***}$ & $-2.850^{***}$ & $0.131^{***}$ & $0.121^{***}$ & $0.127^{***}$ & $3.340^{***}$ & $3.030^{***}$ & $2.770^{***}$ \\
                        & (0.140) & (0.156) & (0.140) & (0.003) & (0.004) & (0.004) & (0.099) & (0.118) & (0.104) \\[0.6em]
Hours until departure   & $-0.045^{***}$ & $-0.048^{***}$ & $-0.055^{***}$ & $-0.001^{***}$ & $-0.001^{***}$ & $-0.001^{***}$ & $-0.015^{***}$ & $-0.019^{***}$ & $-0.027^{***}$ \\
                        & (0.001) & (0.001) & (0.001) & (0.000) & (0.000) & (0.000) & (0.000) & (0.001) & (0.001) \\[0.6em]
Days since posted       & $1.300^{***}$ & $1.420^{***}$ & $1.230^{***}$ & $0.006^{***}$ & $0.008^{***}$ & $0.011^{***}$ & $0.159^{***}$ & $0.216^{***}$ & $0.252^{***}$ \\
                        & (0.016) & (0.024) & (0.017) & (0.000) & (0.000) & (0.000) & (0.006) & (0.009) & (0.008) \\[0.6em]
SNCF strike             & $6.440^{***}$ & $6.280^{***}$ & $6.450^{***}$ & $0.096^{***}$ & $0.136^{***}$ & $0.152^{***}$ & $2.570^{***}$ & $2.960^{***}$ & $2.970^{***}$ \\
                        & (0.426) & (0.492) & (0.654) & (0.010) & (0.013) & (0.020) & (0.291) & (0.338) & (0.487) \\[0.5em]
\midrule
Observations            & 116,900 & 83,596 & 103,374 & 118,319 & 84,382 & 104,178 & 117,141 & 83,482 & 102,904 \\
R$^2$                   & 0.255 & 0.260 & 0.257 & 0.065 & 0.063 & 0.082 & 0.061 & 0.070 & 0.101 \\[0.3em]
Route FE                & Yes & Yes & Yes & Yes & Yes & Yes & Yes & Yes & Yes \\
Time FE                 & Yes & Yes & Yes & Yes & Yes & Yes & Yes & Yes & Yes \\
\bottomrule\bottomrule
\end{tabular}
}
\vspace{0.5em}
\parbox{\textwidth}{\footnotesize \textit{Notes:} OLS estimates. Columns report results for three outcome variables (clicks, seats sold, revenue) separately by driver experience level (number of reviews at time of listing). Additional controls for bio length, car value, and ride description length included but not shown. Standard errors robust to heteroskedasticity in parentheses. $^{***}$ p$<$0.01, $^{**}$ p$<$0.05, $^{*}$ p$<$0.1.}
\end{table}

The results reveal a clear pattern: the minority penalty declines monotonically with experience. Table~\ref{tab:ols_summary_experience} summarizes the minority coefficients across outcomes and experience levels.

\begin{table}[htbp]
\centering
\caption{Summary: Minority Penalty by Experience Level}
\label{tab:ols_summary_experience}
\bigskip
\resizebox{0.65\textwidth}{!}{
\begin{tabular}{lccc}
\toprule\toprule
                & 0--5 reviews & 6--15 reviews & 40+ reviews \\
\midrule
\textit{Panel A: Coefficients} \\[0.3em]
Clicks          & $-0.702^{***}$ & $-0.138$       & $0.001$ \\
                & (0.184)        & (0.225)        & (0.182) \\[0.5em]
Seats sold      & $-0.023^{***}$ & $-0.016^{***}$ & $-0.004$ \\
                & (0.004)        & (0.006)        & (0.006) \\[0.5em]
Revenue         & $-0.598^{***}$ & $-0.543^{***}$ & $-0.366^{***}$ \\
                & (0.119)        & (0.158)        & (0.136) \\[0.8em]
                \midrule
\textit{Panel B: Percent of sample mean} \\[0.3em]
Clicks          & $-4.2\%$       & $-0.8\%$       & $0.0\%$ \\
Seats sold      & $-11.5\%$      & $-6.4\%$       & $-1.3\%$ \\
Revenue         & $-11.8\%$      & $-6.9\%$       & $-2.9\%$ \\
\bottomrule\bottomrule
\end{tabular}
}
\vspace{0.5em}
\parbox{\textwidth}{\footnotesize \textit{Notes:} Panel A reports minority coefficients from Table ~\ref{tab:ols_experience}. Panel B expresses these coefficients as percentages of the outcome mean within each experience group. Standard errors in parentheses. $^{***}$ p$<$0.01, $^{**}$ p$<$0.05, $^{*}$ p$<$0.1.}
\end{table}

For revenue, the minority penalty falls from EUR~0.60 among entrants to EUR~0.37 among experienced drivers---a reduction of nearly 40 percent. Expressed relative to sample means, the gap narrows from 11.8 percent to 2.9 percent. For clicks and seats sold, the pattern is even more pronounced: the minority coefficient becomes statistically indistinguishable from zero for experienced drivers.

These findings are consistent with statistical discrimination driven by incorrect prior beliefs. Passengers initially underestimate minority driver quality, but revise their beliefs as reviews accumulate and reveal that minority drivers are far better than the pessimistic prior implies; their realized ratings are close to those of majority drivers and well above the quality the market attributes to them at entry, even though a small true-quality gap between groups remains. The small revenue gap that persists among experienced drivers is consistent with the statistical-discrimination floor implied by that residual quality difference (a post-burnout type gap of about $0.108$ grades (Section~\ref{results_supply})) and may additionally reflect taste-based discrimination or unobserved heterogeneity; the substantial attenuation indicates that belief-based mechanisms dominate.

\section{Panel Estimates}\label{panel}

Our earlier specifications exploit cross-sectional variation, comparing outcomes across drivers at a point in time. As a robustness check, we construct a panel of drivers observed multiple times during our sample period. Of the drivers in our data, 89,614 appear at least twice with complete covariates, yielding an unbalanced panel of 356,454 driver-listing observations. The median driver in the panel is observed 3 times.

We estimate the following specification:
\begin{align}
y_{it} = \alpha + X_{it}\beta + Z_i\gamma + c_i + \tau_{t} + \epsilon_{it},
\end{align}
where $y_{it}$ is revenue for driver $i$ at time $t$, $X_{it}$ contains time-varying listing characteristics, $Z_i$ contains time-invariant driver attributes (including minority status), $\tau_t$ denotes time fixed effects, $c_i$ is an unobserved driver-specific component, and $\epsilon_{it}$ is an idiosyncratic error term. We report three estimators that differ in how they treat $c_i$: pooled OLS folds $c_i$ into the error term, the between estimator regresses driver-level means (so $c_i$ enters the cross-driver residual), and the random-effects estimator models $c_i$ as a random draw uncorrelated with the regressors. 

\begin{table}[htbp]
\centering
\caption{Panel Estimates: Minority Revenue Gap by Experience}
\label{tab:panel}
\bigskip
\resizebox{0.6\textwidth}{!}{
\begin{tabular}{lccc}
\toprule\toprule
                              & \multicolumn{3}{c}{Revenue (EUR)} \\
                              \cmidrule(lr){2-4}
                              & Pooled OLS & Between & Random Effects \\
\midrule
Minority                      & $-0.367^{***}$ & $-0.583^{***}$ & $-0.706^{***}$ \\
                              & (0.123)        & (0.132)        & (0.129) \\[0.8em]
Entrant                       &                & $-2.380^{***}$ &  \\
                              &                & (0.093)        &  \\[0.8em]
Minority $\times$ Entrant     & $0.126$        & $0.174$        & $0.171$ \\
                              & (0.199)        & (0.227)        & (0.205) \\[0.5em]
\midrule
Observations                  & 356,454        & 89,614         & 356,454 \\
Drivers                       & 89,614         & 89,614         & 89,614 \\
R$^2$                         & 0.059          & 0.011          & 0.044 \\
Driver controls               & Yes            & Yes            & Yes \\
Listing controls              & Yes            & Yes            & Yes \\
Route FE                      & Yes            & No             & No \\
Time FE                       & Yes            & No             & Yes \\
\bottomrule\bottomrule
\end{tabular}
}
\vspace{0.5em}
\parbox{0.9\textwidth}{\footnotesize \textit{Notes:} Dependent variable is revenue in EUR. Sample restricted to drivers observed at least twice. ``Entrant'' indicates drivers with 15 or fewer reviews. Pooled OLS and random effects specifications use listing-level observations; between estimator uses driver-level means. Driver controls include gender, age, platform seniority, posts per month, bio length, car value, and photo indicator. Listing controls include auto-accept indicator, hours until departure, days since posted, ride description length, and SNCF strike indicator. Standard errors clustered by driver in parentheses. $^{***}$ p$<$0.01, $^{**}$ p$<$0.05, $^{*}$ p$<$0.1.}
\end{table}

Table~\ref{tab:panel} reports the results. Across all three estimators, minority drivers earn significantly less revenue than observationally equivalent majority drivers. The minority penalty ranges from EUR~0.37 (pooled OLS) to EUR~0.71 (random effects), consistent with the cross-sectional estimates in Table~\ref{tab:ols_outcomes}.

The interaction between minority status and entrant status is positive but not statistically significant at conventional levels. This pattern is directionally consistent with our main finding, that the minority penalty is larger for inexperienced drivers, but the panel sample lacks sufficient power to detect differential effects by experience. The reduction in precision reflects both the smaller sample size and the limited within-driver variation in reputation: most drivers in the panel do not transition from entrant to experienced status during our observation window.

These panel estimates address concerns about time-invariant unobserved heterogeneity across drivers. The persistence of the minority penalty across estimators that difference out or average over driver-specific factors suggests that the cross-sectional results are not driven by systematic differences in unobserved driver quality.

\section{Augmented Inverse Propensity Weighting}\label{app:propensity}

\begin{figure}[t]
\centering
\caption{Propensity Score Distributions by Experience Group}
\label{fig:propensity_overlap}
\includegraphics[width=0.85\textwidth]{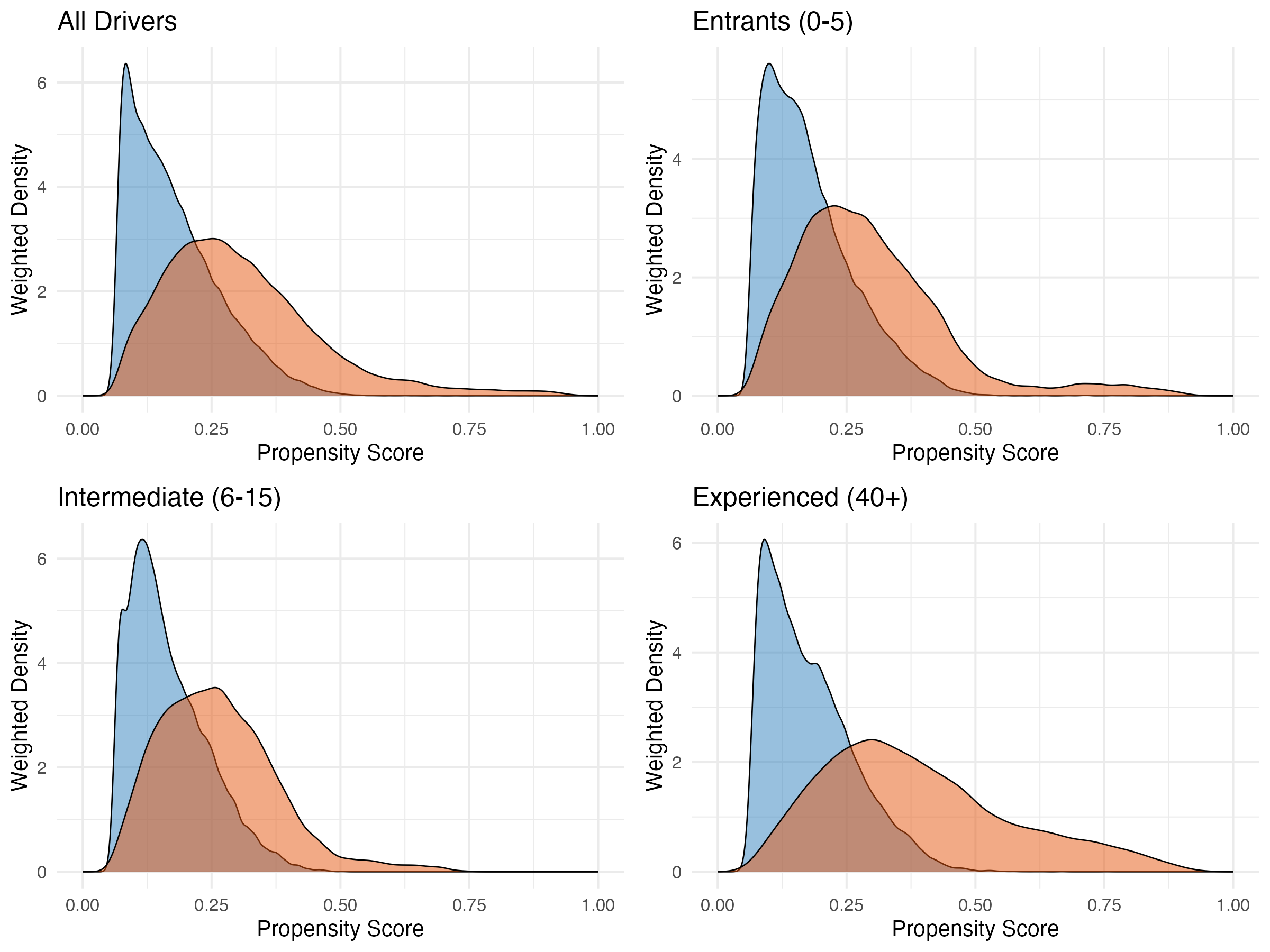}
\vspace{0.3em}
\caption*{\footnotesize\textit{Notes:} Overlap-weighted propensity score densities for nonminority drivers (blue) and minority drivers (orange). Propensity scores are estimated using causal forests. The sample is trimmed to observations with estimated propensity scores in $[0.05, 0.95]$. Distributions are weighted by $\hat{e}(X)(1-\hat{e}(X))$ to reflect the overlap weighting used in the ATE estimates.}
\end{figure}

The validity of propensity score methods depends on adequate overlap between treatment groups across the distribution of estimated propensity scores. Figure~\ref{fig:propensity_overlap} displays the overlap-weighted propensity score distributions for minority and nonminority drivers, separately for each experience group. These distributions correspond to the AIPW estimates reported in Table~\ref{tab:aipw}, where observations are weighted by $\hat{e}(X)(1-\hat{e}(X))$ to emphasize regions of good covariate balance.

Several features of the distributions merit discussion. First, there is substantial common support between minority and nonminority drivers across all experience groups. Both distributions span a similar range of propensity scores, with considerable mass in overlapping regions. This overlap supports the credibility of the AIPW estimates.

Second, the propensity score distributions are concentrated at relatively low values, reflecting the fact that minority drivers constitute approximately 14\% of the sample. The modal propensity score for nonminority drivers lies near 0.10, while for minority drivers it lies near 0.20. Despite this concentration, the overlap weighting ensures that the ATE estimates are identified primarily from regions where both groups are well represented.

Third, the distributions are broadly similar across experience groups, suggesting that the composition of minority and nonminority drivers does not change dramatically with reputation accumulation. This stability supports our interpretation that the attenuation of the minority outcome gap reflects belief updating by passengers rather than differential selection out of the sample.

\section{Drivers' Exit}\label{appendix_exit}

To provide further evidence that selection is not the mechanism behind the reduction of the disparity, in December 2018, we revisited profiles of drivers that appeared in our dataset earlier and collected their newly received reviews. The new data allow us to analyze usage intensity. We define two variables to measure the inactivity of drivers. Variable \textit{exit} takes the value one if no new reviews were received between the last time a given driver appeared in the dataset and December 2018 and is zero otherwise. We also introduce a variable called \textit{disaffection}, which takes the value one if the driver gathered fewer than five new reviews. Table \ref{exits} shows the results of the estimation of a logit model.
\begin{table}[!htbp] \centering
  \caption{Minority entrants are not more likely to exit the platform \label{exits}}
\resizebox{0.65\textwidth}{!} {
\begin{tabular}{@{\extracolsep{5pt}}lcc} 
\\[-1.8ex]\hline 
\hline \\[-1.8ex] 
 & \multicolumn{2}{c}{\textit{Dependent variable:}} \\ 
\cline{2-3} 
\\[-1.8ex] & exit & disaffection \\ 
\hline \\[-1.8ex] 
Minority & $-$0.129$^{***}$ (0.028) & $-$0.097$^{***}$ (0.030) \\ 
Entrant & 1.350$^{***}$ (0.024) & 1.419$^{***}$ (0.025) \\ 
Minority*Entrant & 0.079 (0.065) & 0.065 (0.066) \\ 
 Age & $-$0.005$^{***}$ (0.001) & $-$0.003$^{***}$ (0.001) \\ 
 Male & $-$0.098$^{***}$ (0.018) & $-$0.084$^{***}$ (0.019) \\ 
  Seniority (number of months) & $-$0.005$^{***}$ (0.0003) & $-$0.005$^{***}$ (0.0004) \\ 
  Posts per month & $-$0.731$^{***}$ (0.010) & $-$0.736$^{***}$ (0.011) \\ 
  Bio (number of words) & $-$0.007$^{***}$ (0.001) & $-$0.007$^{***}$ (0.001) \\
  Constant & $-$0.867$^{***}$ (0.053) & $-$1.377$^{***}$ (0.058) \\
   \hline \\[-1.8ex] 
   Other driver characteristics   & X & X \\ 
  Time fixed effects & X & X \\ 
 \hline \\[-1.8ex] 
Observations & 160,923 & 160,923 \\ 
\hline 
\hline \\[-1.8ex] 
\textit{}  & \multicolumn{2}{r}{$^{*}$p$<$0.1; $^{**}$p$<$0.05; $^{***}$p$<$0.01} \\ 
\end{tabular} 
}
\caption*{\footnotesize\textit{Note: Logit regressions, exit and disaffection as dependent variables. }}
\end{table} 

First, minority drivers are more likely to continue using the platform. Second, new drivers are, generally, more likely to quit. However, we find no evidence that minority entrants are leaving the platform more frequently than nonminority entrants.\footnote{The same analysis using the number of listings published (instead of the number of reviews collected) as a proxy for activity on the platform gives similar results.}

\section{Railway Strike: Additional Results}\label{sec:app_strikes}

This appendix provides supporting evidence for the natural experiment analysis in Section~\ref{train}.

\subsection{Covariate Balance}

Table~\ref{tab:strike_balance} compares driver and trip characteristics on strike versus non-strike days during the strike period (April 3--June 28, 2018). The samples are well-balanced across all observable characteristics, supporting the assumption that drivers did not select into treatment based on the strike schedule.

\begin{table}[htbp]
\centering
\caption{Covariate Balance: Strike vs.\ Non-Strike Days}
\label{tab:strike_balance}
\bigskip
\begin{tabular}{lcccc}
\toprule\toprule
                    & Non-Strike Days & Strike Days & Difference & $p$-value \\
\midrule
Minority            & 0.148           & 0.147       & $-$0.001   & 0.72 \\
Male                & 0.726           & 0.729       & 0.003      & 0.45 \\
Reviews             & 28.56           & 28.46       & $-$0.10    & 0.88 \\
Driver age          & 37.59           & 38.09       & 0.50       & 0.31 \\
Car value (000s EUR)& 6.19            & 6.15        & $-$0.04    & 0.67 \\
Platform seniority  & 31.21           & 31.42       & 0.21       & 0.79 \\
Posts per month     & 1.64            & 1.65        & 0.01       & 0.84 \\
Reputation          & 4.640           & 4.640       & 0.000      & 0.99 \\
Trip distance (km)  & 432.1           & 426.8       & $-$5.3     & 0.54 \\
Advance notice (hrs)& 21.52           & 22.30       & 0.78       & 0.38 \\
\bottomrule\bottomrule
\end{tabular}

\vspace{0.5em}
\parbox{0.9\textwidth}{\footnotesize \textit{Notes:} Sample restricted to the strike period (April 3--June 28, 2018). Columns report means for non-strike and strike days, the difference in means, and $p$-values from two-sided $t$-tests. None of the differences are statistically significant at conventional levels.}
\end{table}

\subsection{Alternative Difference-in-Differences Estimates}

As a robustness check, we estimate a traditional difference-in-differences specification of the form:
\begin{align}
y_{it} = \alpha + \beta_1 \text{Treated}_i + \beta_2 \text{Post}_t + \beta_3 (\text{Treated}_i \times \text{Post}_t) + X_{it}\gamma + \epsilon_{it},
\end{align}
where $\text{Treated}_i$ indicates whether driver $i$ drove on at least one strike day --- the same treatment definition as in the main-text analysis --- $\text{Post}_t$ indicates the post-strike period (after June 28, 2018), and $\beta_3$ is the difference-in-differences estimand. Table~\ref{tab:did_traditional} reports results for seats sold and revenue.

\begin{table}[htbp]
\centering
\caption{Traditional Difference-in-Differences Estimates}
\label{tab:did_traditional}
\bigskip
\resizebox{0.75\textwidth}{!} {
\begin{tabular}{lcccccc}
\toprule\toprule
                        & \multicolumn{3}{c}{Seats Sold} & \multicolumn{3}{c}{Revenue (EUR)} \\
                        \cmidrule(lr){2-4} \cmidrule(lr){5-7}
                        & (1) & (2) & (3) & (4) & (5) & (6) \\
\midrule
Treated                 & $-0.042^{***}$ & $-0.024$       & $-0.022$       & $-0.521$       & $-0.438$       & $-0.389$ \\
                        & (0.013)        & (0.015)        & (0.015)        & (0.352)        & (0.401)        & (0.399) \\[0.6em]
Post                    & $-0.154$       & $-0.151$       & $-0.163$       & $-3.594$       & $-3.805$       & $-3.998$ \\
                        & (0.135)        & (0.148)        & (0.148)        & (3.671)        & (4.020)        & (4.018) \\[0.6em]
Treated $\times$ Post   & $0.062^{***}$  & $0.050^{*}$    & $0.050^{*}$    & $1.284^{**}$   & $1.156^{*}$    & $1.178^{*}$ \\
                        & (0.023)        & (0.026)        & (0.026)        & (0.612)        & (0.687)        & (0.685) \\[0.6em]
Minority                & $-0.012^{***}$ & $-0.006^{*}$   & $-0.003$       & $-0.556^{***}$ & $-0.382^{***}$ & $-0.316^{***}$ \\
                        & (0.003)        & (0.004)        & (0.004)        & (0.085)        & (0.097)        & (0.097) \\[0.5em]
\midrule
Observations            & 300,636        & 243,407        & 243,407        & 297,006        & 240,473        & 240,473 \\
R$^2$                   & 0.032          & 0.033          & 0.035          & 0.040          & 0.042          & 0.043 \\[0.3em]
Listing controls        & No             & Yes            & Yes            & No             & Yes            & Yes \\
Driver controls         & No             & No             & Yes            & No             & No             & Yes \\
Route FE                & Yes            & Yes            & Yes            & Yes            & Yes            & Yes \\
\bottomrule\bottomrule
\end{tabular}
}
\vspace{0.5em}
\parbox{\textwidth}{\footnotesize \textit{Notes:} OLS estimates. ``Treated'' indicates drivers who drove on at least one strike day, matching the definition used in the main-text doubly-robust analysis. ``Post'' indicates observations after June 28, 2018. Listing controls include auto-accept indicator, hours until departure, days since posted, and ride description length. Driver controls include gender, age, platform seniority, posts per month, bio length, car value, and photo indicator. Standard errors robust to heteroskedasticity in parentheses. $^{***}$ p$<$0.01, $^{**}$ p$<$0.05, $^{*}$ p$<$0.1.}
\end{table}

The traditional DiD estimates are consistent with the doubly robust estimates reported in Table~\ref{tab:att_revenue_seats}. Treated drivers sell 0.05 more seats and earn approximately EUR~1.18 more per listing in the post-strike period relative to control drivers. These pooled estimates mask the heterogeneity by experience level documented in the main text, but confirm that the positive treatment effect is robust to estimator choice.

\section{Proofs}

\subsection{Proof of Proposition \ref{opt_effort}}\label{proof1}
$a_t$ is the solution to $\sum_{s = t}^{\infty}\delta^{s}\mathbf{E}_{t+s}\left[\frac{\partial\pi_{t+s}}{\partial a_{t}}\right] - f'(a_{t})=0$.  Because $ f'(a)$ is an increasing function of $a$ and $\frac{\partial\pi_{t}}{\partial a_{t}}=0$, it suffices to show that $\frac{\partial\pi_{t+s}}{\partial a_{t}}> \frac{\partial\pi_{t+s+1}}{\partial a_{t+1}}$ for all $s>0$, i.e. we need to show that $\frac{\partial\pi_{t+s}}{\partial a_{t}}$ decreases in $t$.
Recall the driver's optimization problem at time $t+s$:
\begin{align*}
    \max_{a_{t+s}, p_{t+s}}  \mathbb{E} \sum_{k = 0}^{\infty} \delta^k \left[ \pi_{t+s+k}(u_{ij,t+s+m},u_{\mathbf{-i}j,t+s+m}) - f'(a_{t+s+k}) \right],
\end{align*}
where $u_{\mathbf{-i}j,t+s+k}$ is the vector of all utilities for other alternatives, and profit $\pi_{t+s+k}$ increases in $u_{ij,t+s+k}$. By the envelope theorem, we can focus on the direct effect of $a_t$ on $\pi_{t+s}$:
\begin{align*}
    \frac{\partial\pi_{t+s}}{\partial a_{t}} =\frac{\partial \pi_{t+s} }{\partial u_{ij,t+s}}\frac{\partial u_{ij,t+s} }{\partial  a_{t}},
\end{align*}
 with $\frac{\partial \pi_{t+s} }{\partial u_{ij,t+s}}>0$ by assumption of the proposition. Using Equations (\ref{utility}) and (\ref{posterior}) for s>0, we derive:
\begin{align*}
    \frac{\partial u_{ij,t+s} }{\partial  a_{t}} = \alpha  \frac{\tau_{\epsilon}}{\tau_g +(t+s)\tau_{\epsilon}}
\end{align*}
given that $\alpha>0$, $\frac{\partial u_{ij,t+s} }{\partial  a_{t}}$ decreases in $t$. In turn, $\frac{\partial\pi_{t+s}}{\partial a_{t}}$ decreases in $t$ for all $s>0$, which proves the proposition.

\subsection{Proof of Proposition \ref{prop_prices}}\label{proof_prices}
As a driver receives reviews, passengers update their beliefs about expected quality and its variance,
\begin{align}
\lim_{t\rightarrow\infty}\left\{\mathbf{E}[q_{it}|\mathbf{q}_{it}]\right\} = \lim_{t\rightarrow\infty}\left\{\frac{\tau_{g}\hat{\mu}_{g}}{\tau_{g}+t\tau_{\epsilon}}+\frac{\tau_{\epsilon}}{\tau_{g}+t\tau_{\epsilon}}\sum_{s=1}^{t}(\eta_{i}+\epsilon_{s})\right\} =\eta_{i}\lim_{t\rightarrow\infty}\left\{ \frac{t}{\frac{\tau_{g}}{\tau_{\epsilon}}+t}\right\} =\eta_{i}.
\end{align}

Here the accumulated review noise vanishes: writing $\frac{\tau_{\epsilon}}{\tau_{g}+t\tau_{\epsilon}}\sum_{s=1}^{t}(\eta_{i}+\epsilon_{s})=\frac{t\tau_{\epsilon}}{\tau_{g}+t\tau_{\epsilon}}\big(\eta_{i}+\tfrac{1}{t}\sum_{s=1}^{t}\epsilon_{s}\big)$, the prefactor tends to $1$, while $\tfrac{1}{t}\sum_{s=1}^{t}\epsilon_{s}$ has mean zero and variance $1/(t\tau_{\epsilon})\to 0$ and hence converges to zero in mean square and by the strong law of large numbers, since the $\epsilon_{s}$ are i.i.d.; together with the vanishing prior term $\frac{\tau_{g}\hat{\mu}_{g}}{\tau_{g}+t\tau_{\epsilon}}$, this leaves $\eta_{i}$.

\begin{align}
\lim_{t\rightarrow\infty}\left\{\mathbf{Var}\left[\mathbf{E}[q_{it}|\mathbf{q}_{it}]\right]\right\}= 0
\end{align}
Consequently, demand converges to the complete-information demand \(S(p,\eta)\) and an additional review has an asymptotically negligible effect on beliefs. Therefore, in the limit when $t\rightarrow\infty$, and assuming that the problem admits a unique interior maximizer, types are fully revealed and drivers solve the problem in (\ref{objective_in_limit}).

\subsection{Informational value of a match: belief shift}\label{cor_beliefshift}
Corollary \ref{cor_prices} derives from the fact that increasing an individual's belief about own quality  $\hat\eta=\hat\mu+\tau$ relative to market belief ($\hat\mu$) raises the informational value of a match. This makes raising the price (which lowers match probability) less attractive.
Therefore, we want to show that $\frac{dp_t^*}{d\tau}<0$.

Recall the driver's problem at date $t$:
\begin{align}
\max_{a_t,p_t}\;\mathbb{E}\Bigg[\sum_{s=0}^\infty \delta^s (p_{t+s}-c)\,D(u_{ij,t+s},u_{-ij,t+s})\Bigg].
\label{obj_p_a_beliefshift}
\end{align}

Let $\hat\eta$ be the belief of the driver regarding own quality:
\[
\hat\eta=\hat\mu+\tau,
\]
so that $\tau$ measures how much better the driver believes they are relative to what buyers believe.
Buyers' demand depends only on beliefs $\hat\mu$, not on the true $\eta$, so for any fixed history $\mathbf{q}_{t-1}$,
\[
\frac{\partial D_t}{\partial \tau}=0.
\]

Let $p_t^*(\tau)$ be the optimal price given $\hat\eta=\hat\mu+\tau$.
The first-order condition can be written as
\[
F(p_t,\tau)=\frac{\partial\Pi_t(p_t,\tau)}{\partial p_t}=0,
\]
with second-order condition $F_p(p_t^*(\tau),\tau)<0$.
By the implicit function theorem,
\[
\frac{dp_t^*}{d\tau}
=-\frac{F_\tau(p_t^*(\tau),\tau)}{F_p(p_t^*(\tau),\tau)},
\]
so the sign of $\frac{dp_t^*}{d\tau}$ is the sign of $F_\tau$.
We now expand $F(p_t,\tau)$.  Current profit is
\[
\Pi_t=p_tD_t-cD_t+\delta\,\mathbb{E}_t^\tau[V_{t+1}(\mathbf{q}_t)],
\]
so the FOC is
\[
F(p_t,\tau)
=D_t+(p_t-c)D_{t,p}
+\delta\,\frac{\partial}{\partial p_t}\mathbb{E}_t^\tau[V_{t+1}(\mathbf{q}_t)],
\]
where $D_{t,p}\equiv\partial D_t/\partial p_t$.
Since $D_t$ depends only on beliefs $\hat\mu$ and $p_t$, and $\hat\mu$ is held fixed,
\[
\frac{\partial}{\partial \tau}\bigg(D_t+(p_t-c)D_{t,p}\bigg)=0.
\]
Thus
\[
F_\tau(p_t,\tau)
=\delta\,\frac{\partial}{\partial \tau}
\bigg[ \frac{\partial}{\partial p_t}\mathbb{E}_t^\tau[V_{t+1}(\mathbf{q}_t)] \bigg].
\]

\medskip
\noindent
At time $t$, a match occurs with probability $D_t$ and no match with probability $1-D_t$.  $D_t$ is proportional to the probability to receive a review. Assume a share $\rho$ of passengers leave a review. Let
\[
V_{t+1}^1(\tau)=\mathbb{E}_t^\tau[V_{t+1}\mid\text{match at }t],
\qquad
V_{t+1}^0(\tau)=\mathbb{E}_t^\tau[V_{t+1}\mid\text{no match}].
\]
Then
\[
\mathbb{E}_t^\tau[V_{t+1}(\mathbf{q}_t)]
=\rho D_tV_{t+1}^1(\tau)+(1-\rho D_t)V_{t+1}^0(\tau),
\]
and differentiating w.r.t.\ $p_t$ gives
\[
\frac{\partial}{\partial p_t}\mathbb{E}_t^\tau[V_{t+1}(\mathbf{q}_t)]
=\rho D_{t,p}\big(V_{t+1}^1(\tau)-V_{t+1}^0(\tau)\big).
\]
Define
\[
\Delta V_t(\tau)\equiv V_{t+1}^1(\tau)-V_{t+1}^0(\tau),
\]
the incremental informational value of a match at $t$.
A match generates a review whose expected distribution shifts upward when $\hat \eta=\hat\mu+\tau$ increases.
Hence the expected posterior belief next period is higher, so future demand and expected profits conditional on a match increase.
Therefore
\[
\frac{\partial \Delta V_t(\tau)}{\partial\tau}>0.
\]

\medskip
\noindent
Putting this together,
\[
F_\tau(p_t,\tau)
=\delta\,\rho D_{t,p}\,\frac{\partial \Delta V_t(\tau)}{\partial\tau}.
\]
We know $D_{t,p}<0$ and $\partial\Delta V_t(\tau)/\partial\tau>0$, so
$F_\tau(p_t,\tau)<0$. Since $F_p(p_t^*,\tau)<0$ (by the second order condition),
\[
\frac{dp_t^*}{d\tau}
=-\frac{F_\tau}{F_p}
=\frac{F_\tau}{|F_p|}
<0.
\]

  \subsection{Proof of Proposition \ref{disc_dyn}}\label{disc_dyn_proof}
  Discrimination is defined as $D(p,\mathbf{q}) \equiv \mathbf{E}\left[\mathcal{S}(p,\mathbf{q})|n\right] -\mathbf{E}\left[\mathcal{S}(p,\mathbf{q})|m\right]$, the difference in expected sold seats between a driver $i$ from $n$ and a driver $j$ from $m$ evaluated at the \emph{same} price $p$ and review history $\mathbf{q}$; we show that it vanishes as $t\rightarrow\infty$ when $\eta_{i} = \eta_{j}$. By Equation (\ref{posterior}), for a fixed history the group-specific prior receives vanishing weight, so $\lim_{t\rightarrow\infty} \mathbf{E}_{t+1}\left[\eta_i|\mathbf{q}_{t}\right]=\eta_i$ for each group; two drivers with the same history and the same true type thus share the same limiting expected quality. Evaluated at the common price $p$, the gross utility in (\ref{utility}) then converges to the same value for $i$ and $j$, so

  {{\footnotesize
  \begin{align}
  \lim_{t\rightarrow\infty}D(p,\mathbf{q}) =&\lim_{t\rightarrow\infty} \left\{ M_{t} \frac{\exp \left(\alpha\mathbf{E}_t[q_{it}|\mathbf{q}]+\gamma p\right) }{1 +\sum_{k=1}^{N}\exp(\alpha \mathbf{E}_t[q_{kt}|\mathbf{q}_{kt}] +\gamma p_{kt})}
  - M_{t} \frac{\exp \left(\alpha\mathbf{E}_t[q_{jt}|\mathbf{q}]+\gamma p \right)}{1 +\sum_{k=1}^{N}\exp(\alpha \mathbf{E}_t[q_{kt}|\mathbf{q}_{kt}] +\gamma p_{kt})}\right\} \nonumber \\
  &=0
  \end{align}
  }
  which proves the proposition.

\section{Grades do not depend on prices\label{gradesprices}}
We investigate whether grades depend on the prices. We regress obtained grades on price, reputation, and controls. We find that in the OLS estimation there is a positive impact of prices on grades. However, after instrumenting the prices with cost shocks and controlling for driver-specific unobservable effect, we find that the effect is statistically insignificant.
\begin{table}[!htbp]
\centering
\caption{Impact of prices on grades}\label{tab:grades_prices}
\begin{tabular}{lcc}
\toprule\toprule
 & \multicolumn{2}{c}{\textit{Dependent variable:} grade} \\
\cmidrule(lr){2-3}
 & OLS & Panel IV \\
 & (1) & (2) \\
\midrule
Price      & $0.003^{*}$   & $-0.016$      \\
           & (0.002)       & (0.067)       \\[0.3em]
Reputation & $0.655^{***}$ & $0.483^{***}$ \\
           & (0.021)       & (0.076)       \\
\midrule
Driver FE              &          & Yes     \\
Driver characteristics & Yes      & Yes     \\
Time effects           & Yes      & Yes     \\
Route effects          & Yes      & Yes     \\
Listing effects        & Yes      & Yes     \\
Observations           & 10{,}828 & 1{,}072 \\
\bottomrule\bottomrule
\end{tabular}
\vspace{0.5em}
\caption*{\footnotesize\textit{Notes:} Column (1) reports pooled OLS; column (2) uses within-driver variation in prices, instrumented with cost shocks (time and spatial variation in prices and highway tolls). The dependent variable is the review grade. Standard errors in parentheses. $^{*}$p$<$0.1; $^{**}$p$<$0.05; $^{***}$p$<$0.01.}
\end{table}

\section{Instrumental Variables}\label{app:iv}

To address price endogeneity in the context of discrete choice models, we employ a control function approach \citep{petrin2010control}. The control function method extends two-stage least squares (2SLS) to nonlinear models by explicitly modeling the endogeneity through inclusion of first-stage residuals in the second-stage estimation. This approach is particularly well-suited to our conditional logit framework, where standard IV estimators are computationally intractable with large choice sets.

\subsection{Instrument: SmartStop}

Our primary instrument is an indicator for \emph{SmartStop}, a feature introduced by BlaBlaCar to increase the number of potential matches between drivers and passengers. When a driver posts a trip, the platform can algorithmically generate additional ride offers corresponding to sub-segments of the driver's journey, even if the driver did not explicitly declare these intermediate pick-up or drop-off locations. These automatically generated offers are referred to as \emph{SmartStops}. \cite{AstierBouquetLambin2026}

Importantly, SmartStops are displayed to passengers in exactly the same way as regular rides. However, because drivers did not explicitly create these offers, BlaBlaCar mechanically inflates their price relative to the driver's usual per-kilometer rate in order to increase the likelihood that drivers will accept subsequent booking requests. As a result, SmartStops generate plausibly exogenous variation in prices while remaining unobservable to passengers at the time of booking.

\subsection{Estimation Procedure}

Our estimation follows a two-stage control function approach.

\paragraph{Stage 1: First-Stage Regression}
We estimate a linear regression of ride price on the instrument and exogenous covariates:
\begin{equation}
\text{price}_{im} = \pi_0 + \pi_1 \text{smartstop}_{m} + \mathbf{X}_{im}'\boldsymbol{\pi} + \nu_{im}
\end{equation}
where $i$ indexes drivers, $m$ indexes markets (route-day combinations), $\text{smartstop}_{m}$ indicates the presence of a SmartStop station, and $\mathbf{X}_{im}$ contains the exogenous driver and trip characteristics (minority status, driver age, notice, posting recency, auto-acceptance, profile picture). Crucially, $\mathbf{X}_{im}$ does \emph{not} include $\text{reputation}_{im}$ or $\ln(1+n_{im})$: conditioning on these endogenous quality signals in the first stage absorbs the price variation we want the instrument to pick up and shrinks the residual toward the price itself, which in turn drives the reputation coefficient in the second stage to implausibly large values. We compute heteroskedasticity-robust standard errors and test instrument strength using the Wald $F$-statistic.

\paragraph{Stage 2: Control Function Estimation}
We construct the price residual (control function) as $\widehat{\nu}_{im} = \text{price}_{im} - \widehat{\text{price}}_{im}$ and include it alongside price in the conditional logit:
\begin{equation}
U_{ijm} = \gamma \cdot \text{price}_{im} + \beta_{\text{resid}} \cdot \widehat{\nu}_{im} + \alpha \cdot \text{reputation}_{im} + \delta \cdot minority_{i} + \mathbf{X}_{im}'\boldsymbol{\beta} + \varepsilon_{ijm}
\end{equation}
where $minority_{i}$ is a minority indicator. Under the null of price exogeneity, $\beta_{\text{resid}} = 0$. A significant coefficient rejects exogeneity, confirming that price is correlated with unobserved demand factors.

\subsection{Robustness: Alternative Instrument Sets}

As a robustness check, we augment the smartstop instrument with origin and destination diesel prices on the day of the ride. With smartstop alone the first-stage $F$-statistic is $16{,}796$; adding the two fuel-price instruments reduces this to $6{,}245$ (the denominator of the $F$ scales with the number of instruments). Both are well above conventional weak-instrument thresholds. The second-stage coefficients are close across the two specifications: price moves from $-0.106$ to $-0.094$, reputation from $0.352$ to $0.433$, and the minority interactions are essentially unchanged (see Column~2 of Table~\ref{tab:demand_experience} in the main text). We take the single-instrument specification as the baseline because smartstop-station placement is plausibly more orthogonal to contemporaneous demand shocks than day-level fuel prices, while using the fuel-augmented version to document that the results are not driven by a single source of cost variation.

\subsection{Robustness: 2-piece minority interaction}\label{app:iv_2piece}

The main demand specification splits the minority interaction into three pieces ($n\le 5$, $6\le n\le 20$, $21+$ reference). A simpler 2-piece version collapses the last two bins into a single reference group of $6+$ reviews and keeps only the entry-stage interaction. Table~\ref{tab:demand_experience_2piece} reports the estimates with the same two IV variants used in the main text. The entry-stage penalty is $-0.137$ with smartstop alone and $-0.132$ with smartstop + fuel; the price and reputation coefficients land within a few percent of their main-table counterparts.

\begin{table}[!htbp] \centering
\caption{Demand estimates, 2-piece minority interaction (appendix robustness)\label{tab:demand_experience_2piece}}
\begin{tabular}{lcc}
\toprule\toprule
 & IV: Smartstop & IV: Smartstop + Fuel \\
 & (1) & (2) \\
\midrule
Price (EUR)                         & $-0.106^{***}$ & $-0.094^{***}$ \\
                                    & (0.001) & (0.001) \\[0.3em]
Price residual (control fn.)        & $0.106^{***}$ & $0.094^{***}$ \\
                                    & (0.001) & (0.001) \\[0.3em]
Reputation (0--1)                   & $0.367^{***}$ & $0.448^{***}$ \\
                                    & (0.051) & (0.051) \\[0.3em]
$\ln(1+\text{reviews})$             & $0.110^{***}$ & $0.111^{***}$ \\
                                    & (0.002) & (0.002) \\[0.3em]
Minority $\times$ $\mathbf{1}\{n\le 5\}$ & $-0.137^{***}$ & $-0.132^{***}$ \\
                                    & (0.019) & (0.019) \\
\midrule
Reference category                  & $n\ge 6$ & $n\ge 6$ \\
Controls                            & Yes & Yes \\
First-stage $F$                     & 16{,}796 & 6{,}245 \\
N (choice situations)               & 1{,}949{,}074 & 1{,}949{,}074 \\
\bottomrule\bottomrule
\end{tabular}
\vspace{0.5em}
\caption*{\footnotesize\textit{Notes: Same sample and first stage as Table~\ref{tab:demand_experience}; the only difference is the minority interaction, which is collapsed to a single entry indicator ($n\le 5$) with the rest absorbed into the reference. Standard errors in parentheses. $^{***}$p$<$0.01.}}
\end{table}

Collapsing the middle bin (emerging drivers, $6\le n\le 20$) into the reference category mechanically mixes the emerging-stage penalty of roughly $-0.088$ with a zero penalty for $21+$ reviews; the net effect is a slightly smaller estimated entry coefficient in the 2-piece version because the reference group is now contaminated with residual discrimination. The main text retains the 3-piece specification to keep that middle-stage penalty identified separately.

\section{Estimation of the cost of effort function}\label{costeffort}
We are interested in estimating function $g(a_{i,t})$ that measures the cost of exerting effort. The optimal levels of effort, in our model, are determined by the following relation:
\begin{align}
a_{imt}=\gamma\left(\sum_{s=t}^{n}\beta^{s-t}\frac{h_{\epsilon}}{h_{mk}}\frac{\alpha}{\gamma}\mathbb{E}[M_{k}s_{ik}]\right) +\varepsilon_{ijt}
\end{align}
where $\gamma(\cdot)= g^{-1,'}(\cdot)$. In the baseline case the cost of effort follows a quadratic function: $g(a) = c_1 a + c_2 a^2$. The discounted sum of future profits depends on the discount factor $\beta$, which we calibrate at $\beta = 0.96$. Using the IV demand estimates ($\alpha_{\text{grade}} = 0.088$, $\gamma = -0.106$) and the estimated supply-side parameters ($h_\epsilon = 2.74$, $h_{\text{mino}} = 18.86$, $h_{\text{non}} = 27.32$), we compute the discounted sum of future profit impacts for each driver-review observation and regress observed effort on this incentive measure.

The baseline specification yields $g(a) = 0.0604a + 0.1230a^2$, with an $R^2$ of 7.5\%. The convex cost function is consistent with the theoretical prediction: drivers with stronger reputation-building incentives exert more effort, but at a diminishing rate.

To select the functional form, we fit polynomials of degree one through five on the discounted sum of profits and compare them using ANOVA. The quadratic term significantly improves fit over the linear model ($F = 15{,}794$, $p < 0.001$), and the cubic term provides a further improvement ($F = 7{,}116$, $p < 0.001$). However, the quartic and quintic terms yield negligible gains in $R^2$ ($0.092$ vs.\ $0.091$ for the cubic). We adopt the quadratic specification as our baseline given its parsimony and theoretical motivation from the career concerns model.

\input{appendix_sensitivity_supply}

\section{Oblivious equilibrium: definition, algorithm, and calibration}\label{app:oe}

This appendix states the oblivious equilibrium (OE) formally, calibrates the entry and exit processes, and reports the fixed-point algorithm together with its convergence diagnostics.

\subsection{Formal definition}

The OE follows \citet{WeintraubBenkardVanRoy2008} with one substantive departure: the market's prior beliefs about group types, $\hat\mu_g$, are held fixed at the estimated value rather than being part of the equilibrium fixed point. Drivers play best responses to a long-run distribution of competitors, but the population-level belief about each group's type does not adjust in response to play.

A driver in group $g \in \{\text{min}, \text{non}\}$ is described by the state $x = (g, n, \eta, c, \tilde\mu)$, where $n \in \{0,1,\dots,t^*\}$ is the review count, $\eta$ the post-burnout type, $c$ the driver's marginal cost, and $\tilde\mu$ the market's posterior mean on type given the driver's review history. Marginal cost is a fixed driver characteristic, drawn at entry from the group-specific cost distribution recovered from the static pricing FOC (Section~\ref{identification_pricing}) and held constant over the driver's career. The market state on a route is the marginal distribution $\bar s$ of $(g, n, c, \tilde\mu)$ over active drivers, per route-period (one listing opportunity); $\eta$ does not enter $\bar s$ because, under symmetric information, both the perceived quality passengers use and the equilibrium price $p^*$ depend on the belief $\tilde\mu$ and cost $c$ rather than on the unobserved true type. Demand $\bar D$ depends on $\bar s$ only through the inclusive value of competitors,
\begin{equation*}
\bar D(\bar s) \;=\; \bar N_{\text{route}} \cdot \mathbb{E}_{x' \sim \bar s}\!\left[\exp(V_{\text{nonprice}}(x') + \gamma\, p^{*}(x'))\right] + \exp(V_0),
\end{equation*}
where $\bar N_{\text{route}}$ is the expected number of competitors per route-period, $V_{\text{nonprice}}$ collects the non-price utility components, and $V_0$ is the outside option's index.

\begin{definition}[OE]\label{def:oe}
An oblivious equilibrium on a route is a pair $(\sigma^*, \bar s^*)$ with $\sigma^* = (p^*, a^*)$ such that
\begin{enumerate}
\item For every state $x$, $(p^*(x), a^*(x))$ maximizes the driver's expected discounted profit when she takes $\bar D(\bar s^*)$ as fixed and updates $\tilde\mu$ through the Bayesian rule of Section~\ref{career_concerns}.
\item $\bar s^*$ is the stationary distribution induced by $\sigma^*$, the group-specific Poisson entry rates $\lambda_g$, and the constant exit hazard $\zeta$.
\item At the terminal state $n = t^*$, the price reverts to the static Bertrand-Nash inversion $p^*(x) = c + 1/(|\gamma|(1 - s(x)))$, where $s(x)$ is the share implied by $\bar D(\bar s^*)$ and $c$ is recovered from the static FOC on the experienced subsample.
\end{enumerate}
\end{definition}

The OE adds no parameters beyond the earlier estimation and calibration stages: $\bar N_{\text{route}}$ and the cost distribution are pinned down from the data; the policy follows from the demand coefficients $(\widehat\gamma, \widehat\alpha, \widehat\psi, \widehat{\delta}_1, \widehat{\delta}_2)$, the supply primitives $(\widehat h_\epsilon, \widehat\mu_g, \widehat h_g)$, and the estimated effort-cost schedule $f(\cdot)$ (Appendix~\ref{costeffort}), which the Bellman recursion for the effort policy $a^*$ requires. The discount factor is calibrated at $\delta = 0.96$ (Appendix~\ref{costeffort}), and the entry and exit rates $(\lambda_g, \zeta)$ are calibrated below.

\subsection{Entry and exit calibration}\label{app:oe:entryexit}

\paragraph{Entry rate $\lambda_g$.} The entry rate of group $g$ on a route is the average daily count of listings with $n = 0$ posted by drivers in $g$. We compute this from the panel of listings with a censoring correction for drivers whose first observed listing is not their first on the platform: the profile creation date pins down platform tenure, and listings preceding the scrape window are imputed at the driver's observed listing frequency.

\paragraph{Exit hazard $\zeta$.} The exit hazard is platform-wide. A driver-week is \emph{active} if it contains at least one listing; the driver \emph{survives} from week $w$ to week $w+1$ if her next listing falls within seven weeks. Excluding the last month of the panel to avoid right-censoring, the empirical weekly exit hazard is $\zeta_{\text{week}} = 0.749$. The dynamic stage is solved in listing-periods, with $\bar\ell = 1.56$ listings per active driver-week. Under a constant per-listing hazard, $1 - \zeta_{\text{week}} = (1-\zeta)^{\bar\ell}$, so
\begin{equation*}
\zeta = 1 - (1 - \zeta_{\text{week}})^{1/\bar\ell} = 0.59.
\end{equation*}

\subsection{Algorithm}

We solve Definition~\ref{def:oe} by alternating between drivers' best response and the stationary distribution. Initialize $\bar s^{(0)}$ as the long-run distribution under the static Bertrand-Nash policy and $\sigma^{(0)}$ as the static policy. At outer iteration $k$, compute $\bar D^{(k)} = \bar D(\bar s^{(k)})$, solve the Bellman equation under $\bar D^{(k)}$ on the discrete state grid $(n, \tilde\mu)$ to obtain $(p^{(k+1)}, a^{(k+1)})$, simulate the resulting state transitions to obtain a candidate distribution $\tilde s^{(k+1)}$ that integrates over entry, exit, and Bayesian belief updating, and update with damping $\bar s^{(k+1)} = (1-\omega) \bar s^{(k)} + \omega \tilde s^{(k+1)}$. The loop stops when $\|\bar s^{(k+1)} - \bar s^{(k)}\|_{TV} < \varepsilon_s$ and the maximum price and effort changes fall below $\varepsilon_p$ and $\varepsilon_a$. At convergence, the terminal Bertrand stage at $n = t^*$ is recomputed against $\bar D(\bar s^*)$; this is inside the fixed-point loop, not held at the data inversion.

The implementation uses $\omega = 0.30$, $\varepsilon_s = 10^{-3}$, $\varepsilon_p = 5\times 10^{-2}$, $\varepsilon_a = 5\times 10^{-3}$, with a cap of $30$ outer and $8$ inner iterations. Continuous states are discretized: $\tilde\mu$ on a grid spanning the group priors $\pm 4$ standard deviations, $\eta$ via Gauss-Hermite quadrature with group-specific precisions.

\subsection{Convergence diagnostics}

We solve the OE on $128$ representative routes. The iteration converges within tolerance on every route in at most $30$ outer iterations; the median is well below the cap. 
No market oscillates or returns a non-stationary $\bar s$. The damping parameter $\omega = 0.30$ was chosen after a pilot run in which $\omega = 1$ produced limit cycles on a few high-share routes.

\subsection{Model fit: within-driver prices}\label{app:oe:fit}

We compare the baseline's prices to the data on the $128$ solved routes, among entrants with $n\le 20$ reviews. Each listing is assigned the equilibrium price $p^{*}$ for its $(\text{route},g,n)$ cell and matched to its posted price. Because about $94\%$ of the raw price variance is across routes (trip distance), which the per-route OE normalizes away, we demean both series by driver and study within-driver deviations, which isolates the introductory-discount channel the model targets ($96{,}033$ listings, $31{,}700$ drivers observed at least twice).

\begin{figure}[!htbp]
\centering
\caption{Within-driver price calibration: observed versus predicted deviations}\label{fig:calib_price}
\includegraphics[width=0.95\textwidth]{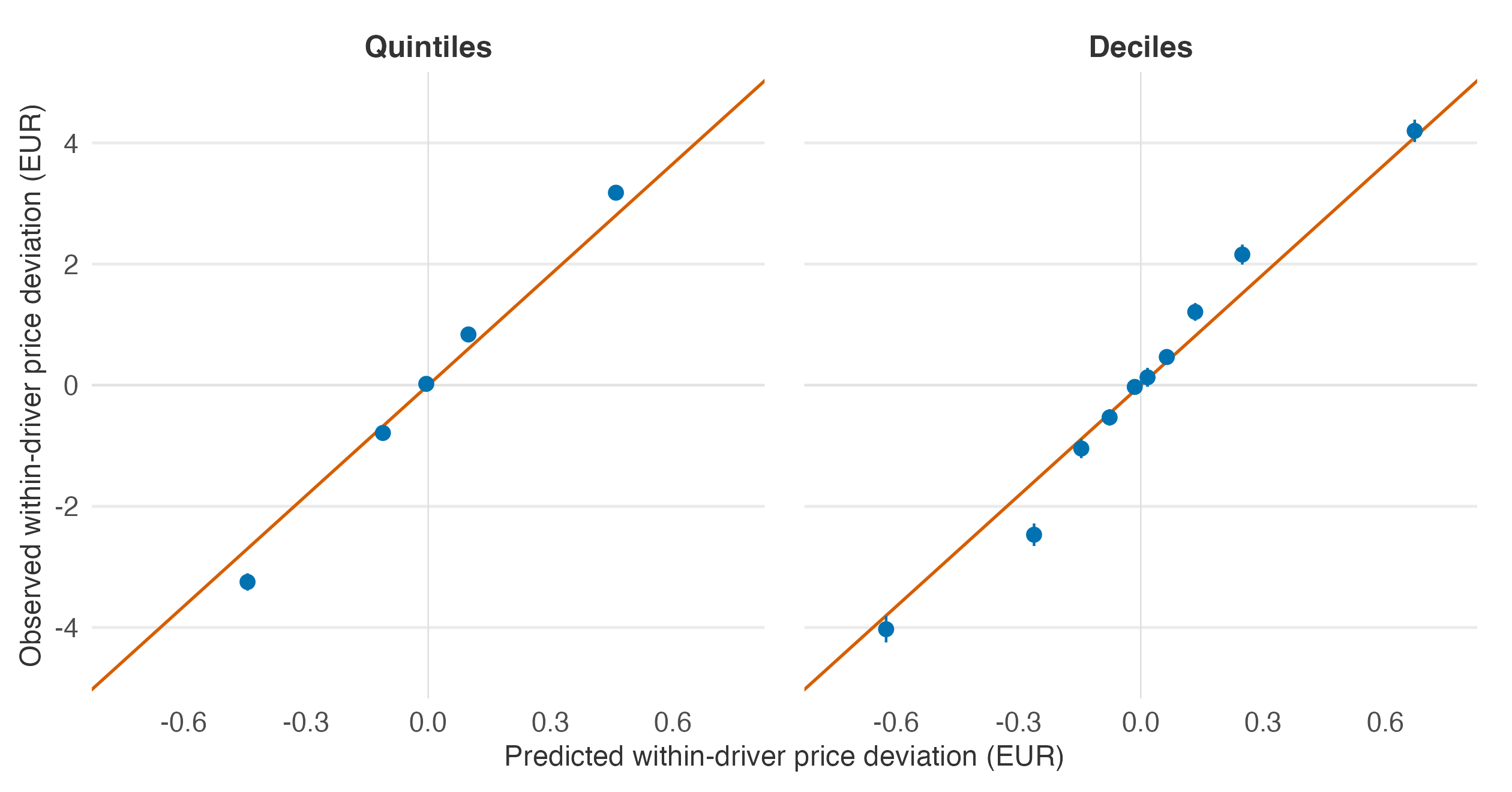}
\caption*{\footnotesize\textit{Note:} Listings on the $128$ solved routes with $n\le 20$ reviews ($96{,}033$ listings from $31{,}700$ drivers observed at least twice). Each listing is assigned the OE-predicted price for its $(\text{route},g,n)$ cell; observed and predicted prices are then demeaned by driver. Points are mean observed within-driver price deviations (\euro), by quintile (left panel) and decile (right panel) of the predicted deviation, with $95\%$ confidence intervals. The line is the OLS fit of the observed deviation on the predicted one.}
\end{figure}

Predicted and observed deviations move together (Figure~\ref{fig:calib_price}): sorted into quintiles and deciles of the prediction, the mean observed deviation rises monotonically, with a within-driver correlation of $0.26$. Levels match, \euro$32.9$ observed against \euro$32.7$.

%% file: appendix_sensitivity_supply.tex
\section{Sensitivity of Supply-Side Estimates to Cutoff Choices}\label{appendix_sensitivity}

The supply-side estimation described in Section \ref{identification} requires two researcher-chosen cutoffs: (i) the burnout period $t^*$, after which driver effort is assumed to be negligible, so that the average grade approximates the driver's intrinsic type $\eta_i$; and (ii) the minimum number of post-$t^*$ reviews required for a driver to be included in the type estimation sample. The baseline specification uses $t^* = 20$ and requires at least 2 post-cutoff reviews. This appendix examines the sensitivity of all supply-side parameter estimates to local perturbations of these cutoffs.

\paragraph{Methodology.} We vary $t^*$ over the grid $\{10, 15, 17, 18, 19, 20, 21, 22, 23, 25, 30\}$ and the minimum post-$t^*$ review requirement over $\{1, 2, 5, 10, 15, 20\}$. For each combination, we re-estimate all supply-side parameters---mean types $\hat{\mu}_g$, standard deviations $\hat{\sigma}_g$, and noise precision $\hat{\tau}_\epsilon$---using the identical estimation procedure as in the baseline. The data and sample restrictions are unchanged.

\begin{figure}[H]
\centering
\caption{Supply-side parameter estimates as a function of the burnout cutoff $t^*$\label{fig:sensitivity_theta}}
\includegraphics[width=0.85\textwidth]{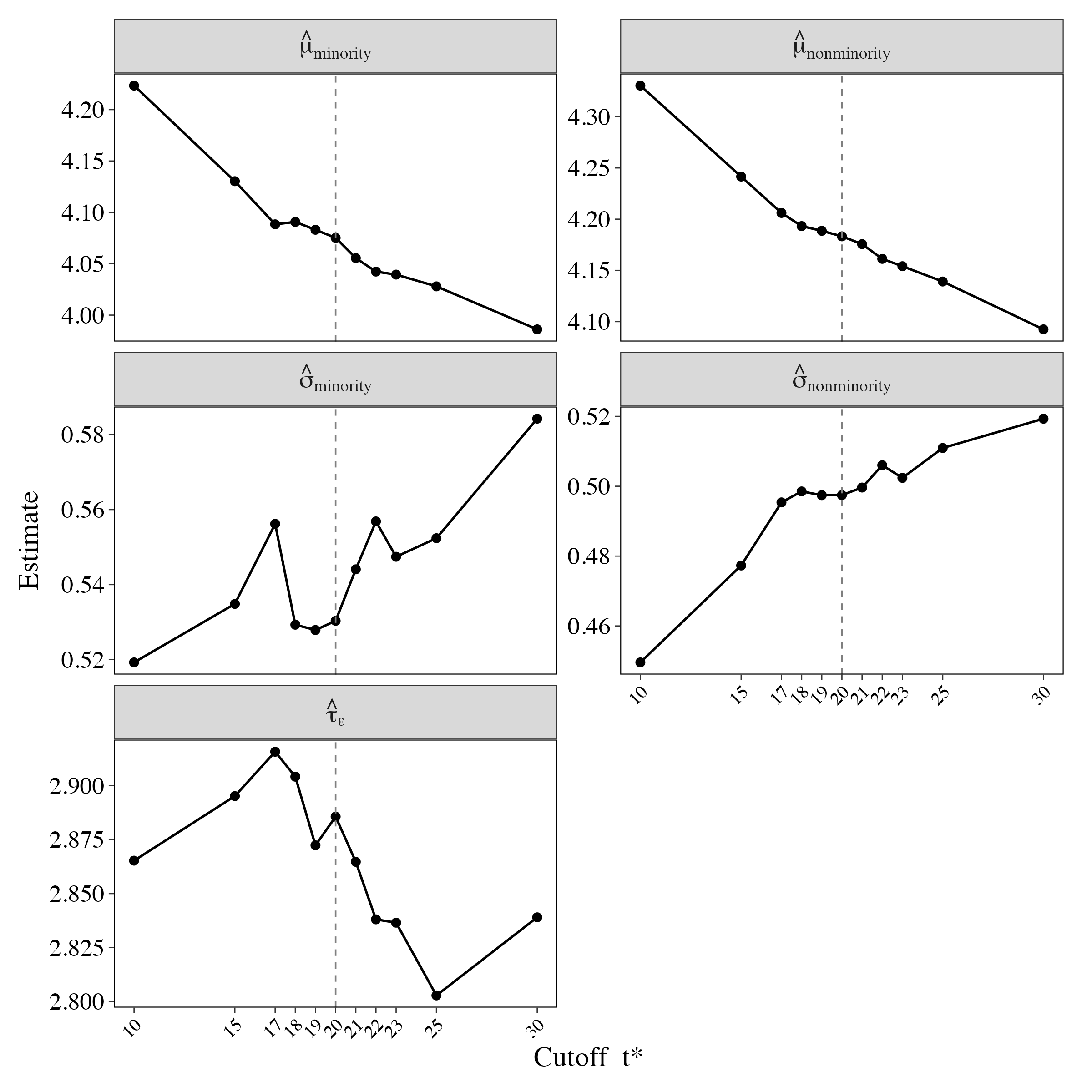}
\caption*{\footnotesize\textit{Notes:} Each panel shows one estimated parameter as a function of the burnout period $t^*$, holding the minimum number of post-$t^*$ reviews at 2. The vertical dashed line indicates the baseline choice of $t^* = 20$.}
\end{figure}

\paragraph{Sensitivity to $t^*$.} Table \ref{tab:sensitivity_theta} reports the parameter estimates for each value of $t^*$, holding the minimum post-cutoff review requirement at the baseline value of 2. The grid is centered on the baseline $t^* = 20$ and includes both aggressive early cutoffs ($t^* = 10$) and conservative late cutoffs ($t^* = 30$). Figure \ref{fig:sensitivity_theta} plots each parameter as a function of $t^*$.

\input{figs/sensitivity_table_theta}

\begin{figure}[H]
\centering
\caption{Type gap as a function of the burnout cutoff $t^*$\label{fig:sensitivity_gap}}
\includegraphics[width=0.65\textwidth]{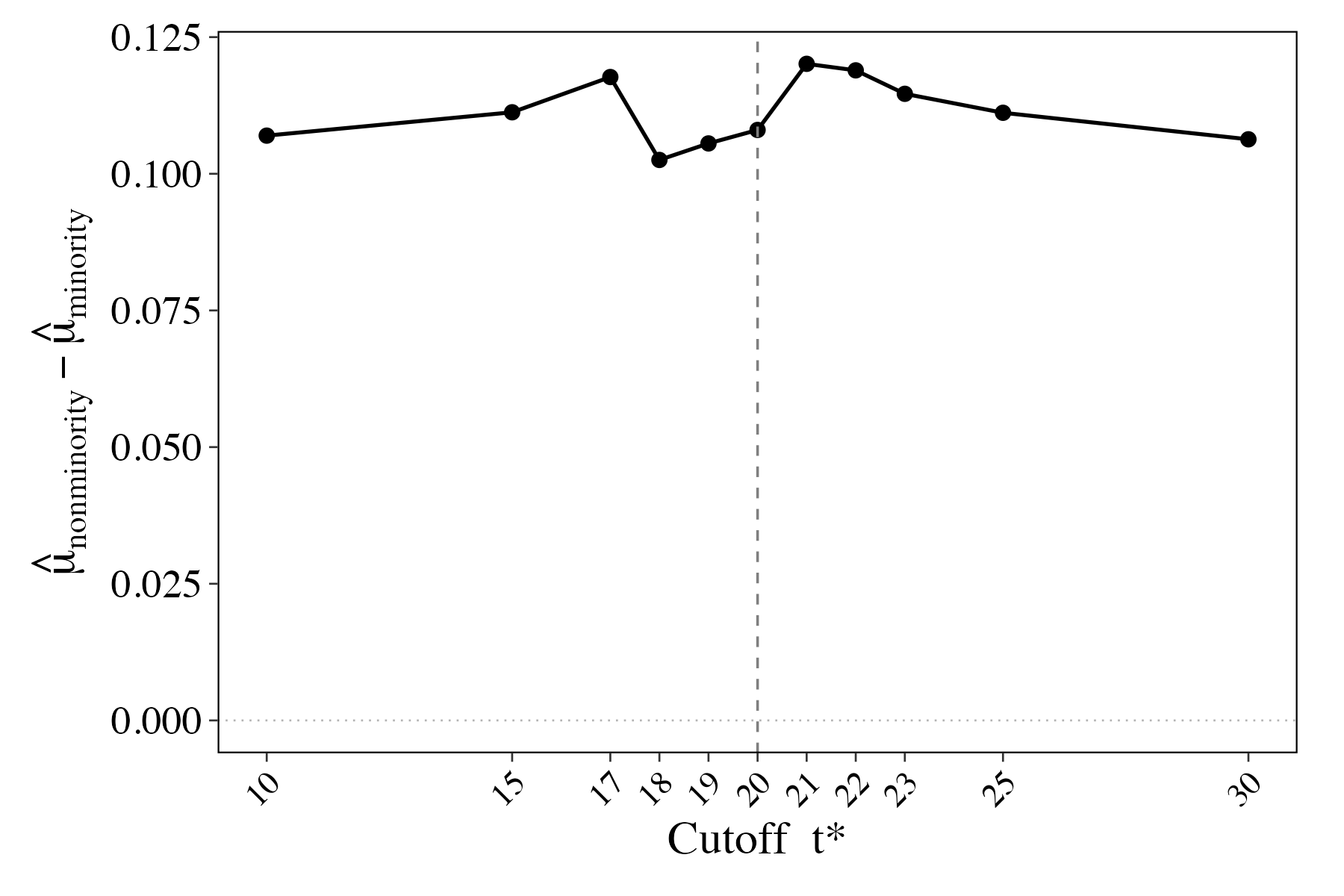}
\caption*{\footnotesize\textit{Notes:} The type gap is $\hat{\mu}_{\text{nonminority}} - \hat{\mu}_{\text{minority}}$. The vertical dashed line marks the baseline $t^* = 20$. Minimum post-$t^*$ reviews fixed at 2.}
\end{figure}

Figure \ref{fig:sensitivity_gap} isolates the type gap $\hat{\mu}_{\text{nonminority}} - \hat{\mu}_{\text{minority}}$, which is the key quantity for the paper's counterfactual analysis.

\paragraph{Sensitivity to minimum post-cutoff reviews.} Table \ref{tab:sensitivity_minpost} reports the parameter estimates for each minimum review requirement, holding $t^*$ fixed at 20. Increasing this threshold restricts the estimation sample to drivers with longer histories on the platform, which may reduce noise in the type estimates at the cost of a smaller sample.

\input{figs/sensitivity_table_minpost}

\begin{figure}[H]
\centering
\caption{Number of drivers with type estimates across cutoff combinations\label{fig:sensitivity_sample}}
\includegraphics[width=0.85\textwidth]{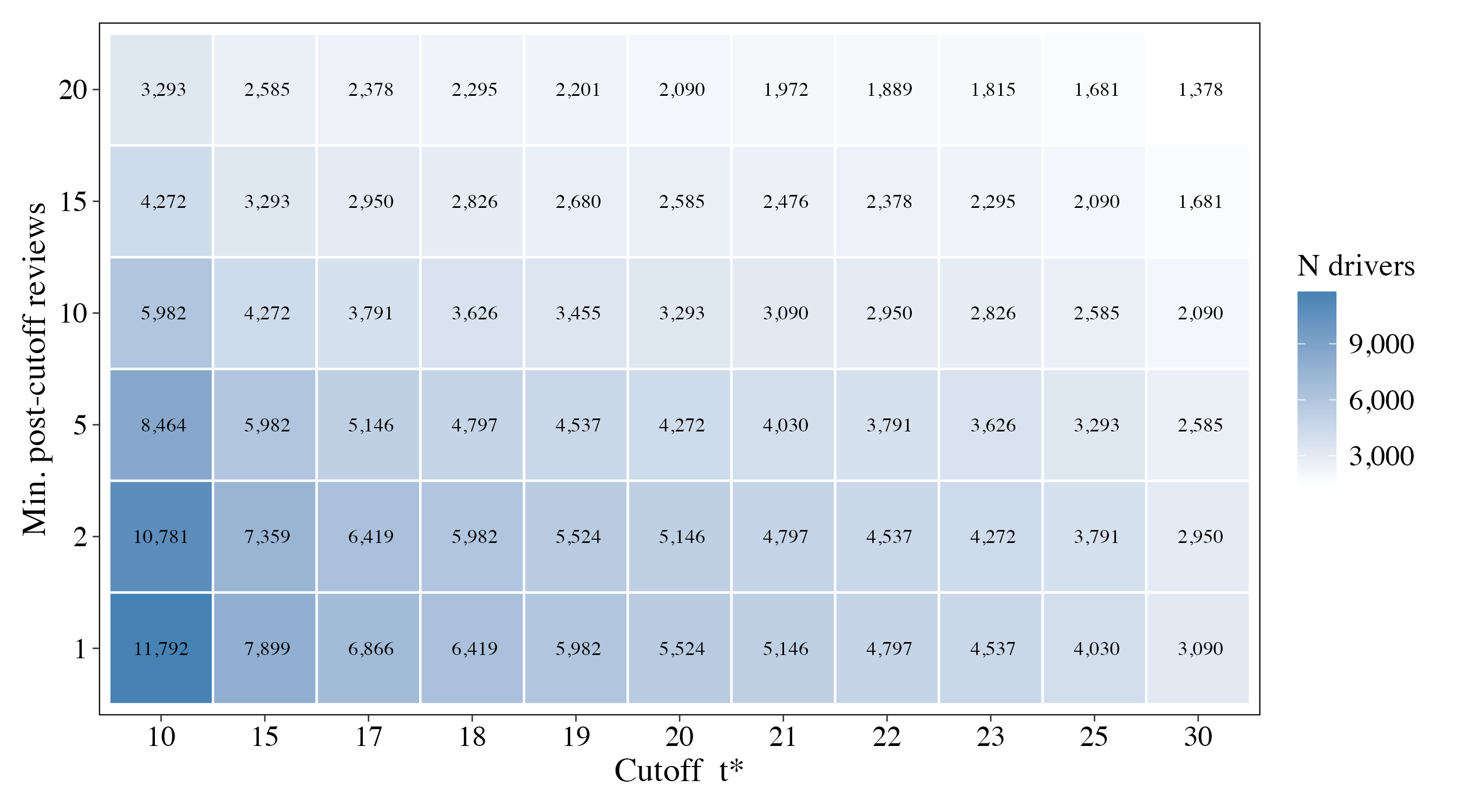}
\caption*{\footnotesize\textit{Notes:} Each cell reports the number of drivers satisfying the indicated cutoff combination. Darker shading corresponds to larger samples.}
\end{figure}

Figure \ref{fig:sensitivity_sample} displays the number of qualifying drivers across the full grid of $(t^*, \text{min. reviews})$ combinations.

\paragraph{Discussion.} The type gap $\hat{\mu}_{\text{non}} - \hat{\mu}_{\text{min}}$ is remarkably stable across the full range of $t^*$: it varies between 0.103 and 0.120, compared to the baseline value of 0.108. This stability is visible in Figure \ref{fig:sensitivity_gap}. The mean types $\hat{\mu}_m$ exhibit a gradual downward trend as $t^*$ increases---from 4.22/4.33 at $t^* = 10$ to 3.99/4.09 at $t^* = 30$---consistent with a small residual effort contamination at lower cutoffs, but the gap between groups remains essentially unchanged. The noise precision $\hat{\tau}_\epsilon$ is stable, ranging from 2.80 to 2.92 across all $t^*$ values (baseline: 2.89). The standard deviations of the type distributions increase modestly with $t^*$, reflecting the smaller samples available at higher cutoffs ($N$ drops from 10,781 at $t^* = 10$ to 2,950 at $t^* = 30$).

Across the minimum post-cutoff review dimension (Table \ref{tab:sensitivity_minpost}), the type gap remains stable at 0.105--0.116 for all thresholds. The mean types shift upward as the minimum review requirement increases---from 4.02/4.13 at 1 review to 4.35/4.47 at 20 reviews---reflecting selection toward drivers with longer platform tenure, who tend to have higher types. The noise precision $\hat{\tau}_\epsilon$ declines from 2.89 to 2.50 at more stringent thresholds. The sample size heatmap in Figure \ref{fig:sensitivity_sample} shows that reasonable combinations (e.g., $t^* \in [15, 25]$, min.\ reviews $\leq 10$) retain between 2,585 and 8,464 drivers, providing adequate statistical power.

Overall, the supply-side parameters that enter the counterfactual analysis---in particular the type gap and noise precision---are robust to reasonable perturbations of the estimation cutoffs.

%% file: figs/sensitivity_table_theta.tex
\begin{table}[H]
\centering
\caption{Supply-side parameter estimates across burnout period cutoffs\label{tab:sensitivity_theta}}
\begin{tabular}{rrrrrrrrr}
\toprule
$t^*$ & $N$ & $\hat{\mu}_{\text{min}}$ & $\hat{\mu}_{\text{non}}$ & $\hat{\sigma}_{\text{min}}$ & $\hat{\sigma}_{\text{non}}$ & $\hat{\tau}_\epsilon$ & Gap \\
\midrule
10 & 10,781 & 4.2233 & 4.3303 & 0.5192 & 0.4496 & 2.8653 & 0.1070 \\
15 & 7,359 & 4.1303 & 4.2416 & 0.5348 & 0.4773 & 2.8951 & 0.1112 \\
17 & 6,419 & 4.0884 & 4.2061 & 0.5562 & 0.4954 & 2.9157 & 0.1177 \\
18 & 5,982 & 4.0907 & 4.1932 & 0.5293 & 0.4985 & 2.9041 & 0.1025 \\
19 & 5,524 & 4.0831 & 4.1886 & 0.5279 & 0.4974 & 2.8723 & 0.1056 \\
\textbf{20} & \textbf{5,146} & \textbf{4.0752} & \textbf{4.1832} & \textbf{0.5303} & \textbf{0.4975} & \textbf{2.8856} & \textbf{0.1080} \\
21 & 4,797 & 4.0555 & 4.1756 & 0.5441 & 0.4996 & 2.8647 & 0.1201 \\
22 & 4,537 & 4.0424 & 4.1613 & 0.5569 & 0.5060 & 2.8381 & 0.1189 \\
23 & 4,272 & 4.0394 & 4.1541 & 0.5474 & 0.5024 & 2.8365 & 0.1146 \\
25 & 3,791 & 4.0280 & 4.1391 & 0.5524 & 0.5109 & 2.8028 & 0.1112 \\
30 & 2,950 & 3.9861 & 4.0924 & 0.5842 & 0.5193 & 2.8390 & 0.1063 \\
\bottomrule
\end{tabular}
\vspace{0.5em}
\caption*{\footnotesize\textit{Notes:} Each row re-estimates all supply-side parameters using the indicated burnout period $t^*$, with a minimum of 2 post-$t^*$ reviews required per driver. $N$ is the number of drivers with type estimates. Gap $= \hat{\mu}_{\text{non}} - \hat{\mu}_{\text{min}}$. The baseline specification ($t^* = 20$) is in bold.}
\end{table}

%% file: figs/sensitivity_table_minpost.tex
\begin{table}[H]
\centering
\caption{Supply-side parameter estimates across minimum post-cutoff review requirements\label{tab:sensitivity_minpost}}
\begin{tabular}{rrrrrrrrr}
\toprule
Min. reviews & $N$ & $\hat{\mu}_{\text{min}}$ & $\hat{\mu}_{\text{non}}$ & $\hat{\sigma}_{\text{min}}$ & $\hat{\sigma}_{\text{non}}$ & $\hat{\tau}_\epsilon$ & Gap \\
\midrule
1 & 5,524 & 4.0229 & 4.1282 & 0.5873 & 0.5607 & 2.8856 & 0.1053 \\
\textbf{2} & \textbf{5,146} & \textbf{4.0752} & \textbf{4.1832} & \textbf{0.5303} & \textbf{0.4975} & \textbf{2.8856} & \textbf{0.1080} \\
5 & 4,272 & 4.1665 & 4.2827 & 0.4539 & 0.4175 & 2.7236 & 0.1162 \\
10 & 3,293 & 4.2686 & 4.3789 & 0.4170 & 0.3554 & 2.6016 & 0.1103 \\
15 & 2,585 & 4.3266 & 4.4352 & 0.3820 & 0.3237 & 2.5347 & 0.1086 \\
20 & 2,090 & 4.3513 & 4.4655 & 0.3543 & 0.3093 & 2.5014 & 0.1142 \\
\bottomrule
\end{tabular}
\vspace{0.5em}
\caption*{\footnotesize\textit{Notes:} Each row re-estimates all supply-side parameters using the baseline burnout period $t^* = 20$, varying the minimum number of post-$t^*$ reviews required per driver. $N$ is the number of drivers with type estimates. Gap $= \hat{\mu}_{\text{non}} - \hat{\mu}_{\text{min}}$. The baseline specification (min.\ reviews $= 2$) is in bold.}
\end{table}

%% file: biblio.bib
@article{bohren2019dynamics,
  title={The dynamics of discrimination: Theory and evidence},
  author={Bohren, J Aislinn and Imas, Alex and Rosenberg, Michael},
  journal={American Economic Review},
  volume={109},
  number={10},
  pages={3395--3436},
  year={2019}
}

@article{Arrow1973,
  title={The theory of discrimination, discrimination in labor markets},
  author={Arrow, Kenneth J},
  journal={Achenfelter, A. Ress (eds.), Princeton--New Jersey},
  year={1973}
}

@article{farajallah2019drives,
  title={What drives pricing behavior in Peer-to-Peer markets? Evidence from the carsharing platform blablacar},
  author={Farajallah, Mehdi and Hammond, Robert G and P{\'e}nard, Thierry},
  journal={Information Economics and Policy},
  year={2019},
  publisher={Elsevier}
}

@article{ge2020racial,
  title={Racial discrimination in transportation network companies},
  author={Ge, Yanbo and Knittel, Christopher R and MacKenzie, Don and Zoepf, Stephen},
  journal={Journal of Public Economics},
  volume={190},
  pages={104205},
  year={2020},
  publisher={Elsevier}
}

@article{AstierBouquetLambin2026,
  title={Riding together: eliciting travelers' preferences for long-distance carpooling},
  author={Astier, Nicolas and Bouquet, Pierre-Fran{\c{c}}ois and Lambin, Xavier},
  journal={Available at SSRN 4360029},
  year={2023}
}

@article{Altonji2001,
  title={Employer learning and statistical discrimination},
  author={Altonji, Joseph G and Pierret, Charles R},
  journal={The Quarterly Journal of Economics},
  volume={116},
  number={1},
  pages={313--350},
  year={2001},
  publisher={MIT Press}
}

@article{Edelman2017,
author = {Edelman, BG; and Luca, Michael; and Svirsky, Dan},
doi = {10.2139/ssrn.2701902},
issn = {1556-5068},
journal = {American Economic Journal: Applied Economics},
number = {2},
pages = {1--22},
title = {{Racial Discrimination in the Sharing Economy: Evidence from a Field Experiment}},
volume = {9},
year = {2017}
}

@article{Zervas2015,
archivePrefix = {arXiv},
arxivId = {1606.07138},
author = {Zervas, G and Proserpio, D and Byers, J},
doi = {10.1080/13683500.2013.827159},
eprint = {1606.07138},
isbn = {1889-0326},
issn = {1368-3500},
journal = {Working paper},
pages = {1--22},
title = {{A First Look at Online Reputation on Airbnb, Where Every Stay is Above Average}},
url = {http://papers.ssrn.com/sol3/papers.cfm?abstract{\_}id=2554500},
year = {2015}
}

@article{Edelman2014,
author = {Edelman, Benjamin G. and Luca, Michael},
doi = {10.2139/ssrn.2377353},
issn = {1556-5068},
journal = {SSRN Electronic Journal},
title = {{Digital Discrimination: The Case of Airbnb.com}},
url = {http://www.ssrn.com/abstract=2377353},
year = {2014}
}

@article{laouenan2022can,
  title={Can information reduce ethnic discrimination? Evidence from Airbnb},
  author={Laou{\'e}nan, Morgane and Rathelot, Roland},
  journal={American Economic Journal: Applied Economics},
  volume={14},
  number={1},
  pages={107--132},
  year={2022},
  publisher={American Economic Association 2014 Broadway, Suite 305, Nashville, TN 37203-2425}
}

@techreport{blablacar2016,
author = {Mazzella, Fr{\'{e}}d{\'{e}}ric and Sundararajan, Arun},
booktitle = {BlaBlaCar},
pages = {23},
title = {{Entering the Trust Age}},
url = {https://www.blablacar.com/wp-content/uploads/2016/05/entering-the-trust-age.pdf{\%}250Ahttps://www.blablacar.com/wp-content/uploads/2016/05/entering-the-trust-age.pdf{\%}255Cnhttps://d2hqsaoq1cfj21.cloudfront.net/blogstatics/files/Entering the Trust Age - 2016.},
year = {2016}
}

@article{Coate1993,
author = {Coate, Stephen and Loury, Glenn C.},
doi = {10.2307/2117558},
isbn = {0002-8282},
issn = {00028282},
journal = {The American Economic Review},
number = {5},
pages = {1220--1240},
pmid = {17746758},
title = {{Will Affirmative-Action Policies Eliminate Negative Stereotypes ?}},
volume = {83},
year = {1993}
}

@article{Holmstrom1999,
  title={Managerial incentive problems: A dynamic perspective},
  author={Holmstr{\"o}m, Bengt},
  journal={The Review of Economic Studies},
  volume={66},
  number={1},
  pages={169--182},
  year={1999},
  publisher={Wiley-Blackwell}
}

@article{Phelps1972,
archivePrefix = {arXiv},
arxivId = {arXiv:1011.1669v3},
author = {Phelps, Edmund S},
doi = {10.2307/1806107},
eprint = {arXiv:1011.1669v3},
isbn = {0002-8282},
issn = {00028282},
journal = {American Economic Review},
number = {4},
pages = {659--661},
pmid = {17746758},
title = {{The Statistical theory of Racism and Sexism}},
volume = {62},
year = {1972}
}

@article{chiappori1999early,
  title={Early starters versus late beginners},
  author={Chiappori, Pierre-Andre and Salanie, Bernard and Valentin, Julie},
  journal={Journal of Political Economy},
  volume={107},
  number={4},
  pages={731--760},
  year={1999},
  publisher={The University of Chicago Press}
}

@article{abrahao2017reputation,
  title={Reputation offsets trust judgments based on social biases among Airbnb users},
  author={Abrahao, Bruno and Parigi, Paolo and Gupta, Alok and Cook, Karen S},
  journal={Proceedings of the National Academy of Sciences},
  volume={114},
  number={37},
  pages={9848--9853},
  year={2017},
  publisher={National Acad Sciences}
}

@article{tjaden2018ride,
  title={Ride with Me -Ethnic Discrimination, Social Markets, and the Sharing Economy},
  author={Tjaden, Jasper Dag and Schwemmer, Carsten and Khadjavi, Menusch},
  journal={European Sociological Review},
  volume={34},
  number={4},
  pages={418--432},
  year={2018},
  publisher={Oxford University Press}
}

@article{alesina2016ethnic,
  title={Ethnic inequality},
  author={Alesina, Alberto and Michalopoulos, Stelios and Papaioannou, Elias},
  journal={Journal of Political Economy},
  volume={124},
  number={2},
  pages={428--488},
  year={2016},
  publisher={University of Chicago Press Chicago, IL}
}

@article{becker1971economics,
  title={The Economics of Discrimination},
  author={Becker, Gary},
  journal={University of Chicago Press Economics Books},
  year={1971},
  publisher={University of Chicago Press}
}

@article{banerjee2004efficiently,
  title={How efficiently is capital allocated? Evidence from the knitted garment industry in Tirupur},
  author={Banerjee, Abhijit and Munshi, Kaivan},
  journal={The Review of Economic Studies},
  volume={71},
  number={1},
  pages={19--42},
  year={2004},
  publisher={Wiley-Blackwell}
}

@article{hjort2014ethnic,
  title={Ethnic divisions and production in firms},
  author={Hjort, Jonas},
  journal={The Quarterly Journal of Economics},
  volume={129},
  number={4},
  pages={1899--1946},
  year={2014},
  publisher={MIT Press}
}

@article{bartovs2016attention,
  title={Attention discrimination: Theory and field experiments with monitoring information acquisition},
  author={Barto{\v{s}}, Vojt{\v{e}}ch and Bauer, Michal and Chytilov{\'a}, Julie and Mat{\v{e}}jka, Filip},
  journal={American Economic Review},
  volume={106},
  number={6},
  pages={1437--75},
  year={2016}
}

@article{agrawal2016does,
  title={Does standardized information in online markets disproportionately benefit job applicants from less developed countries?},
  author={Agrawal, Ajay and Lacetera, Nicola and Lyons, Elizabeth},
  journal={Journal of international Economics},
  volume={103},
  pages={1--12},
  year={2016},
  publisher={Elsevier}
}

@article{cui2019reducing,
  title={Reducing discrimination with reviews in the sharing economy: Evidence from field experiments on Airbnb},
  author={Cui, Ruomeng and Li, Jun and Zhang, Dennis J},
  journal={Management Science},
  year={2019},
  publisher={INFORMS}
}

@article{bohren2025inaccurate,
  title={Inaccurate Statistical Discrimination: An Identification Problem.},
  author={Bohren, J Aislinn and Haggag, Kareem and Imas, Alex and Pope, Devin G},
  journal={Review of Economics \& Statistics},
  volume={107},
  number={3},
  pages={605},
  year={2025}
}

@article{glover2017discrimination,
  title={Discrimination as a self-fulfilling prophecy: Evidence from French grocery stores},
  author={Glover, Dylan and Pallais, Amanda and Pariente, William},
  journal={The Quarterly Journal of Economics},
  volume={132},
  number={3},
  pages={1219--1260},
  year={2017},
  publisher={Oxford University Press}
}

@article{pallais2014inefficient,
  title={Inefficient hiring in entry-level labor markets},
  author={Pallais, Amanda},
  journal={American Economic Review},
  volume={104},
  number={11},
  pages={3565--99},
  year={2014}
}

@article{mcfadden1974frontiers,
  title={Conditional logit analysis of qualitative choice behavior},
  author={McFadden, Daniel},
  year={1974},
  journal={Frontiers in Econometrics},
  publisher={Academic Press}
}

@article{kuznets1955economic,
  title={Economic growth and income inequality},
  author={Kuznets, Simon},
  journal={The American economic review},
  volume={45},
  number={1},
  pages={1--28},
  year={1955},
  publisher={JSTOR}
}

@article{carol2019can,
  title={Who can ride along? Discrimination in a German carpooling market},
  author={Carol, Sarah and Eich, Daniel and Keller, Mich{\`e}le and Steiner, Friederike and Storz, Katharina},
  journal={Population, Space and Place},
  year={2019},
  publisher={Wiley Online Library}
}

@article{tadelis2016reputation,
  title={Reputation and feedback systems in online platform markets},
  author={Tadelis, Steven},
  journal={Annual Review of Economics},
  volume={8},
  pages={321--340},
  year={2016},
  publisher={Annual Reviews}
}

@article{kakar2018visible,
  title={The visible host: Does race guide Airbnb rental rates in San Francisco?},
  author={Kakar, Venoo and Voelz, Joel and Wu, Julia and Franco, Julisa},
  journal={Journal of Housing Economics},
  volume={40},
  pages={25--40},
  year={2018},
  publisher={Elsevier}
}

@article{sant2020doubly,
  title={Doubly robust difference-in-differences estimators},
  author={Sant’Anna, Pedro HC and Zhao, Jun},
  journal={Journal of Econometrics},
  volume={219},
  number={1},
  pages={101--122},
  year={2020},
  publisher={Elsevier}
}

@article{bertrand2004emily,
  title={Are Emily and Greg more employable than Lakisha and Jamal? A field experiment on labor market discrimination},
  author={Bertrand, Marianne and Mullainathan, Sendhil},
  journal={American economic review},
  volume={94},
  number={4},
  pages={991--1013},
  year={2004},
  publisher={American Economic Association}
}

@article{gaddis2017black,
  title={How black are Lakisha and Jamal? Racial perceptions from names used in correspondence audit studies},
  author={Gaddis, S Michael},
  journal={Sociological Science},
  volume={4},
  pages={469},
  year={2017},
  publisher={Society for Sociological Science}
}

@article{fryer2004causes,
  title={The causes and consequences of distinctively black names},
  author={Fryer Jr, Roland G and Levitt, Steven D},
  journal={The Quarterly Journal of Economics},
  volume={119},
  number={3},
  pages={767--805},
  year={2004},
  publisher={MIT Press}
}

@article{pope2011s,
  title={What’s in a Picture?: Evidence of Discrimination from Prosper. com},
  author={Pope, Devin G and Sydnor, Justin R},
  journal={Journal of Human resources},
  volume={46},
  number={1},
  pages={53--92},
  year={2011},
  publisher={University of Wisconsin Press}
}

@techreport{athey2022smiles,
  title={Smiles in profiles: Improving fairness and efficiency using estimates of user preferences in online marketplaces},
  author={Athey, Susan and Karlan, Dean and Palikot, Emil and Yuan, Yuan},
  year={2022},
  institution={National Bureau of Economic Research}
}

@article{zhang2022reducing,
  title={Reducing racial discrimination in the sharing economy: Empirical results from Airbnb},
  author={Zhang, Leihan and Xiong, Shengyu and Zhang, Le and Bai, Lin and Yan, Qiang},
  journal={International Journal of Hospitality Management},
  volume={102},
  pages={103151},
  year={2022},
  publisher={Elsevier}
}

@article{jaeger2020automated,
  title={Automated classification of demographics from face images: A tutorial and validation},
  author={Jaeger, Bastian and Sleegers, Willem WA and Evans, Anthony M},
  journal={Social and Personality Psychology Compass},
  volume={14},
  number={3},
  pages={e12520},
  year={2020},
  publisher={Wiley Online Library}
}

@article{athey2019estimating,
  title={Estimating treatment effects with causal forests: An application},
  author={Athey, Susan and Wager, Stefan},
  journal={Observational studies},
  volume={5},
  number={2},
  pages={37--51},
  year={2019},
  publisher={University of Pennsylvania Press}
}

@article{robins1994estimation,
  title={Estimation of regression coefficients when some regressors are not always observed},
  author={Robins, James M and Rotnitzky, Andrea and Zhao, Lue Ping},
  journal={Journal of the American statistical Association},
  volume={89},
  number={427},
  pages={846--866},
  year={1994},
  publisher={Taylor \& Francis}
}

@article{wager2018estimation,
  title={Estimation and inference of heterogeneous treatment effects using random forests},
  author={Wager, Stefan and Athey, Susan},
  journal={Journal of the American Statistical Association},
  volume={113},
  number={523},
  pages={1228--1242},
  year={2018},
  publisher={Taylor \& Francis}
}

@article{tibshirani2025grf,
  title={grf: Generalized random forests},
  author={Tibshirani, Julie and Athey, Susan and Friedberg, Rina and Hadad, Vitor and Hirshberg, David and Miner, Luke and Sverdrup, Erik and Wager, Stefan and Wright, Marvin},
  year={2025}
}

@article{crump2009dealing,
  title={Dealing with limited overlap in estimation of average treatment effects},
  author={Crump, Richard K and Hotz, V Joseph and Imbens, Guido W and Mitnik, Oscar A},
  journal={Biometrika},
  volume={96},
  number={1},
  pages={187--199},
  year={2009},
  publisher={Oxford University Press}
}

@article{petrin2010control,
  title={A control function approach to endogeneity in consumer choice models},
  author={Petrin, Amil and Train, Kenneth},
  journal={Journal of marketing research},
  volume={47},
  number={1},
  pages={3--13},
  year={2010},
  publisher={SAGE Publications Sage CA: Los Angeles, CA}
}

@article{WeintraubBenkardVanRoy2008,
  title={Markov Perfect Industry Dynamics with Many Firms},
  author={Weintraub, Gabriel Y. and Benkard, C. Lanier and Van Roy, Benjamin},
  journal={Econometrica},
  volume={76},
  number={6},
  pages={1375--1411},
  year={2008}
}

@article{WeintraubBenkardVanRoy2010,
  title={Computational Methods for Oblivious Equilibrium},
  author={Weintraub, Gabriel Y. and Benkard, C. Lanier and Van Roy, Benjamin},
  journal={Operations Research},
  volume={58},
  number={4-Part-2},
  pages={1247--1265},
  year={2010}
}

@article{luca2026evolution,
  title={The evolution of discrimination in online markets: How the rise in anti-Asian bias affected Airbnb during the pandemic},
  author={Luca, Michael and Pronkina, Elizaveta and Rossi, Michelangelo},
  journal={Marketing Science},
  volume={45},
  number={1},
  pages={108--122},
  year={2026},
  publisher={INFORMS}
}

@article{cook2021gender,
  title={The gender earnings gap in the gig economy: Evidence from over a million rideshare drivers},
  author={Cook, Cody and Diamond, Rebecca and Hall, Jonathan V and List, John A and Oyer, Paul},
  journal={The Review of Economic Studies},
  volume={88},
  number={5},
  pages={2210--2238},
  year={2021},
  publisher={Oxford University Press}
}

@article{cobb2024ride,
  title={Ride-hailing technology mitigates effects of driver racial discrimination, but effects of residential segregation persist},
  author={Cobb, Anna and Mohan, Aniruddh and Harper, Corey D and Nock, Destenie and Michalek, Jeremy},
  journal={Proceedings of the National Academy of Sciences},
  volume={121},
  number={41},
  pages={e2408936121},
  year={2024},
  publisher={National Academy of Sciences}
}

@article{coffman2021role,
  title={The role of beliefs in driving gender discrimination},
  author={Coffman, Katherine B and Exley, Christine L and Niederle, Muriel},
  journal={Management Science},
  volume={67},
  number={6},
  pages={3551--3569},
  year={2021},
  publisher={INFORMS}
}

@article{barron2025explicit,
  title={Explicit and implicit belief-based gender discrimination: A hiring experiment},
  author={Barron, Kai and Ditlmann, Ruth and Gehrig, Stefan and Schweighofer-Kodritsch, Sebastian},
  journal={Management Science},
  volume={71},
  number={2},
  pages={1600--1622},
  year={2025},
  publisher={INFORMS}
}

@article{bohren2025systemic,
  title={Systemic discrimination: Theory and measurement},
  author={Bohren, J Aislinn and Hull, Peter and Imas, Alex},
  journal={The Quarterly Journal of Economics},
  volume={140},
  number={3},
  pages={1743--1799},
  year={2025},
  publisher={Oxford University Press}
}

@article{coffman2024stereotypes,
  title={Stereotypes and belief updating},
  author={Coffman, Katherine and Collis, Manuela R and Kulkarni, Leena},
  journal={Journal of the European Economic Association},
  volume={22},
  number={3},
  pages={1011--1054},
  year={2024},
  publisher={Oxford University Press}
}

@article{ruzzier2023discrimination,
  title={Discrimination with inaccurate beliefs and confirmation bias},
  author={Ruzzier, Christian A and Woo, Marcelo D},
  journal={Journal of Economic Behavior \& Organization},
  volume={210},
  pages={379--390},
  year={2023},
  publisher={Elsevier}
}

@inproceedings{filippas2018reputation,
  title={Reputation inflation},
  author={Filippas, Apostolos and Horton, John Joseph and Golden, Joseph},
  booktitle={Proceedings of the 2018 ACM Conference on Economics and Computation},
  pages={483--484},
  year={2018}
}

@article{botelho2025scale,
  title={Scale dichotomization reduces customer racial discrimination and income inequality},
  author={Botelho, Tristan L and Jun, Sora and Humes, Demetrius and DeCelles, Katherine A},
  journal={Nature},
  volume={639},
  number={8054},
  pages={395--403},
  year={2025},
  publisher={Nature Publishing Group UK London}
}

@article{alyakoob2026market,
  title={Market design choices, racial discrimination, and equitable microentrepreneurship in digital marketplaces},
  author={Alyakoob, Mohammed and Rahman, Mohammad},
  journal={Management Science},
  volume={72},
  number={3},
  pages={1878--1903},
  year={2026},
  publisher={INFORMS}
}

@article{park2023fighting,
  title={Fighting bias with bias: How same-race endorsements reduce racial discrimination on Airbnb},
  author={Park, Minsu and Yu, Chao and Macy, Michael},
  journal={Science Advances},
  volume={9},
  number={6},
  pages={eadd2315},
  year={2023},
  publisher={American Association for the Advancement of Science}
}

@article{li2020buying,
  title={Buying reputation as a signal of quality: Evidence from an online marketplace},
  author={Li, Lingfang and Tadelis, Steven and Zhou, Xiaolan},
  journal={The RAND Journal of Economics},
  volume={51},
  number={4},
  pages={965--988},
  year={2020},
  publisher={Wiley Online Library}
}

@article{son2025does,
  title={Does greater visibility benefit minority businesses? Evidence from an online review platform},
  author={Son, Yoonseock and Wowak, Kaitlin D and Angst, Corey M},
  journal={Production and Operations Management},
  volume={34},
  number={4},
  pages={711--724},
  year={2025},
  publisher={SAGE Publications Sage CA: Los Angeles, CA}
}

@article{lepage2021endogenous,
  title={Endogenous learning, persistent employer biases, and discrimination},
  author={Lepage, Louis Pierre},
  journal={Persistent Employer Biases, and Discrimination (March 2, 2021)},
  year={2021}
}

@article{benson2024learning,
  title={Learning to Discriminate on the Job},
  author={Benson, Alan and Lepage, Louis Pierre},
  journal={Available at SSRN 4155065},
  year={2024}
}

@article{komiyama2026statistical,
  title={On statistical discrimination as a failure of social learning: A multiarmed bandit approach},
  author={Komiyama, Junpei and Noda, Shunya},
  journal={Management Science},
  volume={72},
  number={1},
  pages={442--455},
  year={2026},
  publisher={INFORMS}
}

@article{li2026hiring,
  title={Hiring as exploration},
  author={Li, Danielle and Raymond, Lindsey and Bergman, Peter},
  journal={Review of Economic Studies},
  volume={93},
  number={2},
  pages={1200--1240},
  year={2026},
  publisher={Oxford University Press UK}
}

@article{donkor2026leveling,
  title={Leveling Down: Competition and Discrimination in Service Markets},
  author={Donkor, Kwabena},
  journal={Available at SSRN 6171007},
  year={2026}
}
